\documentclass[twoside]{report}
\usepackage{psfig}

\usepackage{natbib}
\bibpunct{(}{)}{,}{a}{}{;}

\begin{document}


\thispagestyle{empty}
\begin{center}

~\vspace{2.5cm}

\Huge 
{\bf Modelling the Far-Infrared Emission

in Spiral Galaxies}

\Large
\vspace{1cm}
by 
\vspace{1cm}

{\bf \huge Simone Bianchi}

\vfill

\large
A thesis submitted to the 

University of Wales

for the degree of

Doctor of Philosophy 

\vspace{1cm}
{\bf September 1999}

\end{center}

\newpage
\thispagestyle{empty}
~\newpage


\chapter*{Acknowledgements}
\thispagestyle{empty}

It is a pleasure to thank my supervisor, Jon Davies, for the help
he continuously gave me during the three years of my PhD project,
not to speak about his contagious enthusiasm. I have benefit a lot
from the discussions and comments of the past and current members of our 
research group, Paul Alton, Lea Morshidi, Matthew Trewhella and 
Alexandros Kambas. A particular thank goes to Rodney Smith, for not
having shouted once during my continuous reports of real (and sometimes
fictitious) problems with the computer, usually followed by a request of
immediate solution; and to Judy Haynes, for having spared me months of 
data reduction. Among all the other people that have helped me during these
three years, I would like to remember Andrea Ferrara, for
suggesting new directions of investigation, and Spyros Kitsionas,
Phillip Gladwin and Neil Francis, for the numerous hints they gave me.

\vspace{0.5cm}
\noindent Un grazie particolare va ai molti amici italiani, espa\~noles,
latinoamericanos, 
$\grave{\,}\! E\lambda\lambda\eta\nu\epsilon\varsigma$,
portugueses e di molti altri paesi, che mi hanno aiutato a passare
felicemente questi tre lunghi anni di permanenza in Galles.
Vorrei infine dedicare questa tesi alla mia famiglia e ai miei amici
di sempre.  {\em Cu!}

\newpage
\thispagestyle{empty}
~\newpage


\chapter*{Summary}
\thispagestyle{empty}

The dust distributions observed in spiral galaxies play a major role in
Astrophysics. Dust very effectively extinguishes UV and optical 
starlight.  Therefore it may alter considerably our view of the galaxy 
itself and of the distant universe in its background. The dust opacity in 
spiral galaxies is still a debated issue. Since the energy absorbed by 
dust grains from starlight is re-emitted at longer wavelengths, mainly
in the Far-Infrared (FIR) and Sub-millimetre ($\lambda >$ 60-$\mu m$),
observations of dust emission can help to constrain the parameters of
the dust distribution.

I have developed an original model for the FIR emission in spirals,
starting from an existing radiative transfer code \citep*{BianchiApJ1996}.
The model's main features are: a complete treatment of multiple 
scattering within geometries appropriate for spirals; a full coverage of
spectral range of stellar emission; the use of empirically determined dust 
properties (some of which are derived in the present work); the production
of maps of the dust temperature distribution, together with simulated optical,
FIR and sub-millimetre images. The model has been applied to
observations of stellar and dust emission in the galaxy NGC~6946.

It is found that optically thick models (central face-on optrical depth
$\tau_\mathrm{V}\sim5$) are 
necessary to explain the observed FIR output. For such models, almost
30-40\% of the intrinsic starlight is absorbed. The observed ratio of
FIR and optical scalelengths can be explained if the dust distribution
is more extended than the stellar. However, because of the less steep
gradients of optical emission in optically thick cases, a very extended
dust distribution is needed ($\alpha_\mathrm{d}\sim 3\alpha_\star$). The
distribution of atomic gas in NGC~6946 has a similar extent.

I discuss the approximations in the modelling (mainly the use of smooth
distributions against the observed clumpiness of the interstellar
medium) and the implications of the results.

\newpage
\thispagestyle{empty}
~\newpage

\pagenumbering{roman}
\tableofcontents
\chapter{Introduction}
\pagenumbering{arabic}

\label{intro}

\begin{figure}[b]
\centerline{\psfig{figure=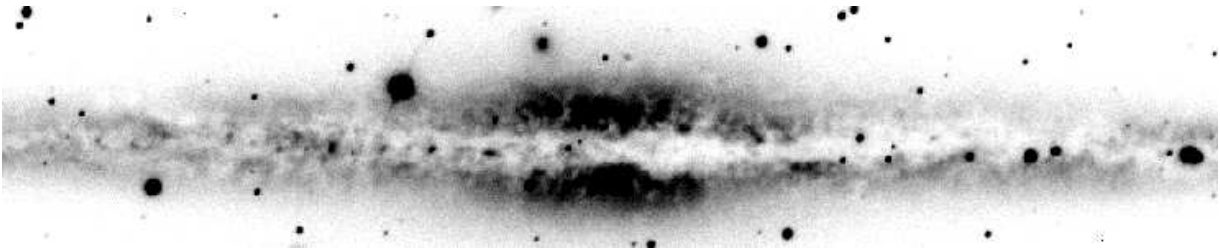,height=3.2cm}}
\caption{B-band image of the dusty edge-on spiral galaxy NGC~891. 
The extinction caused by dust is clearly visible in the 
regions of lower surface brightness (coded with brighter tones)
along the galactic plane.
Image taken in November 96 (Appendix~\ref{unlucky}).}
\label{b_intro}
\end{figure}

Observations of our Galaxy, as well as of other spirals (Fig.~\ref{b_intro}) 
reveal the presence of regions of sky darker than the surroundings. This 
apparent decrease in the number of stars is caused by one of the constituent 
of the Inter-Stellar Medium (ISM), dust. Dust is made of small (mean
radius 0.1$\mu$m, \citealt{HildebrandQJRAS1983}) solid grains possibly
made of silicates and graphite \citep{DraineApJ1984}. It constitutes
only a tiny fraction of the ISM: for instance, in the Solar 
neighbourhood, the mass of dust is less than 1\% of the mass of the gas 
(Sect.~\ref{assu_ext}).
Despite its relatively small abundance, dust plays a major role in 
astrophysics. Dust grains are very effective in extinguishing Ultra-Violet
and Optical ($\lambda<1\mu$m) starlight because the radiation wavelength
is of the same order of the grain size (Sect.~\ref{assu_ext}).
Our view of the universe, being mainly based on optical observation, can
therefore be severely biased by dust extinction.

Dust's ability to extinguish radiation is usually quantified by the 
{\em extinction} $A_\lambda$, i.e. the ratio between the observed and the 
intrinsic unextinguished luminosity, in a magnitude scale. In the case of 
dust laying between a light source and the observer, the extinction is 
approximately equal to the {\em optical depth} $\tau_\lambda$, the inverse 
of the mean free path of light in a dusty medium (See Sect.~\ref{extinction} 
for mathematical definitions).
Extinction and optical depth depend on the wavelength $\lambda$, the effect 
of dust being larger for smaller wavelengths (Sect.~\ref{assu_ext}).
The higher transparency of a dusty medium for radiation at large wavelength 
rather than for short wavelengths, goes under the name of {\em reddening}.
A medium is defined as {\em optically thin} for a radiation of
wavelength $\lambda$ if $\tau_\lambda< 1$, the amount of dust not reducing 
drastically the source radiation, or {\em optically
thick} otherwise. Within the Galaxy, an extinction in the V band
$A_V=0.1-0.2$ mag is observed in the direction of the poles, while in
the direction of the Galactic centre it reaches $A_V\approx 30$
\citep{WhittetBook1992}.
While the relative transparency out of the Galactic plane has permitted
us to observe extragalactic objects, the high opacity along the plane
has severely biased the first determination of shape and dimension of
the Galaxy \citep{WhittetBook1992}. 

Dust extinction in the Galaxy is directly assessed through studies of 
the obscuration of individual stars of known intrinsic luminosity. 
This is not possible in external galaxies. 
For these objects, estimates of the extinction of galactic light by dust
in its own ISM (usually referred to as {\em internal} extinction) rely
on a comparison of the observed luminosity profiles with models of
radiative transfer. Realistic models are necessary, to avoid 
misinterpretations and mutually exclusive results
\citep*{DisneyMNRAS1989}. Among the requirements of realistic models,
the choice of stellar-dust geometries appropriate to galaxies and the
inclusion of light {\em scattering} by dust in the radiative transfer 
are vital (Sect.~\ref{modelet}). A brief review of extinction studies is
presented in Sect.~\ref{extstud}.

The stellar radiation absorbed by dust is re-emitted at infrared
wavelengths, mainly in the Mid-Infrared (5$\mu$m-60$\mu$m; MIR) and in 
the Far-Infrared (60$\mu$m-300$\mu$m; FIR) spectral ranges. Dust emission 
has been observed in our Galaxy as well as in the other spirals
(Sect.~\ref{missing} and Sect.~\ref{colddust}). In our Galaxy,
10-30\% of the total Galactic bolometric luminosity is emitted by dust
\citep{WhittetBook1992}. The Infrared Astronomical Satellite (IRAS)
has revealed that the Galactic dust emission is characterised
by regions of Star-Formation, with dust at higher temperatures because
of the closeness to the radiation sources (Sect.~\ref{emission}),
and diffuse, thin clouds (often denoted as {\em cirrus}) of colder dust, 
heated by a diffuse Inter-Stellar Radiation Field (ISRF)
\citep{BeichmanARA&A1987}. The cold diffuse {\em cirrus} dust, due to its
ubiquity, is responsible of the interstellar extinction.
As for external galaxies, studies of dust emission are limited by the
instrument resolution, sensitivity and spectral range observed
(Sect.~\ref{missing}). However, the recent technological development
(mainly in the Sub-millimetre and millimetre spectral ranges for
$\lambda>300\mu$m) has permitted the observation of cold dust
responsible for extinction in external galaxies as well 
(Sect.~\ref{colddust}).

Because of the direct link between dust emission and extinction, it is 
possible in principle to derive the quantity of dust in a spiral galaxy 
by comparing the observed stellar luminosity with the FIR emission, if an 
accurate radiative transfer model is used. For this Thesis, I have modified 
an existing radiative transfer code for spiral galaxies 
\citep*{BianchiApJ1996} to model dust emission in the FIR. The observed 
Spectral Energy Distributions (SEDs) of stellar and dust emission, as well 
as their spatial distribution, will be compared to the model output to gain
clues about the galaxy dust content and star-dust relative geometry.

In this Chapter I will describe observations of extinction and FIR
emission, introducing the main topics that will be discussed throughout
the rest of the Thesis. A plan of the Thesis is presented at the end of
this Chapter.
A brief discussion of the relevance of dust studies to the understanding
of spiral galaxies and distant universe is given in the next Section.

\section{Relevance of dust studies to extra-galactic astronomy}

Dust plays a very important role in many astrophysical processes,
from the formation of molecular gas, that is believed to combine 
on grain surfaces, to the obscuration of the distant universe.
Without the pretension of being complete, I discuss now a few 
problems that may benefit from a proper knowledge
of dust distribution and amount.

\begin{description}
\item[Galactic Properties:] The study of galactic morphology depends,
obviously, on the observed radiation. Galactic properties, like
luminosity and dimension, may be severely biased by dust extinction.
Objects with the same intrinsic properties but different dust
distribution properties may look  of different type, thus prejudicing
any morphological classification based on the optical aspect.
As an example, \citet{TrewhellaMNRAS1998} found that an Sc galaxy
observed in the B-band, NGC~6946, is similar to an Sb when a
correction for extinction, from his model, is applied.
Because of the selective extinction with the wavelength, ages of distant
objects inferred from broad-band colours may be biased by the
reddening introduced by dust \citep{CimattiMNRAS1997}.

\item[Dark Matter:] Rotation curves of spiral galaxies derived from
atomic Hydrogen observation can be used to infer the galactic mass.
The mass of a galaxy derived from the luminosity, assuming a constant 
mass-to-luminosity ratio, is always smaller than that derived from the 
gravitational studies:
large amounts of dark matter are present. At present, dark matter is
unexplained. Although it is improbable that dark matter is due to a large 
underestimate of the stellar content due to extinction, dust emission may 
trace a possible extended halo component of cold gas, that can account
for some of the unseen mass \citep{GerhardApJ1996}.

\item[Obscuration of the distant Universe:] Extinction due to dust in
foreground objects may be able to explain the fall-off in the number of
detected objects at large redshifts \citep{OstrikerApJ1984,FallApJ1993}.
If extended distributions of dust are present (Sect.~\ref{extdust}) the
effect may be stronger than believed.

\item[Tully Fisher relation:] the relation between the HI line width at
21cm and the galactic luminosity is used to derive the object distance
from its apparent magnitude. Because of dust, corrections are necessary
to bring the luminosities of objects with different inclination to a
common face-on value (Sect.~\ref{extstud}). The Tully-Fisher relation in
the optical band presents a large scatter, mainly because of dust 
extinction. Although the scatter is significantly reduced using
luminosities in the less extinguished Near-Infrared (NIR), extinction
correction may still be necessary \citep*{MoriondoA&A1998}.

\item[Star Formation:] ultraviolet and blue fluxes are used to
derive the star-formation history of the Universe \citep{MadauMNRAS1996}. 
Star-formation rates are therefore greatly dependent on correct estimates of
extinction. On the other hand, if the rate is to be derived from dust
FIR emission, a knowledge of the dust heating mechanism is necessary
(Sec.~\ref{heatmech}).

\end{description}

\section{Studies of dust extinction}
\label{extstud}

A derivation of the extinction in an astrophysical object is relatively
easy only in the case when dust lies between the source of radiation 
and the observer (a {\em screen model}; Sect.~\ref{modelet}), as for stars 
in the Galaxy. Even in 
this case, a knowledge of the intrinsic luminosity of the source is necessary 
to assess the opacity of the dust screen. This is not the case for spiral
galaxies, where the intrinsic properties of unextinguished objects are
unknown and the dust distribution is co-spatial with the stars.

One method used to infer the opacity of spirals is the study of the
variation of some observables, like surface brightness and magnitude,
with the inclination of the object. For a simple model where dust and
stars are homogeneously distributed in an infinite plane-parallel
geometry (a {\em slab} model; Sect.~\ref{modelet}), the surface
brightness (magnitude per unit solid angle) of the object will increase
with the inclination in the optically thin case, because lines of sight
closer to the model plane intersect a larger portion of the galaxy. In
the optically thick case, instead, only the radiation coming from a
region of the dimension of the mean free path for a photon is observed,
and this is independent of the inclination. Opposite behaviours are
expected for the total magnitude, that will be constant in the optically
thin case, because all the light emitted by stars can be seen from any
direction, while it will decrease in an optically thick case, as a
result of the decrease of the object's area projected on the sky.
These results for the {\em slab} model constitute two limiting case, an
object with more realistic dust geometry will have intermediate
behaviours. Statistical studies have been conducted on large samples of 
galaxies of supposedly similar properties as a function of the inclination 
with which they are observed.

The first study of this kind was that conducted by \citet{HolmbergMeLu1958}
analysing the variation of the projected surface brightness with the
inclination for a sample of 119 spirals (53 of types Sa-Sb, 66 of type Sc). 
Using a model derived from the variation of the Galactic extinction with 
the latitude (a {screen} model, as he recognised later;
\citealt{HolmbergBook1975}) he inferred a substantial transparency for 
spiral galaxies. This result was widely accepted, until
\citet{DisneyMNRAS1989} showed how it depended heavily on the assumed
model. They were able to fit \citeauthor{HolmbergMeLu1958} data with an 
optically thick model, provided the dust distribution was internal to
the stellar one, as inferred from observation of edge-on galaxies. For an
infinite opacity, such a model would behave as a transparent one,
because of the unextinguished dust-free layer of stars above the dust
distribution. They showed how more realistic models are necessary to
ascertain the opacity of a galaxy (See Sect.~\ref{modelet} for a
description of the problems involved in producing realistic models
of internal extinction in a spiral galaxy).

Unfortunately statistical studies, even within the framework of 
proper models, can be severely biased by selection effects.
For instance, two works suggesting high optical depths up to large
distances from the galactic centre \citep{ValentijnNature1990,BurteinNat1991}
are shown to be affected by the object selection criteria, by selecting
galaxies with similar surface brightness independently of the
inclination \citep{DaviesMNRAS1993} or lying in a too small space volume
\citep{DaviesMNRAS1995}.

\citet{GiovanelliAJ1994} analysed the photometric properties of a sample
of 1235 Sbc-Sc galaxies observed in the I band. The derived laws relating
galactic photometric properties to the inclination are then compared to
the results from a TRIPLEX model \citep[Sect.~\ref{modelet}]{DisneyMNRAS1989}.
Observations are compatible with a galactic disk having a central face-on
optical depth, $\tau_I<5$
($\tau_V<10$, using the Galactic extinction law in Table~\ref{optpar}).
A similar analysis was conducted by \citet{MoriondoA&A1998}, on a sample
of 154 spirals observed in the Near-Infrared (1$\mu$m-5$\mu$m; NIR) band 
H (68 of which with I band data). Each galaxy is decomposed into its structural 
parameters and their variation with inclination were studied. The effects of 
internal extinction were detected, especially the increase of disk scalelengths
and I-H colour with inclination. Simulations from a modified version of 
the TRIPLEX model, allowing for larger dust scalelengths 
\citep[Sect.~\ref{extdust}]{XilourisA&A1997,XilourisA&A1998}, lead to results 
compatible to the observations if the central face-on optical depth is
$0.3<\tau_H<0.5$ ($1.5<\tau_V<3$). For such optical depths the galactic
disk would be moderately opaque, becoming optically thin, for face-on
inclinations, at about 1 disk radial scalelength from the centre.


Extinction studies have also been carried out more directly on single
objects. One of the dust properties that is frequently exploited in
extinction studies is the selective extinction at different
wavelengths. If intrinsic variations of starlight colour, like those
due to different stellar populations being present in different parts of 
a disk, are not present, a {\em reddening} of the radiation will reveal
the dusty regions. Therefore, comparing images in the optical,
where extinction should be present, with observation in the NIR, which
are far less affected by dust ($\tau_B/\tau_K\approx14$; 
Table~\ref{optpar}) a map of the extinction can be produced. Using a
radiative transfer model, the optical depth can be finally retrieved.
However, even in the hypothetical case of no intrinsic colour variation, 
the method is not of easy application, since it is difficult to derive 
the intrinsic unextinguished colour for the stellar population.
\citet{BlockA&A1994} used optical-NIR colours to derive the extinction 
in two mid-inclined spiral galaxies, NGC~4736 and NGC~4826. They estimate 
the intrinsic colour from regions that look free of extinction or by using 
synthetic models of stellar populations. With a radiative transfer code, 
they retrieve a relation between the colour and the dust optical depth.
In NGC~4736 they detect a dust component demarcating the spiral arms,
with optical depth 2$<\tau_V<$4, and diffuse interarm dust with
$\tau_V\approx 0.75$. In NGC~4826 it is distributed in a foreground screen
of $\tau_V\approx 2$. From the optical depth they deduced the dust mass
and found values an order of magnitude higher than those inferred by IRAS 
observations of dust emission (I'll show in Sect.~\ref{missing} and 
\ref{colddust} how IRAS observations are not able to detect the more
massive dust component in a galaxy).
\citet*{ReganApJ1995} applied the same technique to NGC~1530, using six
optical-NIR colours. A peak face-on optical depth $\tau_V\approx 4$
is derived in a nuclear dust ring.

The same line of reasoning is used by \citet{PeletierA&A1995}. They
measured the radial scalelength in B and K and studied their ratio as a
function of inclination. Assuming that the colour radial gradient is 
due mainly to dust \citep[see][for an opposite view]{DeJongA&A1996b},
the variation of the scalelength ratio can be described by an extended
distribution of dust (Sect.~\ref{extdust}) with face-on optical depth of
order unity in the V band.
\citet{BeckmanApJ1996} measure radial scalelength in B, V, R, I for
three face-on galaxies, over the whole disk and over selected regions, 
to separate the arm contribution from the interarm. The increase of the
radial scalelength with the decreasing wavelength of the observation,
is modelled with a radiative transfer code for face-on galaxies. Higher
central face-on optical depths are found for the arm region than for the
interarm. When optical depths are derived from the mean scalelength of
each whole galaxy, values of $\tau_V=3-4$ are found.

\citet{KuchinskiAJ1998} study several optical-NIR colours of 15 highly
inclined spiral galaxies. Colour-colour plots for different position
along the minor axis of each galaxy are compared with analogous data
from a complex Monte Carlo radiative transfer model (Sect.~\ref{modelet}) 
inclusive of scattering, stellar and dust disk, stellar spheroidal 
distribution and clumping. The observed trajectories in the
colour-colour plots can be explained by models with optical depths
in the range $0.5<\tau_V<2.0$.

In all the optical depth determinations from radiative transfer models
presented above, assumptions where made about the parameters describing
the stellar and dust distribution. In edge-on galaxies, where the
gradients due to dust extinction are maximised, resulting in evident
extinction lanes, it should be possible to derive both the optical depth
and the distribution parameters from an image of the galaxy.
This is achieved by the fitting procedure of \citet{XilourisSub1998}.
A mean V-band face on optical depth $\tau_V=0.5$ is derived from models
of 7 edge-on galaxies.

Alternatively, light from background objects can be used to derive the
extinction, in the simple screen scheme, without any knowledge or
assumption required for the relative distribution of dust and stars in 
the galaxy. The first application of this technique is made by 
\citet{ZaritskyAJ1994}, to detect a possible dust halo in two
spiral galaxy. I will discuss this more extensively in 
Sect.~\ref{extdust} and in Appendix~\ref{unlucky}.
\citet{GonzalezApJ1998} studied the colour and number counts of background 
galaxies through the disks of the spiral galaxy NGC 4536 and the irregular 
NGC 3664, using images from the Wide Field Planetary Camera 2 of the Hubble 
Space Telescope. They found an extinction $A_I$=0.74-1.07 for the arm region 
of NGC 4536 (the value depending on the method used), and of $A_I<$0.5
for the interarm region. The disk of NGC 3664 shows an extinction
$A_I$=1. These results convert to $\tau_V\approx$ 2 for the arm region
of NGC 4536 and the central part of NGC 3664.

Another technique that makes use of background objects is the method of
overlapping galaxies \citep{WhiteNat1992,BerlindAJ1997}. 
To separate the radiation of the foreground object from the attenuated 
radiation of the background one, smooth and symmetric objects are required. 
If this is the case, it is possible to estimate each intrinsic 
individual flux in the region of overlap from the flux in the regions were 
the objects are seen separately. Subtracting the foreground flux from
the overlap, only the attenuated flux of the background object is left.
Comparing it to the intrinsic flux the extinction can be derived. Both
the works of \citet{WhiteNat1992} and \citet{BerlindAJ1997}, conducted
on two different galactic pairs, give higher extinction for the arm
region ($\tau_V\approx$ 1) than for the interarm ($\tau_V\approx$ 0.5).

Most of the recent work listed here seem to suggest that galaxies have
a moderate extinction, with face-on central optical depth $\tau_V$ of
order unity. However, the methods devised to ascertain the opacity in a 
spiral galaxy suffer from a lot of uncertainties and assumptions. 
Further constraints on dust can be obtained analysing its FIR emission. 

\section{FIR emission from spiral galaxies: missing dust?}
\label{missing}

As a consequence of the principle of energy conservation, the radiation
absorbed by dust from stars must be re-emitted. Using simple calculations
for a dust grain immersed in the local ISRF,
\citet{VanDeHulstRAOU1946} derived a grain equilibrium temperature 
$T \approx 16$K.  The peak of dust emission would therefore occur at
$\lambda\sim 170 \mu$m, in the FIR (Sect.~\ref{emission}). 
Dust in the proximity of stars would be heated to a larger temperature
and therefore emit at shorter wavelengths \citep{WhittetBook1992}.
With the launch of IRAS in 1983
\citep{NeugebauerApJL1984}, dust emission was revealed to span over
a wide wavelength range, from the MIR to the FIR, in our Galaxy 
\citep{BeichmanARA&A1987,CoxA&AR1989} as well as in other spirals
\citep{RiceApJS1988}.

IRAS observed in four broad filters centred at 12, 25, 60 and 100
$\mu$m, covering a spectral range from $\sim 5\mu$m to $\sim 120\mu$m.
If dust is heated preferentially by the ISRF to temperatures lower than
30K, the peak of dust emission is not observed. Because of its spectral
range, IRAS is therefore more likely to pick up regions where dust has
higher temperature, like in the proximity of star forming regions, where
the radiation field is higher than in the diffuse medium. Furthermore,
it was discovered that emission at $\lambda\le 60 \mu$m is mostly due to
small grains heated stochastically and not at the thermal equilibrium
(Sect.~\ref{sec_desert}). Because of this excess emission at shorter
wavelengths, dust temperatures derived from IRAS flux ratios under the 
hypothesis of thermal equilibrium are biased towards higher values.

The luminosity emitted by dust depends on a high power of the temperature 
($\propto T^{4+\beta}$, where $\beta\ge1$; Sect.~\ref{emission}). A small 
amount of warm dust can therefore emit more radiation than a large amount 
of cold dust, which could pass undetected, unless observations cover the
spectral range where cold dust emission peaks. Using IRAS data only,
the bulk of the dust in a spiral galaxy may be overlooked. This is
evident in the first determinations of the gas-to-dust mass ratio in
our Galaxy and in other spirals. From the correlation between the local 
column density of interstellar hydrogen (atomic + molecular) 
and the colour excess E(B-V) found by \citet{BohlinApJ1978} towards a
sample of 96 stars (Sect.~\ref{assu_ext}) in the Solar neighbourhood, it 
is straightforward to derive the local gas-to-dust mass ratio. A value of 
130 is found.  The first determinations based on IRAS data gave higher 
values than that, thus implying a substantially smaller amount of dust than 
that derived from the extinction in the local interstellar medium.

\citet{SodroskiApJ1987} analysed the Galactic IRAS FIR emission at 60 
and 100 $\mu$m.  After correcting for zodiacal light and smoothing over 
discrete sources, the FIR emission from the galactic plane was compared to 
CO, HI and 5Ghz surveys, to study the similarities between dust emission 
and the main three phases of the gas, molecular, atomic and ionised.
The longitude profiles of 60 and 100$\mu$m are quite similar to
the CO and 5Ghz emission, while the HI is broader. The latitude
distribution suggests a significant contribution from the dust
associated with the atomic gas, the 100$\mu$m emission being broader
and following closer the HI warp than the molecular.
Temperatures derived from the ratio of 60 and 100$\mu$m fluxes are quite
constant, with a mean value of 24 K (using an emissivity law with
$\beta=2$; Sect.~\ref{emission} and \ref{qemi_new}), decreasing by less 
than 10\% from the inner to the outer Galaxy. This was unexpected, if the
ISRF is the main contributor to dust heating: dust
temperature should be higher in the centre, where the ISRF is higher.
The constancy of T is ascribed to stochastically heated grains, whose
apparent temperature (i.e. the temperature as measured from the flux
ratios under the assumption of thermal equilibrium, which is not the
case for small grains) depends very weakly on the IRSF.
The derived dust masses lead to a gas-to-dust ratio that is twice
the value for the Solar neighbourhood in the inner Galaxy, and
6 times higher in the outer Galaxy.
The larger gas-to-dust ratio can be explained if a cold dust component 
from 1.5 to 6 times more massive than the warm dust is introduced. 
The cold component would contribute only to the 20\% of IRAS emission.

\citet{SodroskiApJ1989} decompose the IRAS Galactic plane emission
at 60 and 100 $\mu$m into three emission components, associated with the 
molecular (H$_2$), neutral atomic (HI) and ionised (HII) phase.
For several position along the galactic plane they derive temperatures 
(under the assumption of a singular temperature component along the line
of sight), optical depths and gas-to-dust ratios for each of the three 
components. The assumption of a single T along the line of sight biases 
the results towards higher values of T.  Using an emissivity law with
$\beta$=2, they derived T=24 K for the HI component, a warmer T=32K for 
the HII component, consistent with OB stars heating the dust, while the
molecular component is colder, T=18K. As in their previous work, the small
variations of the temperature with the galactic longitude are not
compatible with ISRF heating, but rather betray the presence of small
transiently heated grains. The gas-to-dust ratio of the HI component is
higher than the Solar neighbourhood value, as already seen in 
\citet{SodroskiApJ1987}.  The value for the HII component, instead, is 
closer to that expected, because the temperature of dust associated with
HII regions is higher then for dust in the mean ISRF, and therefore
less affected by the IRAS bias on high T and the small grain emission.
The gas-to-dust ratio for the molecular gas has large uncertainties.

A similar trend was observed in other spiral galaxies.
\citet{DevereuxApJ1990} derive the dust mass from 60 and 100$\mu$m IRAS
fluxes, for a sample of 58 spiral galaxies with available HI and H$_2$
data.  A good correlation is found between the mass of gas and the mass of
dust. The dispersion in the correlation is reduced when only data for the
inner disk (R$<$D$_{25}$) are used. Since molecular gas is always
concentrated in the central part of the galaxy, this suggested that the
outer part of the HI disk does not contribute significantly to the FIR
emission. Nevertheless, the derived gas-to-dust ratio is higher than the
Galactic (a mean value 1080). The observed value can be explained if
90\% of the total dust mass has T$\sim$15K, too cold to be detected by IRAS.

It is interesting to note that cold dust at T$<30$ K was not detected 
even when IRAS fluxes of spiral galaxies were integrated with 
sub-millimetre/millimetre observations \citep*{EalesApJ1989,ClementsMNRAS1993}.
\citet{ClementsMNRAS1993} explain this with the different beam sizes of the 
IRAS telescope (FWHM=120'') with respect to the observation at longer 
wavelengths ($\sim10-20''$). Directly comparing the fluxes in the two spectral
ranges is equivalent to assuming that both the emissions come from a
region smaller than the smaller beam size. If instead dust emission is
extended, the flux from mm/sub-mm observations will be
underestimated. Correcting the mm/sub-mm fluxes for this effect,
\citet{ClementsMNRAS1993} retrieve a colder dust temperature of 20K.

\section{FIR emission from spiral galaxies: cold dust}
\label{colddust}

As already foreseen, the picture changed with the availability of
observations at wavelengths longer than the range observed by IRAS.
\citet{SodroskiApJ1994} repeat the same analysis as in
\citet{SodroskiApJ1987,SodroskiApJ1989}, but using the 140 and 240
$\mu$m observation of the Galactic plane from the Diffuse Infrared Background
experiment (DIRBE) aboard the Cosmic Background Explorer (COBE)
satellite. The observations are more sensitive than IRAS to cold
dust, and the contamination from small grain emission is avoided.
A mean gas-to-dust ratio of 160 is found, now compatible with the local
value derived from extinction, and a mean temperature of 19 K. 
The longitudinal trend of T suggests that the dust temperature decreases
with the galactocentric distance, compatible with dust being heated by
a general ISRF.  The gas-to-dust ratio increases with longitude,
suggesting a lower metallicity in the external part of the galaxy, or the 
presence of a colder dust component, too cold to be detected even by DIRBE.
As in \citet{SodroskiApJ1989}, the FIR emission is then decomposed into 
the contribution of the three gas phases. The temperature of dust associated 
with HI is consistent with ISRF heating and similar to the previous IRAS 
determination for the other components.

In \citet{SodroskiApJ1997} a similar data set is used to produce a
three-dimensional model of the Galactic FIR emission. The properties
of the dust component associated to each gas phase are retrieved
as a function of the Galactocentric distance for 3 rings in the inner
galaxy and for the outer galaxy out from the distance of the Sun, after 
adopting a
rotation curve. Temperatures are still derived fitting a blackbody to the 
140 and 240$\mu$m images, using a $\beta=2$ emissivity.
For the HI component, T decreases with the galactocentric distance as for 
dust heated by the ISRF. Apart from the position of the molecular ring, the 
main contributor to the FIR is dust associated with HI (55-65\% of Galactic
FIR emission).  The temperature is T$\approx$21 K.
The gas-to-dust ratio for the component associated with HI increases
outward (consistent with the decrease in the metallicity gradient), 
with a value of 130$\pm$40 at the Sun Galactocentric distance.
A similar gas-to-dust ratio is retrieved for the other gas phases.
Assuming an emissivity law (Sect.~\ref{emission} and \ref{qemi_new})
and using the optical depth at 240$\mu$m for the
HI and H$_2$ dust component, they find that the radial distribution of
the face-on optical depth of the Galactic disk is quite flat, with
$0.5<\tau_B<1$. If seen from a face-on direction, the Galaxy
would look transparent, with a total extinction $A_B<$0.2. 
There is no evidence in DIRBE data to support the idea
that a large fraction of the hidden mass in spirals may be due to unseen
cold gas and stars obscured by intervening dust.

\citet{ReachApJ1995} 
fit several models to the Galactic spectrum from
104$\mu$m to 4.5mm, observed by the Far-Infrared Absolute Spectrometer
(FIRAS) on board of the COBE satellite. Data are best fitted by a two 
temperatures model, with a warm component with 16K$<$T$<$23K and a very 
cold component with 4K$<$T$<$7K. High signal to noise spectra in the inner 
Galactic plane need an intermediate component, with T$\approx$14K.
The warm dust produces the Galactic spectrum between 100 and 
300 $\mu$m and is identified as produced by large grains in equilibrium 
with the ISRF. It is identical to the dust detected
by \citet{SodroskiApJ1994}.
The very cold component gives an important contribution to the
spectrum only for $\lambda>650\mu$m and shows very little variation with
position in the Galaxy. The optical depth of the cold component
correlates well with the warm component and this suggests a Galactic
origin. It is difficult to explain this component with dust shielded from
the ISRF in the core of very opaque clouds: the high optical depths
required and the ubiquity of the cold component would produce a
Galactic extinction much higher than that observed. Transiently 
heated grains would have a very small temperature between each
temperature fluctuation. Nevertheless, dust models \citep{DesertA&A1990} 
predict a contribution of very small grains to the
spectrum in this wavelength range that is smaller than that observed:
an increased amount of small grains to match the FIR-Submm spectrum
would produce an excess of emission in the NIR, that is not observed.
Other possible explanations require the presence of grains with unusual 
optical properties, as for example fractal grains with high emissivity, 
emissivity enhancements, like spectral features of the grains responsible
for the warm component, or very large grains, although they should 
have a dependence on the ISRF, while the cold component has a quite
constant T in the inner and outer galaxy.
The third dust component observed in the inner galaxy is associated with
the molecular gas, as indicated by the rough correlation between the 
variation of its brightness and that of the CO line. Dust in molecular
clouds shielded from the mean ISRF would be heated to similar
temperatures. It is very weak, contributing only $\sim$2\% to the emission 
at 200$\mu$m.

\citet{BoulangerA&A1996} studied the correlation between the FIR
emission from dust as measured from DIRBE and FIRAS and the atomic gas
emission at high galactic latitude. They found a very tight correlation
for atomic hydrogen column densities below 5.5 10$^{20}$ H-atoms
cm$^{-2}$. Above this threshold there is an excess of FIR emission that
is interpreted as the increasing contribution of dust associated
with molecular clouds: FIR emission associated with this dust is observed,
while the H$_2$ is not in a HI survey. In the limit for zero HI column 
density there is a residual FIR emission, that is considered as due 
in part to an isotropic cosmic FIR background \citep{PugetA&A1996} and in 
part to warm ionised gas uncorrelated to the atomic component. After
removing the residual, a mean spectrum for the low column density
regions is computed, characterised by a temperature of 17.5$\pm$0.2 K
($\beta=2$). No evidence for the \citet{ReachApJ1995} very cold
component is found, thus suggesting that it was an artifact caused by
the cosmic FIR background.

\citet{LagacheA&A1998} analysed high latitude observation from DIRBE and
FIRAS, after subtracting the cosmic FIR background of
\citet{PugetA&A1996}. Using the 60$\mu$m DIRBE image as a template 
of the diffuse Galactic emission, they isolate regions with excess 
emission at longer wavelengths. These regions at colder temperatures are 
associated with dense molecular clouds and appear as positive excesses in 
the FIR/HI correlation of \citet{BoulangerA&A1996}. The drop in
temperature may be due to the attenuation of the radiation field in a
dense cloud, but also to a change in the dust properties with the
environment. Some regions have negative excesses, with dust 
hotter than in the mean interstellar medium because of the proximity to
young stars, like in HII regions. The mean FIRAS spectrum for regions
without FIR excess can be fitted with a modified blackbody ($\beta=2$)
of T=17.5$\pm$2.5 K. Temperature fluctuations can be converted to
variations of a 30\% around the mean intensity of the radiation field.
For regions of sky (3.4\% of the celestial sphere) with a significant FIR 
excess, a two component spectrum is required, with a warm temperature
T=17.8$\pm$1.2 K, consistent with the one derived in the other regions
without FIRE excess, and T=15.0$\pm$0.8K for the cold dust, associated
with the molecular clouds. The coldest temperature detected is 13K.
Again, no evidence for the \citet{ReachApJ1995} very cold component is 
found, whose detection is shown to be an artifact of the unsubtracted 
cosmic FIR background. Regions without FIR excess are further analysed in
\citet{LagacheA&A1999}. Decomposing the FIR spectrum in dust associated
with HI gas and with Warm Ionised Medium (WIM), they derive a temperature for
the dust in the second gas component of T=20K. Dust properties in WIM are
quite similar to those in the neutral gas (See Sect.~\ref{emival}) and
consistent with those derived by \citet{BoulangerA&A1996}.

With newly available observations in the FIR at $\lambda>100\mu$m and in
the mm/sub-mm range, large amounts of cold dust have been finally observed 
in other spiral galaxies.  \citet{ChiniA&A1993} mapped the 1.3mm
emission of three galaxies. They find that the dust emission spatial extent
is comparable to the optical size of the galaxies. Because of the
spatial information of the emission, they can safely compare the new
observations with IRAS data, without being affected by the beam size
problem described by \citet{ClementsMNRAS1993}. A cold dust component with 
T=17 K is necessary to explain the spectra for $\lambda>100\mu$m. 

\citet{GuelinA&A1993} observed NGC~891 at 1.3mm, using a bolometer array
at the IRAM telescope. The measured flux is nine times stronger than
what would have been expected on the bases of IRAS fluxes and
temperature. After discharging other possible contributions to
the observed emission, as CO lines or free-free emission, they conclude
that the bulk of the emission must arise from dust at T$<20$ K.
Comparing IRAS fluxes with a 1.3mm image of NGC 3627, \citet{SieversA&A1994} 
concludes that emission at $\lambda>100\mu$m can be explained by a dust
component with T= 19.5K.

\citet{ChiniA&A1995} mapped 32 non-active spirals at 1.3mm, observing
7 of them also at 450 and 800$\mu$m. Integrating the data with IRAS
fluxes, they find that the coldest dust component necessary to fit the
spectrum at larger wavelengths has an average temperature in the range
10-20 K. 

Cold dust is found by \citet{GuelinA&A1995} in M51; they combine fluxes
from a 1.2mm image with FIR observation between 55 and 320 $\mu$m 
from the Kuiper Airborne Observatory. The millimetre image is smoothed
to the poorer resolution of the KAO observation. The spectrum at
$\lambda>100\mu$m can be fitted by dust at T=18 K.
\citet{NeiningerA&A1996} observed NGC4565 at 1.2mm. The emission is
seen to follow the molecular gas in the inner part of the galaxy and
the HI at the periphery. The radial gradient at 1.2mm is shallower than
those previously observed in the range 50-200$\mu$m from IRAS and KAO.
This is a clear signature of dust heated by the ISRF.
The dust temperature in the centre of the galaxy is T=18K. Colder dust 
temperatures (T=15K) are observed in a plateau at a distance of 12 kpc
from the galactic centre, in correspondence with the peak of HI emission.
\citet{DumkeA&A1997} observed NGC 5907 at 1.2mm. The dust emission follows
the gas, but is also present at larger distances from the centre, where no 
CO is observed. Comparing the total flux with IRAS data, a cold dust
component with a mean temperature of 18K is necessary to fit the spectrum,
the warmer dust detected by IRAS being unable to explain the observed
emission at 1.2mm. From an analysis of the radial profiles, a slight 
temperature gradient is inferred, with T dropping from 20K in the centre 
to 16K in the outer  disk.

As suggested by \citet{ChiniA&A1995} more precise temperature estimates 
than using IRAS and millimetre fluxes are possible when data around the
peak of dust emission are available. Data in this spectral range have
been made available by the ISOPHOT instrument \citep{LemkeA&A1996} on board
of the Infrared Space Observatory \citep[ISO;][]{KesslerA&A1996}.
\citet{KrugelA&A1998} observed three quiescent and three active galaxies
with ISOPHOT, obtaining data between 60 and 200$\mu$m. In combination
with fluxes at 1.3mm, they found evidence for large amounts of cold dust 
in the inactive galaxies, with temperatures T=10K or smaller. Compared
to estimates made without ISO data, the mass of dust is increased by a
factor of three.

\citet[][ see also \citealt{DaviesMNRAS1999} for NGC~6946]{AltonA&A1998}
present resolved images of a sample of 8 nearby galaxies, observed with
ISOPHOT at 200$\mu$m. Apart from consideration of the extent of the
dust emission, which are reported in Sect.~\ref{extdust}, they infer a
mean grain temperature between 18 and 20K, using the IRAS 100$\mu$m 
data together with their 200$\mu$m fluxes. Temperatures are about 10K
lower than those based on IRAS data only. Consequently, the dust mass is 
increased by an order of magnitude. Using literature values for the gas
masses, they derived a mean gas-to-dust mass ratio of 220, much closer to the
Galactic value than those derived by \citet{DevereuxApJ1990}. The results hold
even when a possible error of 30\% in the ISO photometric calibration
is taken into account.

Similar results come from ISOPHOT observation at 175$\mu$m of M31, the
Andromeda galaxy \citep{HaasA&A1998}. A dust temperature of 16$\pm$2 K
is fitted to the ISO data and to DIRBE data at 140 and 240$\mu$m
reported by \citet{OdenwaldApJ1998}. The dust mass is boosted up by an
order of magnitude with respect to the IRAS value, thus bringing the
dust-to-gas ratio to 130, close to the determination of
\citet{SodroskiApJ1994}. If the dust is assumed to be distributed in a
thin slab for the inner 10 kpc, a mean face-on optical depth $\tau_V=0.5$ can 
be derived. This agrees with the mean values inferred by \citet{XuApJ1996b}, 
derived from an energy balance method (Sect.~\ref{firmod}).

\citet{AltonApJL1998} observed NGC~891 with the sub-mm camera SCUBA
at 450 and 850$\mu$m. After smoothing the images to a resolution common
with 60 and 100 $\mu$m High Resolution (HiRes) IRAS images, they
find a cold dust component of 15K, together with a hot component of 30 K
necessary to fit the 60 $\mu$m flux. An approximate distribution of cold
dust is retrieved fitting the two temperatures model to the spectra for
each of 6 radial bins at different distances from the centre of the
galaxy. It is found that the cold component contributes increasingly with
the galactocentric distance to the dust mass in each bin.

\citet{OdenwaldApJ1998} searched in the COBE DIRBE all-sky survey (with
a beam size of 0.7$^\circ$x0.7$^\circ$), for all galaxies with locations
listed in the IRAS Catalog of Extragalactic Objects and the Centre for
Astrophysics Catalog of Galaxies. They found 57 galaxies, of which only 7
had available fluxes for $\lambda>100\mu$m. Their spectra could be
fitted by a cold component of T=20-25 K, and a possible weak very cold
component of T=10-15 K. The very cold component contributes usually only up 
to 15\% of the total dust mass. Only two of the seven galaxies are not
compatible with a cold component spectrum only.

Most of the work on spiral galaxies presented in this section makes use
of observations coming from different broad-band instruments to derive flux 
ratios and dust temperatures. Temperatures would be better determined from
spectra of FIR emission (like in the \citealt{ReachApJ1995} work on the
FIRAS Galactic spectrum). The Long Wavelength Spectrometer (LWS)
\citep{CleggA&A1996}, aboard the satellite ISO, cover the spectral range
between 40 to 200 $\mu$m. Although a more extended coverage of the long
wavelength range would be desirable, dust temperatures as cold as 15K
can be derived from the spectrum shape.

\citet{BraineA&A1999} observed the centre of NGC4414, within
the 100'' LWS aperture. A temperature T=24.5 K for the cold component is 
derived. Comparing the LWS data with IRAS fluxes and $\lambda$1.3mm images
covering a larger area, they infer a gradient in the dust temperature,
with colder dust at larger radii.
\citet{TrewhellaPrep1999} observed five galaxies positioning the LWS
aperture in the centre and at different position along the galactic
disk. Although work is in progress, the spectra on the centres suggest 
temperatures T= 30-35 K, with emission peaking at 100$\mu$m, while for
the outer regions the spectra are flat, or still rising out to the
maximum LWS wavelength. Spectra of the outer regions are compatible with
T$<20$K.

As shown in this Section, the problem of the lack of dust resulting from
the use of IRAS data only is solved when observations at $\lambda>$ 100
$\mu$m are available. A cold (T$\approx20$K), massive ($\sim$ 90\%
of the total dust mass) component is necessary to explain the FIR
and sub-mm emission, in the Galaxy as well in other external spirals.
The measured dust temperatures and the gradient of its radial distribution
(observed in the Galaxy and also in other spiral, thanks to new
high resolution and sensitivity sub-mm instruments) indicates that
the cold component is heated by a diffuse ISRF. Since diffuse dust is
the main contributor to extinction, FIR observation around the
peak of dust emission (100$\mu$m-300$\mu$m) can be used to assess the
opacity of a galaxy \citep{ReachApJ1995}.

\section{Heating mechanisms}
\label{heatmech}

While it is accepted that starlight is the major source of dust heating
in normal non-active spiral galaxies, 
there is controversy about which stellar population is the main
contributor to the process. If young, high-mass stellar objects are the main
contributor, it would be possible to derive the rate of recent star 
formation directly from the FIR emission. This would be highly desirable, 
since observations of FIR emission from spiral galaxies are more readily
available than those of other tracers of star formation, like H$\alpha$
emission \citep{DevereuxApJ1991}. Instead, if the diffuse ISRF from an older 
stellar population is the main source of heating, estimates of star formation 
rates from FIR emission would be severely biased.

\citet{DevereuxApJL1990} claim that the correlation between the FIR and 
the molecular gas, as well as the correlation with the non-thermal radio 
emission support the first hypothesis. 
They compare FIR and H$\alpha$ luminosities for a sample of 124 spiral 
galaxies bright in the IRAS bands. L(H$\alpha$) and L(40-120$\mu$m), 
derived from 60 and 100$\mu$m IRAS fluxes, correlates, with a mean 
L(H$\alpha$)/L(40-120$\mu$m)$ \sim 5\cdot 10^{-3}$.
If all the radiation from the star in a HII region is supposed 
to be absorbed and re-emitted in the FIR, the FIR luminosity is
analogous to the bolometric luminosity of the heating star
L$_\mathrm{bol}$. The ratio L(H$\alpha$)/L$_\mathrm{bol}$ can be
therefore computed for any spectral type of ionising stars, after
assuming standard conditions for the HII regions. The measured
L(H$\alpha$)/L(40-120$\mu$m) in their sample of spirals is consistent 
with stars of spectral type O9 being the main source of dust heating.

In two successive works, they analyse FIR and H$\alpha$ emission in two
galaxies, M51 \citep{DevereuxAJ1992} and NGC 6946 \citep{DevereuxAJ1993}.
A good correlation is found in the two objects between H$\alpha$ emission, 
H$_2$ column density and FIR emission 
(at 170$\mu$m for M51 and 160$\mu$m for NGC 
6946), while the atomic hydrogen presents a central depression.
Temperatures derived from 60 and 100 $\mu$m IRAS fluxes are quite
constant all over the galaxy (32-33K and 27-28K, for $\beta=1$ and 2,
respectively). 
A total FIR luminosity is obtained from the IRAS fluxes and from the flux
in the FIR image at $\lambda$=160-170$\mu$m (they derive the temperature of
the warm component from 60 and 100$\mu$m fluxes and assume that 90\%
of dust is cold at T=14-16K, \citealt{DevereuxApJ1990}.
The FIR luminosity is dominated by the warm component.
The ratio between the FIR luminosities and the H$\alpha$ for several
location within the galaxies is consistent with O9-B0 stars being 
responsible for the heating of dust. 

\citeauthor{DevereuxAJ1992} also claim that the absence of a radial
gradient for the temperature can be explained by the fact that the 
typical temperature of dust in HII regions is not expected to depend on
the position in the galaxy. \citet{SodroskiApJ1989} show that the
temperature derived from the ratio of 60 and 100 $\mu$m IRAS fluxes
on the Galactic plane is constant even when the FIR emission
associated with the neutral gas only is considered. This suggests
that stochastically heated grains provide a better explanation for the 
shallow temperature gradient.

For the galaxy M51, \citet{RandApJ1992} come to radically different
conclusions about the source of dust heating. They found that the 
infrared excess, i.e. the ratio between the FIR flux and the flux of
Ly$\alpha$ photons (derived from the free-free continuum at 21 cm)
is higher than that derived in Galactic HII regions, indicating a FIR
emission not due to dust heated by photons originating in massive 
star-forming regions. Furthermore, they found that the arm-interarm
contrast in IRAS images is always lower than for H$\alpha$ images
convolved to the same resolution. This second test too suggests
that the FIR emission does not arise only from dust in star-forming regions.

Another contrasting view is offered by the two works on the Andromeda
galaxy by \citet{DevereuxAJ1994} and \citet{XuApJ1996a}.
In \citet{DevereuxAJ1994} the same arguments as in
\citet{DevereuxAJ1992,DevereuxAJ1993} are brought in favour of the 
FIR luminosity originating mainly from dust in star-forming regions,
i.e. close resemblance of H$\alpha$ and FIR images and the ratio between FIR
and H$\alpha$ luminosities similar to that observed in Galactic HII
regions. The star-forming ring is supposed to contribute 70\% of
the FIR radiation.
\citet{XuApJ1996a} measure the ratio between the IRAS 60$\mu$m
and H$\alpha$ fluxes from bright FIR-resolved sources in M31. Using the
total H$\alpha$ luminosity and the \citet{DesertA&A1990} dust model,
they extrapolate the fraction of the total FIR luminosity that is associated
with HII regions and star-formation. A value of 30$\pm$14\% only
is found.

\citet{WalterbosApJ1996} model the FIR emission of spiral galaxies under
the assumption that dust is heated by the ISRF. The amount of dust is
estimated from the HI column density, assuming a constant ratio between
the two, while the intensity of the ISRF is derived from the blue profile,
after correcting for the inclination and the dust internal extinction.
The ISRF model is then scaled on the Galactic local value and the
60 and 100$\mu$m fluxes are then computed using the values tabulated by
\citet{DesertA&A1990} for their dust model heated by fractions or
multiples of the Galactic local ISRF. On a sample of 20 galaxies, the
modelled FIR fluxes can account for, on average, half of the observed
fluxes. They conclude that the role of the ISRF in heating the dust
should not be ignored.

From their decomposition of the Galactic FIR emission \citet{SodroskiApJ1997}
conclude that the main contributor is dust associated with atomic gas 
(55-65\%).
The dust associated with the  molecular  phase contributes 30-35\%
of the FIR emission. It is believed that FIR emission within the H$_2$
gas comes in part from dust heated by the ISRF and in part by OB stars. 
The HII component contributes only to 5-10\% of the total FIR. They conclude 
that since 55-85\% of FIR emission is not associated to HII or to OB
stars heating within the H$_2$ component, it is not safe to use the FIR to 
derive star-formation rates.

Among the papers presented in this Section, the main evidence in favour 
of the dust heating by high-mass stars is the correlation between the FIR
and H$\alpha$ emissions. This may simply be a reflection of the local 
density of the interstellar medium, both star formation (and H$\alpha$
emission) and FIR emission being stronger in regions of high density.
The problem could be solved by decomposing the FIR emission into the different 
sources of heating. Because of the lack of high-resolution observations,
this is possible only on large objects. Observations of the Galaxy and
M31 favour the hypothesis of a FIR emission arising mainly from ISRF
heated dust. It is then interesting to note that the works in favour of
the high-mass star hypothesis presented here are mainly based on IRAS 
observations.  As shown in Section~\ref{colddust}, dust emitting for
$\lambda>100\mu$m is likely to be heated by a diffuse IRSF.

\section{Evidence for extended dust distributions}
\label{extdust}

In recent years new evidence has emerged that shows how galactic
disks extend beyond the dimensions inferred from the luminous stellar 
distribution. Molecular clouds and associated HII regions has been
observed in our Galaxy, at distances between 18 and 28 kpc, more than
twice the Sun galactocentric distance \citep{DeGeusApJL1993,DigelApJ1994}.
HII regions beyond the optical radius ($R_{25}$, the radius corresponding
to the 25 mag arcsec$^{-2}$ isophote in the B-band) have been observed 
in deep H$\alpha$ images of three spiral galaxies
\citep{FergusonApJL1998,FergusonAJ1998}.

Because of the tight correlation between B-I colour excess and gas column
density (Sect.~\ref{assu_ext}), dust might be expected to be present at large
distances as well.
A few studies of extinction in spiral galaxies suggest larger exponential 
scalelengths for the dust distribution with respect to the stellar.
\citet{PeletierA&A1995} analyses the ratio between the B and K radial
scalelength for a sample of 37 galaxies. Assuming an intrinsic ratio
1.2, due to stellar population gradients, a larger observed ratio can be
caused by dust extinction, that makes the B profile flatter, leaving the
K scalelength substantially unaltered, because of the small opacity at
larger wavelengths. Comparing the observed ratios of 1.3 for face-on
galaxies and 1.7 for edge-on with the results of an absorption (without
scattering) model for dust and stellar exponential distributions, they
find consistent results for exponential dust scalelengths larger then
twice the stellar.
A similar model is compared to J- and V-band images of edge-on spirals by
\citet{OhtaPASJ1995}: for one of three galaxies (NGC 4565), the dust
scalelength is found to be twice the stellar. 
A more complete radiative transfer model, inclusive of scattering and 
spheroidal distributions for stellar bulges is fitted to edge-on galaxies 
surface photometry by \citet{XilourisSub1998}. For a sample of six objects
observed in several optical bands (and two of them also in the Near
Infrared), they derived a mean dust/star scalelength ratio of
$\approx$ 1.5.

A close correlation has been found for the V-I colour of background galaxies
seen in projection along the disk of M31 and the local hydrogen column
density, on a field at a distance of 23kpc from the centre, outside the optical 
radius of the galaxy \citet{LequeuxProc1996}. The result implies a 
substantial dust reddening ($A_V=0.4$ mag). The colour-magnitude
diagram of the galaxy's stars in the field reveals blue stars and
therefore massive star formation.

FIR observations of dust emission confirm the presence of extended dust
distributions.
\citet{DaviesMNRAS1997} use observations at 140$\mu$m and 240$\mu$m from
the DIRBE instrument aboard COBE to model the dust emission. Adopting a
double exponential model for the dust and a dust temperature spatial 
distribution inferred from observations, they produce maps of FIR
emission and temperature as a function of Galactic latitude and
longitude. The observed emission and temperature can be matched by the
model only for a dust disk with a radial scalelength 1.5 times the
stellar. The vertical scalelength of dust is found larger than the
stellar as well. The \citeauthor{DaviesMNRAS1997} model suggest a dust layer
twice as thick as the stellar disk.

The most striking evidence for a large dust distribution comes from ISO
observations at 200$\mu$m \citep{DaviesMNRAS1999,AltonA&A1998}.
\citet{AltonA&A1998} compared the optical (B-band) and FIR scalelength 
at 60 and 100$\mu$m derived from high resolution HiRes IRAS images with
200$\mu$m images of seven resolved spiral galaxies, 
observed by the instrument ISOPHOT aboard ISO. To be sure to compare 
emission coming from the same galactic structures, all the images have 
been smoothed to the poorer ISO resolution (FWHM=117''). It is found
that IRAS scalelengths are generally smaller than the optical.
On the contrary, the 200$\mu$m ISO profiles are shallower, with an 
exponential scalelength $\approx$ 1.3 times the one measured in B. The
dust temperature in the
centre of a galaxy is warmer, because of the higher Interstellar
Radiation Field. Therefore, for a given dust scalelength, the
scalelength of the FIR emission should be smaller, because of the 
steep dependence of the emission on the temperature
(Sect.~\ref{emission}). An emission scalelength larger than the 
optical can thus be produced only if the dust spatial distribution is
more extended than the stellar (as traced by the B-band emission).
A proper analysis would require a detailed modelling of the 
heating by the ISRF at any position along the dust distribution.
This is actually the purpose of this Thesis.

A confirmation of the large extent of dust disks in spiral galaxies
comes from the work of \citet*{NelsonAJ1998}. They selected 
isolated spiral galaxy using the 100$\mu$m IRAS SKY Survey.
Three samples are
defined, classifying the objects on the basis of their optical radius
(10'-30',5'-10',2.5'-5').
For each sample, a coadded image was produced, after rescaling each
object, rotating it according to its position angle, subtracting the 
sky and normalising the flux. An image of the PSF is constructed coadding 
images of two control samples of stars and unresolved galaxies. 
After subtracting the PSF, they find residuals of 100 $\mu$m extended
emission in the two samples of galaxies of larger dimensions. Using
literature data for stellar disks, they derived a mean 100 $\mu$m
scalelength equivalent or up to a factor of 2 smaller than the 
stellar scalelength. A simple model shows that this implies a dust spatial
distribution less concentrated than the stellar.

As well as along the disk, there are suggestions for extended
distribution above the galactic plane. 
\citet{ZaritskyAJ1994} reports a preliminary detection 
of a dusty halo, through the colour excess of background object in fields
close to two galaxies, with respect to fields at larger distances.
Though, the measured colour excess is only twice as large as the rms
coming from the intrinsic dispersions of the background object colours.
A large number of objects is necessary to produce a statistically
convincing result.  \citet{LequeuxA&A1995} and \citet{LequeuxA&A1996} 
apply the same technique to two galaxies, but without producing statistically
robust results, the largest colour difference being always smaller than
3$\sigma$. The only positive detection of a colour excess is the one 
reported for in the already cited work of \citet{LequeuxProc1996}, but
for a field along the galactic plane and thanks to a corroborating
correlation with the gas column density.

As I have shown in this Section, detection of star-formation at large
distances from the galactic centre, models of dust extinction and FIR
emission, observation of FIR dust scalelengths, all point towards the
existence of extended distributions of dust in spiral galaxies, more
extended than the stellar disk. if this is the case, the study of the
distant universe might be severely biased by dust extinction.
However, the only direct detection of extended dust through the
extinction of background objects is not statistically robust.

\section{Plan of the Thesis}

The work presented in this Thesis consists of a self-consistent model 
for the radiative transfer and the dust emission in spiral galaxies.
I will adopt the {\em energy balance} method (Sect.~\ref{firmod}),
within which the FIR emission is directly compared to the stellar
emission to derive the extinction. The  simulation also gives
the dust temperature distribution and the emission at different FIR 
wavelengths. Results of the models will be compared to observation, 
to address some of the topics presented in this Introduction, mainly: 
i) are spiral galaxies optically thin or thick? ii) Can the SED of FIR 
emission be explained advocating only the ISRF as the source of dust
heating? iii) What kind of extended distributions of dust are needed 
to explain the observed spatial distribution of FIR emission?

\vspace{0.5cm}
Apart from Chapter~\ref{intro}, the present {\bf Introduction}, this 
Thesis is organised in the following Chapters:

\begin{enumerate}
\setcounter{enumi}{1}
\item {\bf Dust properties: extinction and emission}
Basic theory of dust extinction and emission is presented, especially
those formulae that are used elsewhere in the Thesis. The adopted 
dust parameters are presented, together with an original determination
of the dust emissivity, based on Galactic extinction and FIR emission.
The fraction of absorbed energy re-emitted in the Mid-Infrared is
estimated.
\item {\bf The radiative transfer and dust emission model}
First I review radiative transfer and FIR emission models available in
the literature. I then describe the adopted distribution for the stars
and the dust in the model of the galaxy. Finally, the radiative transfer
and dust emission model developed for this Thesis is presented. The
adopted procedure is described with the help of example program
outputs.
\item {\bf Modelling NGC 6946}
The model is finally applied to the spiral galaxy NGC 6946. Optical and
FIR observation from the literature are used to derive the stellar and dust
Spectral energy Distribution (SED). Observation of NGC 6946 carried out at the
James Clark Maxwell Telescope (JCMT), using the Sub-millimetre Common User
Bolometer Array (SCUBA) in June 1998 are presented here. Several models
for the dust distribution are discussed, in the quest for a match with 
the observed FIR SED and spatial distribution of emission.
\item {\bf Conclusions}
\end{enumerate}

During the period of my PhD, I have also worked on two other projects, that
are presented in the Appendices: Appendix~\ref{n7331} {\bf SCUBA imaging
of NGC 7331 dust ring} is dedicated to sub-mm observation of the spiral 
galaxy NGC~7331, that I have carried out in October 1997 at the JCMT
using SCUBA. A dust ring, also detected in an optical-NIR colour image, has
been revealed. 

In Appendix~\ref{unlucky} {\bf Search for dust in the halos of spiral
galaxies} I report on the attempt to measure the colour differences of 
objects in the background of two nearby edge-on spiral galaxies, following 
the \citet{ZaritskyAJ1994} technique, as outlined in
Sect.~\ref{extdust}. Unfortunately both the observing runs I attended
at the Isaac Newton Telescope
(November 1996 and December 1997 - January 1998) were undermined by bad 
weather and it was not possible to detect a sufficiently large
number of objects to produce sound results.
I have detected a faint extended luminous halo around the galaxy NGC 891
(Sect.~\ref{lumhalo}), similar to that observed in NGC 5907
\citep{SackettNat1994,LequeuxA&A1996}.

\chapter{Dust properties: extinction and emission}
\label{chapdu}
In this chapter I give a brief description 
of dust properties: the main aim is that of introducing the definitions 
and the parameters used later on in the radiative transfer model.
A particular emphasis is given to the derivation of the dust emissivity
presented in Sect.~\ref{qemi_new}, an original contribution of this work.
\section{Dust extinction}
\label{extinction}

If a light ray travels through a length $ds$ in a dusty medium with 
a grain number density $n_{\mathrm{d}}$, its specific intensity 
$I_\lambda$ (energy emitted per unit time, area, wavelength band and 
solid angle, also called surface brightness) is attenuated by a quantity
\begin{equation}
dI_\lambda=-n_{\mathrm{d}}\sigma_{\mathrm{ext}}(\lambda) I_\lambda\;ds,
\label{eq_diffrt}
\end{equation}
where $\sigma_{\mathrm{ext}}(\lambda)$ is the extinction cross section
of dust for radiation at wavelength $\lambda$.
The extinction cross section is usually written as
\begin{equation}
\sigma_{\mathrm{ext}}(\lambda)=\pi a^2 Q_{\mathrm{ext}}(\lambda),
\end{equation}
the product between the geometric cross section of a grain of radius
$a$ and the \emph{extinction efficiency} $Q_{\mathrm{ext}}(\lambda)$.

Dust extinction involves two different processes, absorption and
scattering: in the former, photons are actually absorbed by the dust
grain and goes into its heating (and the related emission, 
Sect.~\ref{emission}); in the latter radiation is re-directed along 
directions different from the incident one. Therefore, the extinction 
efficiency can be split in two terms describing the contributions 
from each of these two processes:
\begin{equation}
 Q_{\mathrm{ext}}(\lambda)= Q_{\mathrm{abs}}(\lambda)+
 Q_{\mathrm{sca}}(\lambda).
\end{equation}
The fraction of the extinguished radiation that is diffused by scattering
is given by the albedo
\begin{equation}
\omega=\frac{Q_{\mathrm{sca}}(\lambda)}{Q_{\mathrm{ext}}(\lambda)}.
\end{equation}
The angular distribution of the scattered radiation is described by the
phase function $\phi$, that, for spherical grains, is a function of the
scattering angle $\theta$ only, i.e. the angle between the incident and the
scattered direction. The directionality of the phase function is characterised
by the asymmetry parameter $g$, that is the mean of $\cos\theta$ over $\phi$,
\begin{equation}
g=\frac{\int_{0}^{\pi}\phi(\theta)\cos\theta\sin\theta\;d\theta}
{\int_{0}^{\pi}\phi(\theta)\sin\theta\;d\theta}.
\end{equation}
A general solution for the extinction efficiency, albedo and phase
function can be found in principle for any grain shape and $\lambda$, 
if the optical
properties (i.e refractive index) of the material are known
\citep{VanDeHulstBook1957,BohrenBook1983}. The solution for spherical 
grains goes under the name of Mie Theory \citep{MieAnnPhys1908}.

For a point source, the solution of the radiative transfer 
(Eqn.~\ref{eq_diffrt}) is
\begin{equation}
I_\lambda=I_\lambda^0 e^{-\tau_\lambda},
\label{i=io}
\end{equation}
where $I_\lambda^0$ is the intrinsic intensity of the source and 
$\tau_\lambda$ the optical depth of the dusty medium between the 
source and the observer (along the line of sight), given by
\begin{equation}
\tau_\lambda=\int_{\mathrm{l.o.s.}} n_{\mathrm{d}} 
\sigma_{\mathrm{ext}}(\lambda)\;ds.
\end{equation}
Under the assumption that grain properties do not change along
the line of sight, 
\begin{equation}
\tau_\lambda=N_{\mathrm{d}}\pi a^2 Q_{\mathrm{ext}}(\lambda),
\label{tauex}
\end{equation}
with $N_{\mathrm{d}}$ the dust column density. 
A common parameter used to describe the attenuation properties of dust
is the \emph{extinction} $A_\lambda$,
\begin{equation}
A_\lambda=-2.5\log_{10} \frac{I_\lambda}{I_\lambda^0}=1.086\tau_\lambda.
\label{extA}
\end{equation}
Eqn.~(\ref{i=io}) and (\ref{extA}) are only valid under the assumption
of a point source hidden by a layer of dust. In the case of extended
sources with intermixed dust, like in the common geometries used to
describe galaxies, radiation can be scattered into the line of sight, 
thus adding a positive term to Eqn.~(\ref{eq_diffrt}). In this case,
the simple relation between optical depth, intrinsic and observed 
radiation as in~(\ref{extA}) will not hold.
It is difficult to solve analytically the radiative transfer equation 
in spiral galaxies including scattering, unless there is a 
simplification of the geometry \citep{BruzualApJ1988} or approximations are
made in the treatment of the scattering \citep{ByunApJ1994}. 
An exact treatment of scattering is possible, in principle, for any
geometry by using Monte
Carlo methods~\citep[See also Sect.~\ref{modelet}]{WittApJ1992,BianchiApJ1996}.

\section{Assumed parameters for extinction}
\label{assu_ext}
It is possible to measure the \emph{extinction law}, i.e. the
variation of extinction with the wavelength $\lambda$,
comparing the extinction towards stars of the same spectral type. 
In Fig.~(\ref{fig_desert}) (data points) the mean Galactic extinction 
law from \citet{WhittetBook1992} is plotted in the form of $A_\lambda$
normalised to $A_V$ versus 
$1/\lambda$\footnote{Usually
the \emph{colour excess} $E(\lambda-V)=A_\lambda-A_V$ is plotted, 
normalised to $E(B-V)$. For $\lambda\rightarrow\infty$ the normalised
colour excess tends to the value $-R=-A_V/E(B-V)$ (for the mean Galactic
extinction law $R=3.1$). The ratio $A_\lambda/A_V$ can be derived from the 
normalised colour excess using the measured value for $R$.
Alternatively, the ratio between optical depth and hydrogen column
density is plotted. \citet{BohlinApJ1978} found a correlation between
the column density of interstellar hydrogen (atomic + molecular)
as measured in UV absorption spectra towards a sample of 96 stars and
the colour excess E(B-V),
\begin{equation}
N(\mathrm{H})=5.8\; 10^{21} E(B-V) \mbox{H atoms cm$^{-2}$}.
\label{nh_ebv}
\end{equation}
Using the mean Galactic extinction law, Eqn.~(\ref{nh_ebv}) gives
\begin{equation}
A_V/1.086 =\tau_V=4.9\; 10^{-22} N(\mathrm{H}).
\label{t_nh}
\end{equation}
The ratio $A_\lambda/A_V$ can then be converted to 
$\tau_\lambda/N(\mathrm{H})$.
The ratio $N(\mathrm{H})/N_{\mathrm{d}}$ can be easily found from 
Eqn.~\ref{t_nh} and Eqn.~\ref{tauex}, and hence the gas-to-dust mass
ratio, if $Q_{\mathrm{ext}}$, the grain
dimension $a$ and density $\rho$ are known. Assuming
$Q_{\mathrm{ext}}=1.5$, $a=0.1\mu$m and $\rho=3$ g cm$^{-3}$
\citep{HildebrandQJRAS1983}, the local gast-to-dust mass ratio is
$\approx$ 130.
}. 
The mean Galactic
extinction law is measured along several line of sight through the Galaxy.
Its main characteristics are the linear (versus $1/\lambda$) growth in the 
optical, a bump at 2175\AA\ and a steeper rise at shorter wavelength
in the far-UV
(a more detailed description can be found in Sect.~(\ref{sec_desert})).
There are local variations of the mean extinction law, mainly consisting
in the change of strength of the 2175\AA\ bump and in the Far-UV slope.

As for external galaxies, extinction laws have been directly measured
only towards stars of the Magellanic Clouds. In the LMC the curve is quite 
similar to the Galactic one, apart from the star forming region 30 Doradus,
where the bump is weaker and the far-UV rise steeper. The SMC extinction
law is characterised by the absence of the bump and the steep far-UV
rise \citep[][and references therein]{WhittetBook1992,GordonApJ1997}.
The lack of the 2175\AA\ bump has been noted in starburst galaxies
\citep{CalzettiApJ1994,CalzettiProc1997}. 
The weakening of the bump can be due to the differential effects of 
scattering and absorption \citep{CimattiMNRAS1997}, but 
\citet{GordonApJ1997} argue that the absence in starbursts is due to a 
real absence of bump-carrier grains. The bump has been observed in
high-redshift Mg$_2$ absorbers \citep{MalhotraApJL1997}.
Extinction laws in the optical show smaller differences. Applying a 
radiative transfer model to seven edge-on galaxies, \citet{XilourisSub1998} 
find an extinction law similar to that of the Galaxy longward of the U-band. 
Since I am interested in modelling normal galaxies, rather than
starbursts, in this work I use the Galactic extinction law, as given by
\citet{GordonApJ1997}.

The scattering properties of dust, i.e. albedo, asymmetry parameter and
phase function, can be in principle derived from dust extinction models 
\citep[see, for example][]{DraineApJ1984,BianchiApJ1996}. For the sake of 
simplicity and due to the uncertainties in current models, I prefer to
use the empirical determination of albedo and asymmetry parameter
for Milky Way dust in reflection nebulae, given by \citet{GordonApJ1997}. 
\citet{WittApJ1996} point out that the presence of clumps in dust may
bias the derived albedos toward lower values, since observational data 
are always analysed in the framework of homogeneous radiative transfer models.

For the phase function, I use the analytical expression
derived by \citet{HenyeyApJ1941} relating the angular distribution 
of the scattering angle $\theta$ to the asymmetry parameter $g$:
\begin{equation}
\phi(\theta)=\frac{1}{2}\frac{1-g^2}{(1+g^2-2g\cos\theta)^{3/2}}.
\label{hgphase}
\end{equation}

\begin{figure}[t]
\centerline{\psfig{file=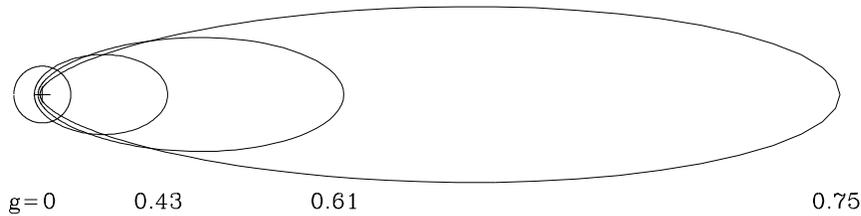,width=14cm}}
\caption{Henyey-Greenstein phase function for different values of the
asymmetry parameter $g$: 0.0 (isotropic), 0.43 (K band), 0.61 (V band)
and 0.75 (EUV, UV1 and UV2 bands).}
\label{fig_hgphase}
\end{figure}

Values for the extinction law, albedo and asymmetry parameters used in
this thesis are given in Tab.~(\ref{optpar}), for the
bands defined in Sect.~\ref{SED}. Values are
taken from \citet{GordonApJ1997} apart from the bands EUV and LMN. 
The extinction law for these two bands 
has been taken from \citet{WhittetBook1992} and \citet{RiekeApJ1985},
respectively. Albedo and asymmetry parameter for the EUV band are from
\citet{WittApJ1993} data at 1000\AA\, while for the LMN band they 
have been assumed equal to those in the K band. 
The directionality of scattering for an Henyey-Greenstein phase function
is shown in Fig.~(\ref{fig_hgphase}), for a few values of the asymmetry
parameter $g$ as in Tab.~(\ref{optpar}) and for the isotropic case
($g$=0).

\begin{table}[t]
\centerline{
\begin{tabular}{lcccc}
\hline
band& $A_\lambda/A_V$ & $\omega$ & $g$ & 
$Q_\mathrm{abs}^\mathrm{SG}/Q_\mathrm{abs}$\\
\hline
EUV & 4.34 & 0.42 & 0.75 & 0.72 \\
UV1 & 3.11 & 0.60 & 0.75 & 0.54 \\
UV2 & 2.63 & 0.67 & 0.75 & 0.44\\
UV3 & 2.50 & 0.65 & 0.73 & 0.47\\
UV4 & 2.78 & 0.55 & 0.72 & 0.58\\
UV5 & 3.12 & 0.46 & 0.71 & 0.65\\
UV6 & 2.35 & 0.56 & 0.70 & 0.54\\
UV7 & 2.00 & 0.61 & 0.69 & 0.43 \\
U   & 1.52 & 0.63 & 0.65 & 0.36\\
B   & 1.32 & 0.61 & 0.63 & 0.28\\
V   & 1.00 & 0.59 & 0.61 & 0.23\\
R   & 0.76 & 0.57 & 0.57 & 0.18\\
I   & 0.48 & 0.55 & 0.53 & 0.14\\
J   & 0.28 & 0.53 & 0.47 & 0.12\\
H   & 0.167& 0.51 & 0.45 & 0.12\\
K   & 0.095& 0.50 & 0.43 & 0.12\\
LMN & 0.04 & 0.50 & 0.43 & 0.12\\\hline
\end{tabular}
}
\caption{Milky Way extinction law ($A_\lambda/A_V$), albedo $\omega$ and 
asymmetry parameter $g$ for the bands as defined in \citet{GordonApJ1997}. 
Two bands (EUV,LMN) has been added (see text for the derivation of the
parameters). The last column gives the ratio between the absorption
efficiency of small grains and the total absorption efficiency, derived
in Sect.~(\ref{sec_desert}).}
\label{optpar}
\end{table}

\section{Dust emission}
\label{emission}
The energy absorbed from photons heats up dust grains and is
successively re-emitted in the infrared, preferentially at $\lambda >
10\mu$m. If the energy of each photon 
impinging on a dust grain is small compared to the internal energy of
the grain itself, the radiation is emitted at the thermodynamic
equilibrium. This is not the case for small grains absorbing high energy
photons: I discuss this in Sect.~\ref{sec_desert}. In this section I
will consider only processes at the thermodynamic equilibrium.

The power emitted by a dust grain at a temperature $T_{\mathrm{d}}$ can
be expressed as
\begin{equation}
W_{\mathrm{em}}=4\pi a^2 \int_0^\infty Q_{\mathrm{em}}(\lambda)\pi
B_\lambda(T_{\mathrm{d}})d\lambda,
\label{eq_wem}
\end{equation}
where $B_\lambda(T_{\mathrm{d}})$ is the Plank function at the
wavelength $\lambda$ and $Q_{\mathrm{em}}(\lambda)$ is the emission 
efficiency (or emissivity). By 
Kirchhoff's law, $Q_{\mathrm{em}}(\lambda)=Q_{\mathrm{abs}}(\lambda)$. 
At long wavelength scattering efficiencies are negligible compared to 
absorption (e.g. for the Mie theory $Q_{\mathrm{abs}}\sim\lambda^{-1}$ 
while $Q_{\mathrm{sca}}(\lambda)\sim\lambda^{-4}$)
and therefore $Q_{\mathrm{em}}(\lambda)\approx Q_{\mathrm{ext}}(\lambda)$.

The emissivity in the infrared, $Q_{\mathrm{em}}(\lambda)$ is usually 
described by a function of the form
\begin{equation}
Q_{\mathrm{em}}(\lambda)=Q_{\mathrm{em}}(\lambda_0)
\left(\frac{\lambda_0}{\lambda}\right)^{\beta}
\label{single}
\end{equation}
where $Q_{\mathrm{em}}(\lambda_0)$ is the value of the emissivity at the
reference wavelength $\lambda_0$, and $\beta$ is the wavelength
dependence index. A more detailed description, together with a new
derivation of $Q_{\mathrm{em}}(\lambda)$ is presented in the next section.

Substituting Eqn.~(\ref{single}) in Eqn.~(\ref{eq_wem}) the emitted
power is
\begin{equation}
W_{\mathrm{em}}=4\pi a^2 Q_{\mathrm{em}}(\lambda_0) K(\beta) 
T_{\mathrm{d}}^{4+\beta}.
\label{eq_wemsol}
\end{equation}
The function $K(\beta)$ is
\begin{equation}
K(\beta)=2\pi\lambda_0^\beta \frac{k^{4+\beta}}{h^{3+\beta} c^{2+\beta}} 
\int_0^\infty\frac{x^{3+\beta} dx}{e^x-1}.
\label{kappabeta}
\end{equation}
The integral in Eqn.~(\ref{kappabeta}) results from the substitution 
$x=hc/k\lambda T_{\mathrm{d}}$ and has the analytical solution
$\Gamma(4+\beta)\zeta(4+\beta)$, with $\zeta$ the Riemann function.

As evident in (\ref{eq_wemsol}), the radiated power is strongly dependent 
on the temperature of dust, thus colder grains radiate at a much smaller
rate. By analogy with the Wien displacement law, the peak of the emitted 
radiation will occur at
\begin{equation}
\lambda_{\mathrm{peak}}\approx\frac{3000}{T_{\mathrm{d}}}\frac{5}{5+\beta} 
\;\mu\mathrm{m}.
\label{wien}
\end{equation}

The rate of absorbed energy can be computed, if both the radiation field
in which the grain is immersed and an expression for the absorption
efficiency are known. For a dust grain in the general interstellar 
radiation field
\begin{equation}
W_{\mathrm{abs}}=4\pi a^2 \int_0^\infty Q_{\mathrm{abs}}(\lambda)
\overline{w}_\star \pi B_\lambda(\overline{T}_\star)d\lambda,
\label{eq_wab}
\end{equation}
where $\overline{T}_\star$ and $\overline{w}_\star$ are the mean 
temperature and dilution factor of the general interstellar radiation
field. These parameters define the interstellar radiation field as the 
field produced by a collection of star with effective temperature
$\overline{T}_\star$ that covers a fraction $\overline{w}_\star\approx
10^{-14}$ of the celestial sphere. A typical interstellar radiation field 
has $\overline{T}_\star=10000$ K and $\overline{w}_\star\approx 10^{-14}$
\citep{DisneyMNRAS1989}.

A crude approximation would be to extend the validity of the 
expression for $Q_{\mathrm{abs}}(\lambda)=Q_{\mathrm{em}}(\lambda)$ as in Eqn.~(\ref{single}) 
to the wavelength range were the interstellar radiation field spectrum peaks 
(i.e. in the optical and UV). This would give as a result 
\begin{equation}
W_{\mathrm{abs}}=4\pi a^2 \overline{w}_\star Q_{\mathrm{em}}(\lambda_0) 
K(\beta) \overline{T}_\star^{4+\beta}.
\label{eq_wabsol}
\end{equation}
At the thermodynamic equilibrium the principle of detailed balance
imposes the rate of absorbed energy to be equal to the rate of emitted
energy: equating Eqn.~(\ref{eq_wabsol}) and Eqn.~(\ref{eq_wemsol}) the dust
temperature is
\begin{equation}
T_{\mathrm{d}}=\overline{w}_\star^{1/(4+\beta)} \overline{T}_\star.
\end{equation}
Using $\beta=1$ (a valid approximation for $Q_{\mathrm{abs}}(\lambda)$ in the
optical and in the infrared shortward of 100 $\mu$m; see next section)
a temperature $T_{\mathrm{d}}\sim$ 16 K is obtained 
\citep{VanDeHulstRAOU1946}. Following Eqn.~(\ref{wien}) it can be seen
that the emission of dust at this temperature peaks at 
$\lambda\sim 170\mu\mathrm{m}$, i.e. in the Far-Infrared.
Similar temperatures are found using more complex models for
$Q_{\mathrm{abs}}(\lambda)$ \citep{DraineApJ1984} and from observations
of the the Galaxy (Sect.~\ref{qemi_new}). 
If the dust grain lies in a region denser in stars than the mean interstellar
field the dilution factor would be larger, resulting in an increased
dust temperature and in an emission peaking at shorter $\lambda$.
Eventually, for circumstellar dust, the temperature could rise to the
sublimation temperature of the grain, causing is destruction.

Dust grains can be heated also by collision with gas atoms: this process
is normally negligible and the dust temperature is almost entirely
determined by radiative processes \citep{SpitzerBook1978}.

If a region of space is occupied by grains of radius $a$, number density
$n_{\mathrm{d}}$ and thermal equilibrium temperature $T_{\mathrm{d}}$,
the emission coefficient $j_\lambda$ (energy emitted per unit time,
volume, solid angle and wavelength) can be written as
\begin{equation}
j_\lambda=
n_{\mathrm{d}} \pi a^2 Q_{\mathrm{em}}(\lambda) 
B_\lambda(T_{\mathrm{d}}).
\label{emeff}
\end{equation}
An external observer would see a specific intensity $I_\lambda$ that,
assuming the region is transparent ($\tau_\lambda\ll 1$)
to radiation in the wavelength range
of dust emission, can be obtained integrating along the line of sight
\begin{eqnarray}
I_\lambda &=& \int_{\mathrm{l.o.s.}}j_\lambda ds \nonumber \\
&=& N_{\mathrm{d}} \pi a^2 Q_{\mathrm{em}}(\lambda) 
B_\lambda(T_{\mathrm{d}}).
\label{ilamb}
\end{eqnarray}
As already said, in the wavelength range of dust emission scattering is
negligible and $Q_{\mathrm{ext}}(\lambda)\approx
Q_{\mathrm{abs}}(\lambda) = Q_{\mathrm{em}}(\lambda)$. Therefore,
using Eqn.~(\ref{tauex}), Eqn.~(\ref{ilamb}) can be rewritten as
\begin{equation}
I_\lambda = \tau_\lambda B_\lambda(T_{\mathrm{d}}).
\label{taubt}
\end{equation}
I will make use of Eqn.~(\ref{taubt}) in the next Section.

\section{A new determination of dust emissivity}
\label{qemi_new}

I have derived the dust emissivity in the Far-Infrared
(FIR) using data available in the literature.
I use two wavelength dependences derived from spectra of 
Galactic FIR emission \citep{ReachApJ1995}. 
A value for the emissivity, normalised to the extinction 
efficiency in the V band, has been retrieved from maps of 
Galactic FIR emission, dust temperature and extinction 
\citep{SchlegelApJ1998}.

The results presented here are similar to other measurements in the Galaxy
but only marginally consistent with the widely quoted values of
\citet{HildebrandQJRAS1983} derived on one reflection 
nebula. The discrepancy with measurements on other reflection 
nebulae \citep{CaseyApJ1991} is higher and suggests
a different grain composition in these environments with
respect to the diffuse interstellar medium.

I measure dust masses for a sample of six spiral galaxies
with FIR observations and obtain gas-to-dust ratios close
to the Galactic value.

 
\subsection{Introduction}

Assessing the quantity of dust in spiral galaxies is of primary 
importance in both understanding the intrinsic properties of 
galaxies themselves and interpreting observations of the distant 
universe: large quantities of dust can modify the optical
appearance of galactic structures like spiral arms
\citep{TrewhellaMNRAS1998}; if the distribution of dust is extended,
a large fraction of the radiation from the distant universe can be
blocked \citep{OstrikerApJ1984}; star formation as determined from UV 
fluxes could be severely underestimated thus altering our knowledge 
of the star formation history of the universe 
\citep{HughesPrep1998}.

Dust mass can be retrieved from extinction or from emission in
the FIR. In the former case information about the star-dust
relative geometry is needed and the method can only be applied to nearby 
edge-on galaxies, where the dust distribution can be
inferred from extinction features \citep{XilourisA&A1997,XilourisA&A1998}.
In the latter case there are no such limitations, and the wealth of
data in the FIR and Sub-mm from instruments like the Sub-mm camera
SCUBA and from the satellites {ISO} and {COBE}, 
can be used to measure dust mass. Unfortunately, the determination of dust 
mass is entangled with that of dust temperature  and they both rely on 
knowledge of the dust emissivity \citep{HildebrandQJRAS1983}, 
the form of which is currently highly uncertain.

Usually the emissivity 
$Q_{\mathrm{em}}(\lambda)$ is described by Eqn.~\ref{single}, with
values of $\beta$ between 1 and 2.
While a value $\beta=1$ seems to be plausible for $\lambda<100$\,$\mu$m 
\citep{HildebrandQJRAS1983,RowanRobinsonMNRAS1992}, there is observational 
evidence for a steeper emissivity at longer wavelengths. 
The difference in emissivity is not unexpected, since emission in the 
Mid-Infrared (25-60 $\mu$m) is dominated by transiently heated grains, 
while at $\lambda>100$\,$\mu$m grains
emit at thermal equilibrium \citep{WhittetBook1992}.
Sub-mm observations of spiral galaxies 
\citep{BianchiMNRAS1998,AltonApJL1998}) 
show that it is not possible to use an emissivity with $\beta=1$
to fit the 450 and 850 $\mu$m emission.
\citet{ReachApJ1995} came to a similar conclusion. 
They used the spectrum of the Galactic plane observed by the 
spectrophotometer FIRAS on board the satellite {COBE}, 
to find that the data are well fitted by an emissivity:
\begin{equation}
Q_{\mathrm{em}}(\lambda)\propto
\frac{
\lambda^{-2}
}{
\left[1+\left(\lambda_1/\lambda\right)^6\right]^{1/6}
},
\label{turning}
\end{equation}
for the range 100 $\mu$m to 1 cm. 
Eqn. (\ref{turning}) behaves like (\ref{single}) with $\beta=1$ at
small $\lambda$ ($\lambda\ll\lambda_1$) and $\beta=2$ at large $\lambda$
($\lambda\gg\lambda_1$) (they set $\lambda_1=200$-$\mu$m).

\citet{MasiApJ1995} measure a value $\beta=1.54$ by fitting a single 
temperature grey-body spectrum to Galactic plane data in four bands between 
0.5 and 2-mm taken by the balloon born telescope ARGO.
\citet{ReachApJ1995} suggest that a single temperature fit 
may bias towards lower values of $\beta$ \citep[see also][]{WrightApJ1991}; 
over the whole FIRAS spectral range, a two temperature grey-body with 
$\beta=2$ at large $\lambda$ provides a significantly better fit than 
a single temperature spectrum with $\beta\approx 1.5$. 
At long wavelengths theoretical calculations for crystalline substances
constrain $\beta$ to be an even integer number \citet{WrightProc1993}.
For amorphous materials $\beta$ depends on the temperature: 
\citet{AgladzeApJ1996} find $1.2<\beta<2$ for amorphous silicate
grains at a temperature of 20~K.

A value for the emissivity at a specific wavelength 
$Q_{\mathrm{em}}(\lambda_0)$ normalised to the extinction efficiency in 
the optical can be determined by carrying out an energy balance in a 
reflection nebula, comparing the energy absorbed from the central star 
with the FIR output from the surrounding dust.  
Alternatively, the extinction measured toward the star can be
directly compared to the optical depth in the FIR 
\citep{WhitcombApJ1981,HildebrandQJRAS1983,CaseyApJ1991}.
These methods are complicated by the unknown nebular geometry and by
temperature gradients in the dust; as an example, \citet{CaseyApJ1991}
found that the extinction method usually retrieves higher values than 
the energy balance.

In this section I use the extinction method comparing the Galactic
extinction to FIR emission: in this case the same column density 
of dust is responsible both for emission and extinction and a
reliable result can be obtained.

\subsection{The method}
\label{emi_method}

\citet*[][hereafter SFD]{SchlegelApJ1998} have presented a new map of Galactic 
extinction. After removing emission from zodiacal light and a cosmic 
infrared background, they have combined the 100 $\mu$m
map of Galactic emission taken by the DIRBE experiment on board the 
{COBE} satellite with the 100 $\mu$m large-area ISSA map from
satellite {IRAS}, to produce
a map of Galactic emission with the quality calibration of DIRBE and the
high resolution of IRAS. The dust temperature has been retrieved using
the DIRBE maps at 100 $\mu$m  and 240 $\mu$m assuming $\beta$=2.
Knowing the temperature, the 100 $\mu$m map has been converted into
a dust column density map and subsequently calibrated to E(B-V) using 
colours and Mg$_2$-index of elliptical galaxies.
I would like to stress that the colour excess has been derived from the 
100 $\mu$m emission without {any} assumption about the value of the 
emissivity at any wavelength. Moreover, the choice of $\beta$ does not 
affect significantly their results: when $\beta$=1.5 is used, the
dust column density map varies only of 1\%, aside from an overall 
multiplicative factor that is taken account of
when calibrating with the colour excess.
I have accessed the electronic distribution of this remarkable dataset 
to retrieve the 9.5 arcmin/pixel  maps of the intensity at 100 $\mu$m, 
I(100 $\mu$m), the temperature and the colour excess E(B-V) for the north 
and south Galactic hemispheres. 

When the same dust grains are responsible for emission and extinction, 
the ratio between the extinction coefficient in the V-band and the 
emissivity at 100 $\mu$m is equivalent to the ratio of the optical depths 
\begin{equation}
\frac{Q_{\mathrm{ext}}(V)}{Q_{\mathrm{em}}(\mbox{100 $\mu$m})}=
\frac{\tau(V)}{\tau(\mbox{100 $\mu$m})}.
\label{ratio_tau}
\end{equation}
The above formula is correct if all of the dust grains are identical.
In a mixture of grains of different sizes and materials, the ratio of
emissivities in Eqn.~(\ref{ratio_tau}) can still be regarded as a mean 
value characteristic of diffuse galactic dust, if the dust composition 
is assumed to be the same on any line of sight.

The optical depth at 100 $\mu$m, in the optically thin case,
is measured using
\begin{equation}
\tau(\mbox{100 $\mu$m})=\frac{I(\mbox{100 $\mu$m})}
{B(\mbox{100 $\mu$m}, T_\mathrm{d})},
\label{taufir}
\end{equation}
where $B(\mbox{100 $\mu$m},T_\mathrm{d})$ is the value of the Planck function at 
$100\mu\mbox{m}$ 
for a dust temperature $T_\mathrm{d}$, both the intensity $I(\mbox{100 $\mu$m})$
and $T_\mathrm{d}$ coming from the maps of SFD\footnote{The assumption
of an optical thin medium for FIR radiation is always valid in
a spiral galaxy. The mean optical depth for the 100$\mu$m radiation
over a region of 20$^\circ$ around the Galactic poles, derived from the SFD 
maps, is 2.5 10$^{-5}$. The maximum on the Galactic plane is 0.14.
An optically thin emission is also derived by
\citet{SodroskiApJ1994,SodroskiApJ1997}.}.
The optical depth in the V-band can be found from the colour
excess E(B-V) maps,
\begin{equation}
\tau(V)=\frac{A(V)}{1.086}=2.85 E(B-V),
\label{tauopt}
\end{equation}
where I have used a mean galactic value $A(V)/E(B-V)$=3.1
\citep{WhittetBook1992}. \citet{ReachApJ1995}
suggest that dust emitting
in the wavelength range 100-300~$\mu$m traces interstellar extinction.
Since the FIR optical depth in Eqn. (\ref{taufir}) has been measured
using data at 100 and 240 $\mu$m, it is then justified to compare it
with extinction as in Eqn. (\ref{tauopt}) to find the ratio of the
extinction coefficient and emissivity.
Knowing the optical depths from~(\ref{taufir}) and~(\ref{tauopt}), I can 
compute a map of the ratio as in Eqn.~(\ref{ratio_tau}); 
I obtain a mean value of
\begin{equation}
\frac{Q_{\mathrm{ext}}(V)}{Q_{\mathrm{em}}(\mbox{100 $\mu$m})}=760\pm60
\end{equation}
for both hemispheres. This value is included, together with other 
multiplicative factors, in the calibration coefficient $p$ as in
Eqn. (22) in SFD. An estimate for 
$Q_{\mathrm{ext}}(V)/Q_{\mathrm{em}}$ can be easily derived from 
that equation, if the DIRBE colour corrections factors, slowly 
depending on T, are omitted. Following this method I obtain a value 
of 765.5.  SFD give an error of 8\% for $p$ and this is the 
value quoted here.
Since most ($\approx$ 90\%) of the elliptical galaxies used to calibrate 
colour excess maps have galactic latitude $b>20^\circ$, one may argue that 
the measured value is characteristic only of high latitude dust. 

\citet{ReachApJ1995} find that the emissivity (Eqn.~\ref{turning})
is best determined by fitting the FIRAS spectrum on the Galactic plane. 
They say that
high latitude data have a smaller signal-to-noise ratio and can be fitted
satisfactory with $\beta=2$ (Eqn.~\ref{single}) although the same 
emissivity as on the plane cannot be excluded. Under the hypothesis that 
the same kind of dust is responsible for the diffuse emission in the whole 
Galaxy, I have corrected SFD temperatures 
using Reach et al. emissivity (Eqn.~\ref{turning}).
The new temperatures are a few degrees higher than those measured with
$\beta=2$ (as an example the temperature passes from a mean value of 18K in a
20$^\circ$ diameter region around the north pole to a new estimate 
of 21K\label{newtemp}). It is interesting 
to note that the difference between the two estimates of temperature is of 
the same order as the difference between the temperatures of warm dust at high 
and low Galactic latitude in \citet{ReachApJ1995} and
this may only be a result of the different emissivity used 
to retrieve the temperature.

When the correction is applied 
\begin{equation}
\frac{Q_{\mathrm{ext}}(V)}{Q_{\mathrm{em}}(\mbox{100 $\mu$m})}=2390\pm190.
\end{equation}
The new ratio is about
three times higher, and this is a reflection of the change of
temperature in the black body emission in (\ref{taufir}): for a
higher temperature, a lower emissivity in the FIR is required to 
produce the same emission. Uncertainties in 
$Q_{\mathrm{ext}}(V)/Q_{\mathrm{em}}(\lambda_0)$
are thus greatly affected by assumptions about the emissivity
spectral behaviour.

\subsection{Comparison with other measurements}
\label{emival}
I now compare the emissivity for $\beta=2$ with literature results
derived under the same hypothesis. Since no emissivity has been derived 
to my knowledge assuming Eqn.~(\ref{turning}), I do not attempt any
comparison with that result. 
All the data are scaled to $\lambda_0=100$ $\mu$m.

Studying the correlation between gas and dust emission from FIRAS and 
DIRBE, \citet{BoulangerA&A1996} 
derived an emissivity $\tau/N_H=1.0 \cdot 10^{-25}$ cm$^2$ at 250 $\mu$m
for dust at high galactic latitude; 
assuming the canonical $N_H=5.8 \cdot 10^{21} E(B-V)$ 
cm$^{-2}$ mag$^{-1}$ and $A(V)/E(B-V)$=3.1 \citep{WhittetBook1992},
this is equivalent to 
$Q_{\mathrm{ext}}(V)/Q_{\mathrm{em}}(\mbox{100 $\mu$m})$=790.
\citet{LagacheA&A1999} have analyzed again the HI/FIR correlation, 
decomposing the FIR emission into two components, associated to neutral
gas and the Warm Ionized Medium. They found $\tau/N_{HI}=8.7\pm0.9 \cdot10^{-26}
$ cm$^2$ and $\tau/N_{H^{+}}=1.0\pm0.2\cdot 10^{-25}$ cm$^2$ at 250 $\mu$m.
These values correspond to $Q_{\mathrm{ext}}(V)/Q_{\mathrm{em}}(\mbox{100 
$\mu$m})$= 900$\pm$90 and 790$\pm$160, if we assume that the same 
$N_H/E(B-V)$ ratio can be used for dust associated with ionized gas
as well as for the atomic. \citet{LagacheA&A1999} argues that the
different emissivities in the two components may be due to a smaller
cutoff in the large grain size distribution for dust within the hotter
ionized medium. Nevertheless, the two values are consistent with each
other.

Quite similar values are found in the \citet{DraineApJ1984} 
dust model, which has a $\beta=2$ spectral dependence in this
wavelength range. At 125 $\mu$m the optical depth is $\tau/N_H=4.6 
\cdot 10^{-25}$ cm$^2$ which corresponds to 
$Q_{\mathrm{ext}}(V)/Q_{\mathrm{em}}(\mbox{100 $\mu$m})$=680. 
\citet{SodroskiApJ1997} find a value for the ratio at 240 $\mu$m, using 
literature data identifying a correlation between B-band extinction and 
100 $\mu$m IRAS surface brightness in high latitude clouds, assuming 
a dust temperature of 18 K. Converted to my notation,
using a standard extinction law, the ratio is 
$Q_{\mathrm{ext}}(V)/Q_{\mathrm{em}}(\mbox{100 $\mu$m})$=990.

The measurement by \citet{WhitcombApJ1981}
on the reflection nebula NGC 7023 is the most commonly quoted value for
the emissivity \citep{HildebrandQJRAS1983}.
Their value derived at 125 $\mu$m for $\beta=2$ is only marginally 
consistent with my result. Following my notation, their result is
equivalent to $Q_{\mathrm{ext}}(V)/Q_{\mathrm{em}}(\mbox{100 $\mu$m})=$ 
220 and 800\footnote{\citet{WhitcombApJ1981} and \citet{CaseyApJ1991}
originally presented values for $Q_{\mathrm{ext}}(UV)/Q_{\mathrm{em}}(FIR)$: 
I have corrected to $Q_{\mathrm{ext}}(V)/Q_{\mathrm{em}}(FIR)$ using the 
provided $\tau(UV)=2\tau(V)$.}, using the energy balance and the extinction 
method, respectively.
The values obtained by \citet{CaseyApJ1991} on a sample of 
five nebulae using the energy balance method are a factor of 3 smaller 
than the ones presented here
(corresponding to $Q_{\mathrm{ext}}(V)/Q_{\mathrm{em}}
(\mbox{100 $\mu$m})=$ 80-400).

\begin{figure}[t]
\centerline{\psfig{file=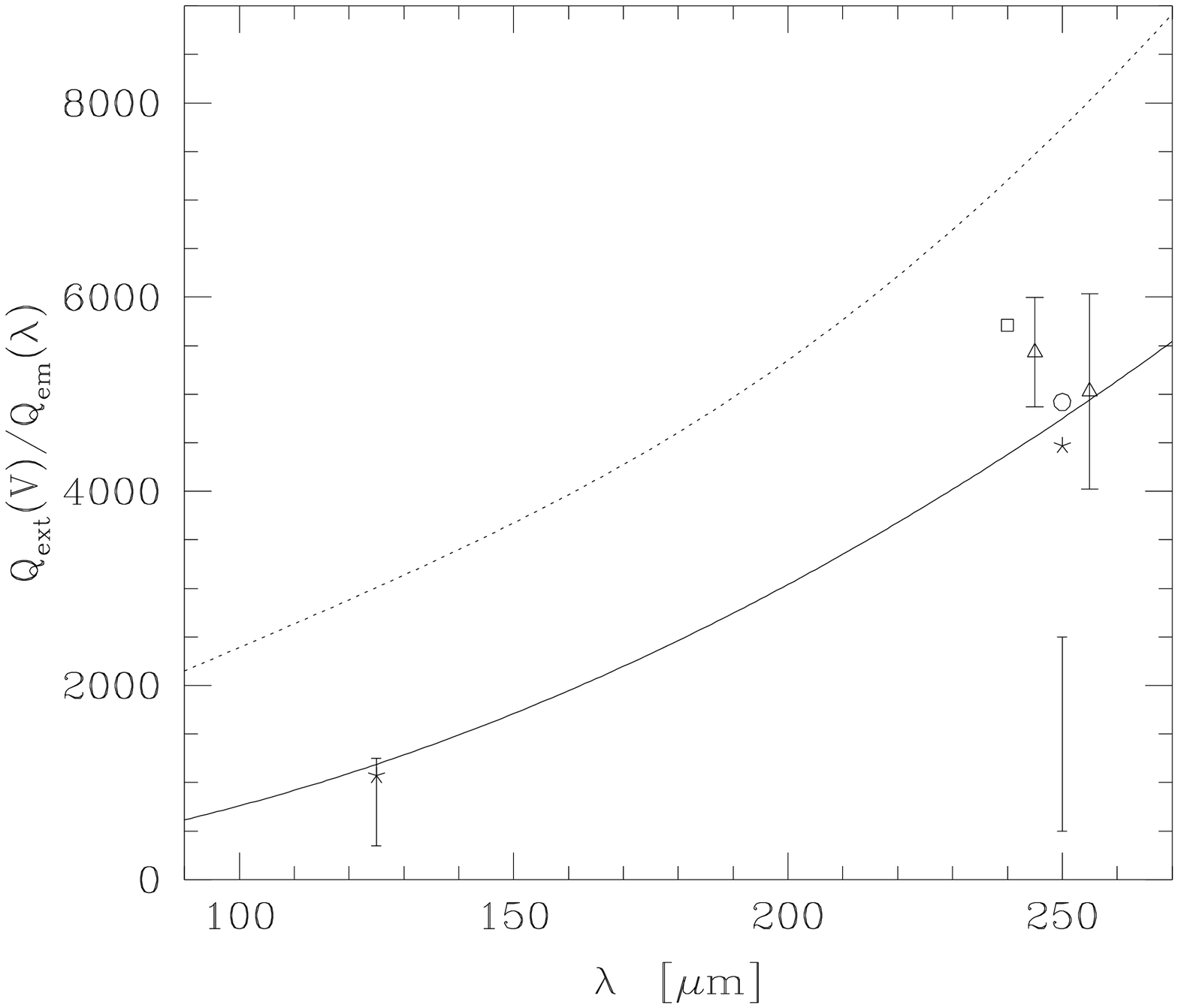,width=12cm}}
\caption{FIR emissivity derived in this work: assuming $\beta$=2 (solid
line) and the \citet{ReachApJ1995} spectral dependence (dotted line).
Error bars at 125 $\mu$m give the range of values of
\citet{WhitcombApJ1981}.  The error bar at 250 $\mu$m gives the range of 
values of \citet{CaseyApJ1991}. 
Data points are from \citet{DraineApJ1984} model (stars), 
\citet{SodroskiApJ1997} (square), \citet{BoulangerA&A1996} (circle)
and \citet{LagacheA&A1999} (triangles). The two values of 
\citet{LagacheA&A1999} have been shifted from $\lambda$=250 $\mu$m
at which they were derived for ease of presentation.
}
\label{fig_emissi}
\end{figure}

In Fig.~\ref{fig_emissi} I show the literature data (plotted at the 
wavelength they have been derived in the original papers) together
with the derived emissivity laws.
I have added the value for \citet{DraineApJ1984} model at 250 $\mu$m.

\subsection{Gas-to-dust ratio of external spiral galaxies}
I now exploit the FIR emissivity derived in this work by
determining dust masses for nearby spiral galaxies.
Following \citet{HildebrandQJRAS1983}
dust masses can be measured from FIR emission using
\begin{equation}
M_{\mathrm{dust}}=\frac{F(\lambda) D^2}{B(\lambda,
T_\mathrm{d})}\cdot \frac{4 a \rho}{3 Q_{\mathrm{em}} (\lambda)},
\label{dmass}
\end{equation}
where $F(\lambda)$ is the total flux at the wavelength $\lambda$, $D$ 
the distance of the
object, $B(\lambda,T_\mathrm{d})$ the Planck function, $a$ the grain radius 
(0.1~$\mu$m) and $\rho$ the grain mass density (3~g~cm$^{-3}$). 
The emissivity $Q_{\mathrm{em}}(\lambda)$ is derived from the ratio
$Q_{\mathrm{ext}}(V)/Q_{\mathrm{em}}(\lambda)$ assuming 
$Q_{\mathrm{ext}}(V)$=1.5 \citep{CaseyApJ1991,WhittetBook1992}.

\citet{AltonA&A1998} provide total fluxes at 100 
$\mu$m and 200 $\mu$m from {IRAS} and  {ISO} for a sample of 
spiral galaxies. I have derived dust temperatures and masses using
$Q_{\mathrm{ext}}(V)/Q_{\mathrm{em}}(\mbox{100 $\mu$m})$=760 and 2390,
for $\beta$=2 and \citet{ReachApJ1995} emissivities,
respectively.
Using literature values for gas masses, I have computed the
gas-to-dust ratios. Values of gas masses, temperatures and gas-to-dust
ratios are presented in Table~\ref{gd_ratio}.

\begin{table}[t]
\begin{center}
\begin{tabular}{lccrcr}
\hline
Galaxy & Gas Mass &  \multicolumn{2}{c}{emissivity~(\protect{\ref{single}})}
&  \multicolumn{2}{c}{emissivity~(\protect{\ref{turning}})}\\
~&10$^{10}$ M$_\odot$& T(K)   & G/D ratio &   T(K)  & G/D ratio \\
\hline
NGC 628     & 1.1  & 16 (17) &  90 (190) & 18 (20) & 100 (200)\\
NGC 660     & 0.91 & 19 (21) & 110 (230) & 23 (26) & 120 (250)\\
NGC 5194    & 0.75 & 18 (20) &  90 (180) & 21 (24) &  90 (190)\\
NGC 5236    & 3.5  & 19 (21) & 240 (500) & 22 (25) & 255 (540)\\ 
NGC 6946    & 3.0  & 17 (19) &  75 (150) & 20 (22) &  80 (160)\\
NGC 7331    & 1.0  & 17 (19) &  70 (145) & 20 (22) &  70 (155)\\
\hline
\end{tabular}
\end{center}
\caption{Sample of galaxies from \citet{AltonA&A1998}. 
Gas masses have been derived from \citet{DevereuxApJ1990}
\citep[for NGC 660]{VanDrielAJ1995} and corrected to 
the distances quoted by \citet{AltonA&A1998}. 
Dust temperature and gas-to-dust ratio are derived using Eqn.~(\ref{single}) 
with $\beta$=2 ($Q_{\mathrm{ext}}(V)/Q_{\mathrm{em}}(\mbox{100 $\mu$m})$=760), 
and Eqn.~(\ref{turning}) 
($Q_{\mathrm{ext}}(V)/Q_{\mathrm{em}}(\mbox{100 $\mu$m})$=2390). 
Values within brackets are derived under
the hypothesis that {ISO} fluxes are overestimated by 30\%.
}
\label{gd_ratio}
\end{table}

The mean value of the gas-to dust ratio for the sample is 100 using 
Eqn.~(\ref{single}), 110 using Eqn.~(\ref{turning}).
Mean temperatures go from 18K with the $\beta=2$ emissivity to 
21K when the \citet{ReachApJ1995} behaviour is assumed 
(as for the north galactic pole in Sect.~\ref{newtemp}). 
\citet{AltonA&A1998} pointed out that {ISO} 
200 $\mu$m fluxes could be overestimated by about 30\%; 
correcting for this I obtain a mean gas-to-dust
ratio of 220-240 (for $\beta$=2 and \citet{ReachApJ1995})
emissivity, respectively).
As shown above, dust masses obtained with the two methods are quite 
similar. This can be explained by substituting eqs.~(\ref{ratio_tau}) and 
(\ref{taufir}) into (\ref{dmass}). For $\lambda=\mbox{100 $\mu$m}$
I can derive
\begin{equation}
M_{\mathrm{dust}}\sim\frac{B(\mbox{100 $\mu$m},T_\mathrm{d}^\mathrm{G})}
{B(\mbox{100 $\mu$m},T_\mathrm{d})},
\end{equation}
where $T_\mathrm{d}^\mathrm{G}$ is the mean temperature of dust in the
Galaxy. From the equation it is clear that the dust mass determination is
insensitive to the emissivity law used, as long as the dust temperature 
in external galaxies and in our own are similar.

The range of values for the gas-to-dust ratio (100--230) encompasses
the Galactic value of 160 \citep{SodroskiApJ1994}. 
As a comparison, the mid-value of \citet{WhitcombApJ1981}
would have given dust-to-gas ratios larger by a factor 1.5.

\subsection{Conclusion}
I have derived the dust emissivity $Q_{\mathrm{em}}$ in the FIR using 
the wavelength dependence derived from the FIR Galactic spectrum 
\citep{ReachApJ1995}. The emissivity has been normalised to the
extinction efficiency in the V band using dust column density maps
calibrated to Galactic extinction (SFD).
$Q_{\mathrm{em}}$ depends strongly on the assumed wavelength dependence.
For a $\beta=2$ emissivity index I obtained
\begin{equation}
Q_{\mathrm{em}}(\lambda)=\frac{Q_{\mathrm{ext}}(V)}{760}
\left(\frac{\mbox{100 $\mu$m}}{\lambda}\right)^2.
\label{qema_b2}
\end{equation}
This result is consistent with other values derived from FIR Galactic
emission \citep{BoulangerA&A1996,SodroskiApJ1997} and with the 
\citet{DraineApJ1984} dust model.
The widely quoted emissivities of \citet{WhitcombApJ1981,HildebrandQJRAS1983}
derived from the reflection nebula NGC 7023 are only marginally consistent
with these values, while the emissivity measured by \citet{CaseyApJ1991}
on a sample of five nebulae are smaller by a factor of 3.
This may suggest a different grain composition for dust in the diffuse
inter-stellar medium compared to reflection nebulae.

When the wavelength dependence derived by \citet{ReachApJ1995}
on the Galactic plane is used, I obtain
\begin{equation}
Q_{\mathrm{em}}(\lambda)=\frac{Q_{\mathrm{ext}}(V)}{2390}
\left(\frac{\mbox{100 $\mu$m}}{\lambda}\right)^2
\frac{2.005}{
\left[1+\left(\mbox{200 $\mu$m}/\lambda\right)^6\right]^{1/6}
}.
\label{qema_used}
\end{equation}
I have used the derived emissivities to measure dust masses from 
100 $\mu$m and 200 $\mu$m fluxes of a sample of six spiral 
galaxies \citep{AltonA&A1998}. 
I have retrieved similar dust masses with both the spectral 
dependences. The gas-to dust ratios of the sample (100-230) are 
close to the Galactic value of 160 \citep{SodroskiApJ1994}.

\vspace{0.5cm}

Since Eqn.~(\ref{qema_used}) has been determined using the most accurate 
information available for the spectrum of dust emission and for 
the extinction in a galaxy, I will use that emissivity law for the FIR
simulation of this thesis.

\section{The MIR correction}
\label{sec_desert}
As seen in Section~\ref{emission}, grains emitting at the thermodynamical 
equilibrium are responsible for the emission in the FIR.  
However, if the dust grain is small, the absorption of a single
high-energy photon can substantially alter the internal energy of the
grain. The dust grain thus undergoes temperature fluctuations of
several degrees and cools by emission of infrared radiation, mainly in
the MIR range \citep{WhittetBook1992}. 
Since in this thesis I want to model the FIR galactic emission only,
I need to exclude from the total absorbed energy the fraction that goes 
into non-equilibrium heating. This can be done if a model
of the absorption efficiency for all the dust components is known.

\begin{figure}
\centerline{\psfig{file=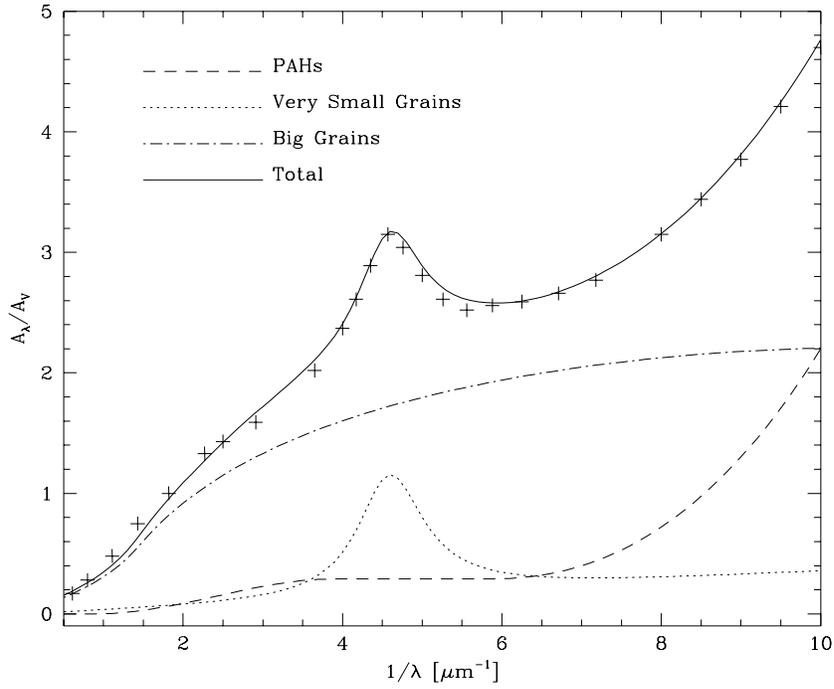,width=12cm}}
\caption{Extinction law from the \citet{DesertA&A1990} model
(solid line) together with the contribution of each dust component.
The observed extinction law (crosses) is from \citet{WhittetBook1992}.
}
\label{fig_desert}
\end{figure}

\citet*{DesertA&A1990} built an empirical dust model to interpret both
extinction and infrared emission in the solar neighbourhood and other 
astrophysical situations. Analysing the features in the extinction curve
and in the infrared emission, they found that three dust components 
are needed. They are:
\begin{description}
\item[Big grains:] classical big grains ($0.015\mbox{$\mu$m}<a<0.11 
\mbox{$\mu$m}$) are needed to explain the rise in NIR-Optical extinction 
curve and the linear rise in the UV. They suggest that big grains could 
be of silicate with a coating of a blacker, carbon-dominated material, 
since silicates alone, because of their large albedo, cannot explain 
extinction. Simple functions are used for the optical properties 
($Q_{\mathrm{abs}}\propto\lambda^{-1}$, $Q_{\mathrm{sca}}\propto\lambda^{-4}$ 
for $\lambda>a$, both constant for larger radii\footnote{These are the
general trends predicted by the Mie theory for spherical grains in the
limits of grains very small or very large compared to $\lambda$.
The oscillating behaviour around these trends \citep{VanDeHulstBook1957}
have been neglected.}). 
Big grains are needed to explain the thermal equilibrium emission in the FIR.
\item[Very small grains:] these grains are responsible for the bump
feature in the extinction curve at 2175\AA. In the model they are
supposed to be small graphite grains 
($0.0012\mbox{$\mu$m}<a<0.015\mbox{$\mu$m}$), pure absorbers and with an
absorption efficiency empirically derived fitting the bump with a Drude
profile \citep{FitzpatrickApJ1988}. These small grains are heated
stochastically and are responsible for the infrared emission in the range 
$25\mbox{$\mu$m}<\lambda<80\mbox{$\mu$m}$.

\item[PAHs:] Polycyclic Aromatic Hydrocarbon molecules are introduced to
explain the continuum emission and the Unidentified Infra-Red emission
features at $\lambda<25\mbox{$\mu$m}$ and the Far UV non linear rise in
the extinction curve. Radii of these molecules are taken to be between
$0.0004\mbox{$\mu$m}$ and $0.0012\mbox{$\mu$m}$. PAHs are pure absorbers
and the absorption
efficiency is derived both empirically (from a fit of extinction in
the FUV \citep{FitzpatrickApJ1988} ) and theoretically.
The presence of PAHs features in the infrared spectra of dusty areas
closely correlates with the FUV non-linear rise, but not with the
presence of the 2175\AA\ bump. This is why two small grain components
are needed. Moreover, PAHs of reasonable physical dimensions cannot produce
the emission in the range $25\mbox{$\mu$m}<\lambda<80\mbox{$\mu$m}$.

\end{description}
Each dust component has a size distribution described by a power law.
The parameters of the model, i.e. size distributions, grain
radii, albedo of big grains and relative abundances of each
component, have been derived comparing the FIR output of the model
heated by the Local Inter Stellar Radiation Field
with observations. Using the information provided
by \citet{DesertA&A1990} I have been able to reproduce their model for
the extinction curve (Fig.~\ref{fig_desert}).

\begin{figure}
\centerline{\psfig{file=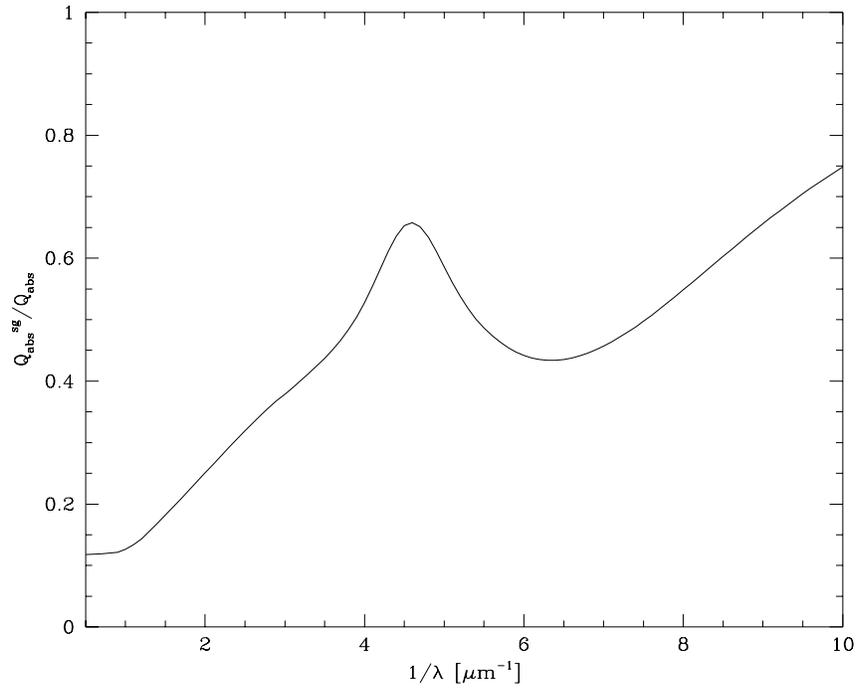,width=12cm}}
\caption{Ratio $Q_\mathrm{abs}^\mathrm{SG}/Q_\mathrm{abs}$ i.e. the
fraction of total absorbed energy that is absorbed by small grains
(PAHs and Very small grains in \citet{DesertA&A1990} model.)}
\label{fig_frac}
\end{figure}

The fraction of starlight that is absorbed by small grains (both PAHs
and very small grains) can be computed using the absorption efficiencies
from the \citet{DesertA&A1990} model.
The extinction efficiency for all the components is
\begin{equation}
Q_\mathrm{ext}=Q_\mathrm{ext}^\mathrm{BG}+Q_\mathrm{ext}^\mathrm{VSG}
+Q_\mathrm{ext}^\mathrm{PAH}.
\end{equation}
Since both PAHs and Very Small Grains are pure absorbers,
\begin{equation}
Q_\mathrm{ext}^\mathrm{PAH}=Q_\mathrm{abs}^\mathrm{PAH}
\;\;\;\;\;\;\;\;\;\;\;\;\;\;\;\;\;\;\;\;\;\;\;\;\;\;\;
Q_\mathrm{ext}^\mathrm{VSG}=Q_\mathrm{abs}^\mathrm{VSG},
\end{equation}
while
\begin{equation}
Q_\mathrm{ext}^\mathrm{BG}=Q_\mathrm{abs}^\mathrm{BG}+
Q_\mathrm{sca}^\mathrm{BG},
\end{equation}
where $Q_\mathrm{sca}^\mathrm{BG}$ is the scattering efficiency.
The absorption efficiency of the dust model is therefore
\begin{equation}
Q_\mathrm{abs}=Q_\mathrm{ext}-Q_\mathrm{sca}^\mathrm{BG}
\end{equation}
and the absorption efficiency of small grains (PAHs + Very Small Grains)
is
\begin{equation}
Q_\mathrm{abs}^\mathrm{SG}=
Q_\mathrm{abs}^\mathrm{VSG}+Q_\mathrm{abs}^\mathrm{PAH}.
\end{equation}
Of the light impinging on a dust grain, a fraction of energy
proportional to $Q_\mathrm{abs}$ is absorbed; the fraction of total 
absorbed energy  that is absorbed by small grains is therefore
\begin{equation}
\frac{Q_\mathrm{abs}^\mathrm{SG}}{Q_\mathrm{abs}}.
\label{mir_frac}
\end{equation}
Values of the fraction of energy absorbed by small grains are plotted
in Fig.~(\ref{fig_frac}) as a function of $1/\lambda$. Tab.~(\ref{optpar}) 
gives the values of the ratio for each of the used wavelength bands. 
In the model, after computing the energy absorbed by dust from each
wavelength band, a fraction as in Eqn.~(\ref{mir_frac}) is excluded and is
not converted into FIR emission (see Chapters~\ref{program} and \ref{model}).

\section{Summary}
In this Chapter I have introduced the dust properties used 
in this work. The mathematical expression for dust extinction and
emission will be used in the next Chapter, to model the galactic FIR
emission. An original derivation for the dust emissivity has been
presented, using the best available data for the Galactic dust
extinction and emission. Finally, I have evaluated the contribution of
the MIR radiation from transiently-heated dust grains to the total dust 
emission. The purpose of this Thesis is modelling the FIR emission by 
dust grains at the thermal equilibrium only. The fraction of energy
absorbed by grains emitting in the MIR correction will be excluded from
the model using the MIR correction derived in the last Section.

\chapter{The radiative transfer and dust emission model}
\label{program}

This chapter is dedicated to a description of the radiative transfer and
dust emission models used in this Thesis. First, a review of previous
models for the solution of the radiative transfer problem and for the
simulation of dust emission in spiral galaxies is presented. I then
describe observations of galactic dust and stellar geometrical distributions 
and I justify the parameters adopted in this work. A brief description
is given of the Monte Carlo radiative transfer code of 
\citet*[][hereafter BFG]{BianchiApJ1996}, the backbone of the project.
Finally I discuss how dust temperatures and FIR emission are derived by
another code, using the map of absorbed energy produced by the Monte
Carlo code.

\section{Radiative transfer models}
\label{modelet}

Several radiative transfer models are available in the literature. I
describe in this section those that have been developed to simulate
the extinction properties of general galactic environment.

A review of simple geometry models is given by \citet{DisneyMNRAS1989}.
They show how the early {\em screen} model with all the dust in front of
the emitting source (in analogy to the first studies of extinction of
Galactic stars through layers of dust), always gives higher extinction
than models with interspersed stars and dust, like the
{\em slab} model, where dust and stars have the same plane parallel
geometry, and the {\em sandwich} model, where the dust is confined to a
thinner distribution internal to the stellar one. They then present
the Triple Exponential model (TRIPLEX), where the simple plane-parallel
scheme is abandoned in favour of more realistic vertical and radial 
exponential distributions for both stars and dust (Sect.~\ref{stardisk} 
and \ref{dustdistri}). The radial scalelength of dust and stars is
assumed to be the same, while the vertical scalelengths of the two
distribution are independent, to be able to simulate the thinner dust
disk. Exact analytical solutions are provided for the face-on case,
and approximations for larger inclination of the line of sight.

The models presented by \citet{DisneyMNRAS1989} include absorption as
the only process leading to extinction. Because of the large observed
values for dust albedo (Sect.~\ref{extinction}) and in view of possible 
large optical depth inside a galaxy, any realistic models should include a 
treatment of multiple scattering of light by dust. Unfortunately, the
scattering term makes the radiative transfer problem more complicated.
A compromise between an exact treatment of scattering and a realistic
galactic geometry must be adopted, if an analytical solution is to be
achieved.

\citet*{BruzualApJ1988} chose to adopt a simple plane parallel geometry. 
The scattering is dealt with in a complete form, through the solution of
a system of integro-differential equations. 
Solutions are found for a {\em slab} geometry and a forward scattering
Henyey-Greenstein phase function (Eqn.~\ref{hgphase}),  or for a {\em
sandwich} geometry and isotropic scattering. For the same geometries,
\citet*{DiBartolomeoMNRAS1995} use a quickly convergent approximation to
the solutions derived with the method of spherical harmonics.
\citet*{CorradiMNRAS1996} present solution for face-on models, allowing
for a dust and stellar exponential distribution but assuming a local
plane parallel geometry. The solutions are therefore exact only when the
mean free path of the photon is small with respect to the scale of radial
variation of the stellar and dust distributions. This is not the case
for optically thin regions. However in these regions the scattering 
events are reduced in number and the results are not greatly affected.

On the other hand, if more realistic geometries are desired, the exact
treatment of multiple scattering has to be sacrificed. 
\citet{KylafisApJ1987} model the surface brightness of the edge-on
galaxy NGC~891 with a a three-dimensional model for dust and stars.
The adopted distributions are radially exponential and behaves as a 
sech$^2$ law (Eqn.~\ref{kyba}) along the vertical direction.
Multiple scattering is treated correctly only for the first order term,
then an approximation is introduced for the higher orders.
\citet*{ByunApJ1994} have generalised the method to deal with any
galactic inclination. A spheroidal stellar distribution has been
included to simulate the bulge. 
\citet{XilourisA&A1997,XilourisA&A1998,XilourisSub1998} successfully
applied it to the fitting of edge-on galaxies surface photometry.

\citet*{TrewhellaAAS1998} are designing an analytical model with a
cellular approach. The three-dimensional space is divided into cubic
cells and radiation is passed from one cell to the other through 26
directions defined by the cell faces, edges and corners. Local optical
depth, albedo and dust phase function regulate the absorption of energy 
and its diffusion through the 26 directions. The model is able to deal
with arbitrary geometries, not necessarily homogeneous, and produces 
results at several wavelengths simultaneously.

The Monte Carlo technique, instead, does not suffer of the limitation of 
the analytical solutions.  Within this technique, each individual photon
is randomly created and its path followed, through scattering and
absorption, until it escapes the dust distribution. The treatment of
scattering is straightforward and no approximations are needed.
Arbitrary geometries can be used in principle. 
With respect to the analytical methods, it has disadvantages in the lack 
of handy solutions to be used in fitting algorithms (like in 
\citealt{XilourisA&A1997,XilourisA&A1998,XilourisSub1998}) and in the large
amount of computational time needed to produce high signal-to-noise
results. 

\citet*{WittApJ1992} present the first application of the Monte Carlo
method for extended distribution of stars and dust. The adopted spherical
geometry is more suitable to describe extinction in the nuclei of active
galaxies or in ellipticals. Several relative distributions of dust and
stars are explored.  BFG have developed a Monte Carlo code for realistic
geometries for spiral galaxies. Since the work of this Thesis is based 
on this model, a more complete description of the original code,
together with the main modification apported for the present
simulations, is given separately  in Sect.~\ref{montecarlo}.
Other models for spiral galaxies based on the Monte Carlo methods
have been developed by \citet{DeJongA&A1996b} and \citet{WoodAPJL1997}.

The cellular approach can be successfully adopted in the Monte Carlo
technique to simulate Dis-homogeneous distributions of dust and stars.
Results for the radiative transfer through clumpy dust distributions have
been recently presented by \citet{WittSub1999} for spherical geometries 
and by \citet{KuchinskiAJ1998} and \citet{BianchiSub1999} for galactic 
disks. The last work includes the possibility of clumpy light emission 
as well.

\section{Models of FIR emission}
\label{firmod}

The model for the FIR emission I am going to present in this Chapter is
a sophisticated version of the {\em energy balance} method. Because of
the conservation of energy, all the starlight absorbed by dust must be
re-emitted. In Chapter~\ref{chapdu} I showed how most of the dust
emission occurs at $\lambda>10\mu$m, in the MIR and FIR. The intrinsic,
unextinguished stellar luminosity is therefore the sum of the
stellar luminosity that escapes dust and is observed in the UV, Optical
and NIR and the infrared luminosity emitted by dust. Using a radiative
transfer model it is then possible to relate the ratio between intrinsic and
emitted starlight to the amount of dust and its geometrical distribution.

\citet{EvansThesis1992} applies the energy balance to a sample of nine
galaxies with available optical and FIR observations. As for the
radiative transfer, he uses a TRIPLEX model with a dust vertical scalelength 
half that of the stellar (Sect.~\ref{dustdistri}). After fitting a 
blackbody to the observed stellar fluxes, the integrated luminosity is 
computed over a few wavebands from the optical to the NIR. 
For a given optical depth in a reference band 
(the V-band), the adopted extinction law (Sect.~\ref{assu_ext}) can be
used to find the optical depth for the central wavelength of each band.
The radiative transfer model is then used to compute, from the 
luminosity observed in each optical band, how much starlight must have 
been originally emitted, for the chosen dust/star geometry and the
band optical depth. Constraining the absorbed energy to equal the
observed energy emitted by dust, a value for the reference optical depth
(and therefore for the amount of dust) can be derived.
A slight modification of this method is used by \cite{TrewhellaMNRAS1998}
to derive the mean absorption $A_B$ on several cells on the galaxy NGC~6946.
A high resolution extinction map in the B band is then derived from 
B-K colour using the relation
\begin{equation}
A_B=(B-K)+\left[A_B-(B-K)\right]_{cell},
\end{equation}
where the quantity between brackets is the mean derived on each cell.

Both \citet{EvansThesis1992} and \cite{TrewhellaMNRAS1998} have a
wide range of optical and FIR observations for their small number of
objects. \citet{XuA&A1995} apply the energy balance to a large 
sample of 135 nearby spiral galaxies with available UV and B fluxes
and 60 and 100$\mu$m IRAS data. Since the spectral coverage is not as
extended as in the other two works, a considerable 
extrapolation is necessary to derive the stellar SED and the total FIR
emission.
The non-ionising emission of the stellar component (f$_\mathrm{star}$) is 
derived for each galaxy from an extrapolation of the UV and the B-band 
monochromatic fluxes, using synthetic spectra for specific Hubble types 
and/or UV-optical colours. 
The total FIR emission is derived from the IRAS data at 60 and 100$\mu$m,
using a relation derived for 13 galaxies with available sub-mm observations.  
In this subsample,
the ratio between the total FIR flux (their f$_\mathrm{dust}$) and the 
FIR flux in the spectral range 40$\mu$m-120$\mu$m, as derived from the  
60 and 100$\mu$m observation (f$_\mathrm{fir}$),
is found to anti-correlate with the ratio between the two IRAS fluxes.
This is expected, since warmer dust (higher 60/100 flux ratio) would emit 
at wavelengths shorter than the 100$\mu$m and most of the radiation 
would be detected by IRAS. For colder dust (lower 60/100 flux ratio), 
the emission would occur out of the IRAS spectral range, and the ratio
f$_\mathrm{dust}$/f$_\mathrm{fir}$ would be larger.
Using this anti-correlation and the available 60 and 100$\mu$m fluxes,
the total dust emission is derived for each galaxy. Ionising photons are 
assumed to be absorbed locally, in the HII regions. The energy absorbed 
from ionising photons is derived from the H$_\alpha$ fluxes of a subsample 
of 34 galaxies and subtracted from the total energy absorbed by dust. 
The remaining absorbed energy is then compared to the stellar emission,
as described for the other two models. The only difference is that 
\citet{XuA&A1995} use a sandwich model inclusive of scattering for the
radiative transfer. 

The energy balance methods presented above have the disadvantage that 
the FIR radiation is treated as a bulk and all the information about the
spectral distribution of the FIR radiation is unused.
The relative geometry of stars and dust affects the ISRF and the way it 
heats dust. It would be desirable to know, from a radiative transfer
code, not only how much energy is absorbed by dust, but also
the distribution of the absorbed energy. After assuming dust emission
properties it would then be possible to know the temperature and the
dust emission as a function of the position in the model.
A comparison between the model and the observed spectrum would
put firmer constraints on the dust distribution. Usually, instead, the
spectrum of dust is modelled {\em assuming} an ISRF.
\citet{DesertA&A1990} and \citet{RowanRobinsonMNRAS1992}, for example,
first derive the properties of their dust models comparing local Galactic 
FIR spectrum with that produce by their dust model heated by a local ISRF. 
When the dust model is determined, spectra of dust heated by fractions
or multiples of the local ISRF are derived, to be compared with the
spectra observed in other environments, internal or external to the
Galaxy.

Few models produce their own ISRF from the stellar model and
apply it to a dust distribution. \citet{SilvaApJprep1998} build a complex
model for the photometric evolution of galaxies. An intrinsic SED is
derived from a spectral synthesis model. Stars, dust and gas are distributed 
in three exponential (both on the radial and vertical direction) 
distributions, to describe: i) a distribution of spherical molecular clouds 
and associated young
stellar objects; ii) free stars that have escaped the molecular clouds;
and iii) diffuse gas and associated dust. Although the three
distributions in the paper have identical scalelengths, independent
parameters can be chosen. Spheroidal galaxies can be modelled as well.
The dust mass is  derived from the residual gas mass in the galaxy
evolution model, assuming a fixed dust to gas ratio. A fraction of the
total mass of dust is ascribed to MCs. A model is used for the dust
properties, slightly different in the molecular clouds, where there is a
decrease of the number of PaHs. 
As for the radiative transfer, they use a cellular approach for the
smooth medium. Scattering is dealt with in an approximate way, using an
effective optical depth that leads to rigorous results only for an
infinite homogeneous medium and isotropic scattering. A separate
treatment is implemented for the objects within molecular clouds.
After computing the ISRF in the grid cells, dust emission is derived.
The stellar and dust spectral energy
distributions are then followed in their evolution with time.

\citet*{SautyA&A1998} present numerical simulations of radiative
transfer in the spiral galaxy NGC~6946. After dividing the galactic
volume into a three-dimensional grid, they represent the interstellar
medium as a two phase medium, with molecular clouds and a constant
density smooth diffuse phase associated with the atomic gas. 
The distribution of molecular clouds is derived from
models of cloud collisions and of the gravitational potential of
the galaxy. OB association are created within the more massive clouds, 
and local fluxes are computed accordingly. The radiative transfer 
through dust is carried out using a modification of the Monte Carlo 
method, but only for wavelengths between 912\AA\ and 2000\AA\, since the
code does not include stars of later spectral types that contribute at
longer wavelengths. In fact, the authors are mainly interested in the 
effects of  the UV ISRF on the H$\alpha$ and C$^+$ emission lines.
The optical and NIR ISRF is scaled from the Galactic, using the 
local surface brightness in a R-band image, but is only used in regions
with no UV flux, like inter-arm regions. Maps of the FIR emission are
produced.

It is difficult, in the two models described before, to disentangle
the effects of dust from the other (numerous) parameters. In
\citet{SilvaApJprep1998} a complex model is adopted for the stellar SED
and for the dust properties while a correct treatment of the radiative 
transfer is sacrificed. Furthermore, only the global dust SED is produced,
but no information on the spatial distribution of emission is present. 
The model of \citet{SautyA&A1998} perhaps deals with the radiative
transfer in a more correct way, but the ISRF is derived only for the UV.
For the spectral regions where the peak of stellar emission is, a
Galactic ISRF is used.

\vspace{0.5cm}

For this Thesis, I have built a self-consistent model for the dust
emission in spiral galaxies. The radiative transfer is treated
correctly. Given an input stellar SED, dust absorbs radiation from an
ISRF that is consistent with the radiative transfer itself.
The temperature along the dust distribution is computed, so that not
only the FIR luminosity and spectrum can be retrieved, but also the
surface brightness distribution  of the FIR radiation for any
wavelength.
A Monte Carlo model with this characteristics has been presented by
\citet*{WolfA&A1998}, but only for star formation environments and not
for galactic distributions of stars and dust.

\section{Stellar disk}
\label{stardisk}
The luminosity density distribution of a galactic disk is usually
described by 
\begin{equation}
\rho=\rho_0 \mbox{exp}(-r/\alpha_\star) Z(z/\beta_\star)
\label{diskequ}
\end{equation}
where r and z are the galactocentric distance and the height above the
galactic plane, respectively, and $\alpha_\star$ and $\beta_\star$ the
relative scalelengths. While there is a consensus for the radial 
exponential behaviour~\citep{DeVaucouleursBook1959,FreemanApJ1970},
a number of expressions for the function $Z(z/\beta_\star)$ have been used in
the literature. 
\citet{VanDerKruitA&A1981}, in their analysis of optical images of edge-on
disks, find that the vertical profile is best fitted by
\begin{equation}
Z(z/\beta_\star)=\mbox{sech}^2(\mid z\mid/z_0)
\label{kyba}
\end{equation}
with $z_0=2\beta_\star$. This function, the solution for a self 
gravitating isothermal sheet, behaves like an exponential of scalelength
$\beta_\star$ at large radii, but has a less sharp peak in the centre.
\citet{PeletierProc1997} suggested that the results of
\citet{VanDerKruitA&A1981} may  have been influenced by
an high dust extinction.
\citet*{WainscoatApJ1989} find the sech$^2$ function unsuitable to
describe the vertical profile of the edge-on galaxy IC 2531 and prefer
a more sharp function, the  exponential 
\begin{equation}
Z(z/\beta_\star)=\mbox{exp}(-\mid z\mid/\beta_\star).
\label{vertexpo}
\end{equation}
This function has the advantage of mathematical simplicity, 
but has no firm physical justification.
An alternative function 
\begin{equation}
Z(z/\beta_\star)=\mbox{sech}(-\mid z\mid/\beta_\star)
\end{equation}
has been proposed by \citet{VanDerKruitA&A1988}. This function has
a peak intermediate between the sech$^2$ and the exp and it is
consistent with measures of the velocity dispersion of stars along the
z-axis.

Analysing the vertical profiles of a sample of 24 edge-on galaxies 
in the relatively dust-free K-band, \citet*{DeGrijsA&A1997b} find 
that a distribution with a peak intermediate between the exp and the 
sech fits better the central peak~\citep[see also][]{DeGrijsA&AS1996}. 
Since a small inclination from the pure edge-on case can produce a 
less sharp profile, these results are compatible with the exponential 
distribution. Therefore in this thesis I will adopt  a radial exponential disk
of star with a vertical exponential distribution as in Eqn.~(\ref{vertexpo}).

In Eqn.~(\ref{diskequ}) it is assumed that the vertical and radial 
behaviour are independent: \citet{DeGrijsA&A1997a} noticed a 
constant increase of $\beta_\star$ with $\alpha_\star$ in a sample of 
spiral galaxies.
This effect is stronger for early-type galaxies, where
$\beta_\star/\alpha_\star$ can increase as much as a factor of 1.5 per 
radial scalelength. In late-type galaxies it is almost zero.
This behaviour has been explained by a thick disk of star with a bigger
scalelength than the ordinary one.
Being dusty late-type galaxies the main concern of this work, 
I use here a constant $\beta_\star/\alpha_\star$.

\section{Observed Galactic scalelengths}
\label{obscales}
In this section I give a list of references to recent determination of
radial and vertical scalelengths of the stellar disk distribution
in our Galaxy.

In the bi-dimensional model of \citet{DeVaucouleursAJ1978} the Galactic disk 
has a radial scalelength $\alpha_\star=3.5$ kpc. The value is obtained 
by comparing the local ($R_\odot$) vertically integrated surface 
brightness in the B-band (derived from star counts) with the mean central 
surface brightness 21.65 mag arcsec$^{-1}$ for disks of spiral
galaxies~\citep[Freeman Law; ][]{FreemanApJ1970}.

\citet{BahcallApJS1980} built a three-dimensional Galactic model to fit
B and V star counts at a Galactic latitude $b>20^\circ$ (where extinction 
is low). The disk component is modelled with an exponential both in the 
radial and in the vertical direction. For $\alpha_\star$ they assume
the \citet{DeVaucouleursAJ1978} value; values for $\beta_\star$ are 
derived from the literature and summarised in a magnitude
(age)-dependent relation: young stars ($M_V<2$) have $\beta_\star=90$ pc,
old disk dwarfs ($M_V>5$) have $\beta_\star=325$ pc, while a linear relation
for stars in the range $2<M_V<5$ is assumed. In this formalism giants 
have $\beta_\star=250$ pc. This model is commonly referred to in the 
literature as the {\em standard model}.
In a subsequent work \citep{BahcallApJS1984} the parameter space is
explored and limits are put on the scalelengths:
$\beta_\star=250\pm100$ pc for giants and $\beta_\star=350\pm50$ pc for 
old disk dwarfs. Although the results depend weakly on $\alpha_\star$
(because the model is compared to star counts at high latitude), data 
can be reasonably fit only for $\alpha_\star>2.5$ kpc.

A modified \citeauthor{BahcallApJS1980} model is used by
\citet{VanDerKruitA&A1986} to fit the integrated blue and red starlight 
observed by the Background Experiment on board of {\em Pioneer 10}.
To avoid regions with high extinction, only data for galactic latitude 
$b>20^\circ$ are used: thus, as shown by the author, only the value 
$\alpha_\star/\beta_\star$ can be derived. Assuming $\beta_\star=325$ pc
a value of $\alpha_\star=5.5$ kpc is derived. As suggested by 
\citet{KentApJ1991} this method is sensitive to local surface brightness 
gradients and may only give a measure of the local exponential scalelength.

\citet*{KentApJ1991} derive the parameters of Galactic structure
from the 2.4-$\mu m$ integrated light, measured from the Infrared
Telescope aboard the {\em Spacelab 2} mission in 1985. 
A radial exponential distribution is assumed for the disk, allowing for
different vertical distributions.
A correction for extinction (small at this wavelength, although not
negligible) is incorporated in the model, evaluated from the atomic 
hydrogen distribution, assuming a constant value for the dust/hydrogen ratio.
Surface brightness profiles as a function of the galactic longitude at 
several latitudes are compared to the data: the {\em standard model} is 
unable to describe the profile on the galactic plane. 
Using an exponential vertical distribution they find $\beta_\star=204\pm
57$ pc and $\alpha_\star=2.7\pm0.5$ kpc. 
The $\beta_\star$ value is small compared to the \citet{BahcallApJS1980} 
value for the giant star distribution (the main contributor to the 
luminosity in the observed wavelength range). 
For this reason they try a different fit allowing for a varying $\beta_\star$. 
Up to 5.3 kpc they keep $\beta_\star=165$ pc then
they assume a gradient of 30 pc/kpc so that $\beta_\star=247\pm 69$ at 
$R_\odot$. The corresponding radial scalelengths is $\alpha_\star=3\pm0.5$ kpc.
I note here that all the works cited up to this point used R$_\odot$=8 kpc,
while in the next references the value 8.5 is preferred. When a
scalelength is derived from a model it should be scaled to the value
of  R$_\odot$~\citep{KentApJ1991}.

\citet{WainscoatApJS1992} fit star counts at 12-$\mu m$ and 25-$\mu m$ 
from the {\em IRAS} Point Source Catalog with a Galactic model with
a disk of radial scalelength $\alpha_\star=3.5$ kpc.
Scaleheights for the vertical exponential distribution are given 
for a a compilation of 87 types of galactic sources responsible for
emission in the NIR bands. Values are similar to those in the 
\citet{BahcallApJS1980} model (e.g. $\beta_\star=325$ pc for 
old dwarfs, 270 pc for giants etc.).
A previous model to fit K-band star count and 12-$\mu m$  IRAS source
count (\citealt{GarwoodPASP1987}, see also \citealt{JonesMNRAS1981})
used slightly smaller values ($\beta_\star=300$ pc for old disk
dwarfs).

In \citet{RobinA&A1992} the scalelength is measured from B and V-band 
star counts up to $m_V=25$ mag in a low absorption window in the 
direction of the Galactic anticenter. They use a complex structure and
evolution Galaxy model (the {\em Besan\c{c}on} model): the disk in
this model follows an {\em Einasto} density law, that behaves similarly
to an exponential in the radial direction and to a $sech^2$ in the 
vertical. A correction for extinction is evaluated from the U-B/B-V plot 
of a sample of stars for which U-band data are available. 
They obtain $\alpha_\star=2.5$ kpc. 

A similar value ($\alpha_\star=2.3\pm0.1$ kpc) is obtained by 
\citet{RuphyA&A1996} using J- and K-band star counts of two
fields in the direction of the Galactic anticenter 
from the Deep Near-Infrared Survey of the Southern Sky. 
The same model as in \citet{RobinA&A1992} is assumed.
Data are corrected for extinction (affecting the J-band)
matching the positions of the observed  and the model J-K colour 
distribution of the stars. The scalelength is obtained by minimising the 
difference in shape between the two distributions.
Like in \citet{RobinA&A1992} their fit is not very sensitive on the 
assumed value for $\beta_\star$ (because data are very close to the 
Galactic plane).

\citet{PorcelA&A1998} measures $\alpha_\star$ from K-band star counts 
in the Two Micron Galactic Survey database. Extinction effects, low in 
the K-band, are further minimised by selecting stars out of the galactic 
plane (at a galactic latitude $b=5^\circ$). A range of galactic longitudes 
($30^\circ<l<70^\circ$) is chosen to minimise the contribution 
to the star distribution from local structures: namely, bulge, ring and bar 
in the centre ($l<30^\circ$) and Local Arm, warp and truncation ($l>70^\circ$).
The contribution from other spiral arms are minimised by the choice of $b$.
In the database, only stars with K-band apparent magnitude $9<m_K<10$
are selected: it can be shown that 80\% of the total light emitted
in this magnitude range is provided by K2-K5III stars, which have 
an absolute magnitude $M_K=-2.5\pm0.6$. Their luminosity function can
thus be conveniently approximated by a Dirac delta function.
They find $\alpha_\star=2.1\pm0.3$ kpc, assuming a vertical scalelength 
$\beta_\star$ of 200 pc (again the result
is not very sensitive on $\beta_\star$).

These recent results based on star counts are confirmed also by 
the kinematical estimate of \citet{FuxA&A1994}, based on the 
asymmetric drift equation, that gives $\alpha_\star=2.5^{+0.8}_{-0.6}$ kpc,
assuming a constant value for $\beta_\star$. If the gradient in 
$\beta_\star$ derived by \citet{KentApJ1991} is included,
$\alpha_\star=3.1$ kpc. 

Recently \citet*{HaywoodA&A1997a,HaywoodA&A1997b} have questioned the
validity of the exponential assumption for the vertical structure, on
the basis of the {\em Besan\c{c}on} model.  
Rather than adopting different values of $\beta_\star$ for different 
galactic sources, the model derives the vertical structure from the 
combined effect of the star formation history and of the secular heating 
of the disk. After tests on B, V, and I star counts (the \citet{RuphyA&A1996}
value of $\alpha_\star$, 2.5 kpc, is used) they claim that the
exponential is unsuitable to properly describe data for $z<500$ pc. 
They find that the vertical density distribution decreases faster than the
$300-350$ pc scaleheight generally assumed for old disk dwarf stars. 
In \citet{RuphyA&A1996} a value of $250$ pc is suggested for
$\beta_\star$.

\section{Adopted disk parameters}
\label{adopted}

A table of vertical scalelengths, B, V, J, H, K-band absolute magnitudes, 
local number density and other values for the main Galactic stellar sources
is provided by \citet{WainscoatApJS1992}. I have computed the mean values 
for the vertical scalelengths averaging over the disk luminosity in each 
waveband. A mean value is computed also averaging over the total disk
luminosity integrated from the B to the K band.
Mean values for $\beta_\star$
are presented in Tab.~(\ref{table_beta}), together with the ratios
$\alpha_\star/\langle\beta_\star\rangle$. I have assumed 
$\alpha_\star=3$ kpc, a compromise between the highest and lowest values
presented in Sect.~\ref{obscales}.

\begin{table}[t]
\begin{center}
\begin{tabular}{ccc}
Band & $\langle\beta_\star\rangle$ [pc] & 
       $\alpha_\star/\langle\beta_\star\rangle$ \\ \hline
B    &  150  &  20.0 \\
V    &  182  &  16.5 \\ 
J    &  256  &  11.7 \\
H    &  268  &  11.2 \\ 
K    &  271  &  11.1 \\
Tot  &  208  &  14.4 \\ \hline
\end{tabular}
\end{center}
\caption{Values of mean Galactic vertical scalelengths $\langle\beta\rangle$ 
obtained averaging over the disk luminosity for spectral bands 
B, V, J, H, K and over the total luminosity integrated over all the
bands. The ratio 
$\alpha_\star$/$\langle\beta_\star\rangle$ is computed assuming 
$\alpha_\star$=3 kpc.}
\label{table_beta}
\end{table}

As shown by Tab.~(\ref{table_beta}) and by the discussion in
Sect.~\ref{obscales}, a model of a spiral galaxy should include
different scalelengths ratios for different stellar components,
and therefore for emission at different wavelengths. 
For simplicity, in this thesis I will mainly use one scalelength
ratio only, the one averaged over the total luminosity,
$\alpha_\star/\langle\beta_\star\rangle=14.4$.
A test simulation has been run using different scale heights for 
different wavelength ranges (Sect.~\ref{test}); for that simulation I
use $\alpha_\star/\langle\beta_\star\rangle=18$ 
($\beta_\star=170$ pc for an $\alpha_\star=3$ kpc) shortward
of the V-band (obtained averaging over the total integrated luminosity
from B to V) and 11.3 ($\beta_\star=265$ pc) at longer wavelengths
(averaging over the total integrated luminosity from J to K).

Similar values for  $\alpha_\star/\langle\beta_\star\rangle$ can be
found in the literature.
Analysing the structure of eight large edge-on galaxies in the UGC
catalog, \citet{DeGrijsA&AS1996} find  a mean ratio of the radial and
vertical exponential scalelength in the I-band of $11.8\pm0.8$.
They argue that looking at edge-on objects, the radial scalelength can 
be overestimated at least by 10\% (in a transparent case) because of
projection effects, thus reducing the ratio. No clear trend of variation
of $\beta_\star$ from B to I-band is observed in their sample.
\citet{VanDerKruitA&A1982} obtained $9.4\pm3.6$ in the J-band for eight
edge-on spiral galaxies, including our own.
Fitting the stellar emission in a sample of seven spiral galaxies with 
their radiative transfer model, 
\citet{XilourisA&A1997,XilourisA&A1998,XilourisSub1998} measure 
$\alpha_\star/\beta_\star\approx$ 14, in B,
12 in V-band and 11-12 in I, J, K.

In the models of this thesis, only one value of the radial scalelength
is used, i.e. all the stellar components have a radial exponential
distribution with the same $\alpha_\star$. A few authors have
investigated the ratios of radial scalelengths in different passbands.
\citet{PeletierA&AS1994} have compared B and K-band disk scalelengths
for a diameter limited sample of 37 Sb-Sc galaxies with a uniform
distribution of orientation respect to the sky. Their aim is to separate in
$\alpha_B/\alpha_K$ the effects of extinction from those of a change in
stellar population. If the ratio $\alpha_B/\alpha_K$ is mainly due to
extinction, it increases with inclination, while it is constant if the
change in stellar population is the dominant contributor to the galactic
colour gradient.  From the metallicity gradients a ratio of 1.17 is
estimated, independent of inclination, while the observed value goes 
from 1.3 for face-on galaxies to 1.7 for edge-on \citep{PeletierA&A1995}.
Moreover they observe from  previous work that in dust-free late type
galaxies the ratio between scalelengths is small:
$\alpha_B/\alpha_I=1.04\pm0.05$, equivalent to $\alpha_B/\alpha_K=1.08\pm0.1$
for almost any stellar population model. Thus they claim that the
change in scalelength with $\lambda$ is mainly due to dust extinction,
rather than being a reflection of different stellar components.
On the contrary, comparing colour-colour plots of a sample of 86 face-on 
galaxies with a Monte Carlo radiative transfer model, 
\citet{DeJongA&A1996b} concludes that dust can't be responsible for the 
gradient in colour. He suggests that the observed ratio
$\alpha_B/\alpha_K=1.22\pm0.23$ is caused by a change in stellar
population in the outer parts of the disk.

The disk models of this thesis are truncated at 6$\alpha_\star$.
To explain star counts of faint sources in our Galaxy, 
\citet{WainscoatApJS1992} introduce a truncation at 4.3$\alpha_\star$, 
\citet{RobinA&A1992} and \citet{RobinApJL1992} at 5.6-6$\alpha_\star$,
\citet{RuphyA&A1996} at $6.5\pm0.9 \alpha_\star$.

\section{Dust disk}
\label{dustdistri}

The parameters for the dust disk are by far more uncertain than
the stellar disk. Usually the same functional form is used
as for the stellar distribution, with independent scalelengths 
\citep{KylafisApJ1987,ByunApJ1994,
BianchiApJ1996,XilourisA&A1997,XilourisA&A1998,XilourisSub1998}.

\citet{DevereuxApJ1990} find a very good correlation between the mass of
dust derived from IRAS fluxes at 60$\mu$m and 100$\mu$m and the total
mass of gas inside the optical radius for a sample of 58 galaxies. 
\citet{XuA&A1997} compared the extinction in the B band derived on a
sample of 79 galaxies using an energy balance model with that obtained 
from the gas column density using the relation measure in the Galaxy
(Eqn.~\ref{nh_ebv}). A good agreement is found. When the correlation is
analysed separately for each gas phase, a tighter correlation is found
for the H$_2$, suggesting that extinction is due essentially to dust
associated with the molecular gas. This is because in their sample,
molecular gas is the dominant component in the inner galaxy.
Indeed, when galaxies with a dominant HI component are selected,
extinction is mainly associated to the atomic gas. Therefore,
it is reasonable to use the gas distribution as a tracer of the
dust distribution.

In luminous, face-on, late-type spirals, H$_2$ peaks in the centre
and falls off monotonically with increasing distance from the centre.
This contrasts markedly with the central HI depression and the nearly
constant HI surface density across the rest of the optical disk
\citep{YoungARA&A1991}. The same behaviour has been observed in some
early type galaxies, although a good fraction of them presents a
 central depression and a flatter ring distribution for the molecular
gas.
Conducting an energy balance on M31 for several cells associated with 
diffuse FIR emission on M31, \citet{XuApJ1996b} found a quite flat 
face-on optical depth with the radius. Because of their selection
against FIR sources associated with molecular clouds and star-forming
regions, the optical depth profile is close to the HI, rather than 
to the H$_2$ component. A dust ring in emission has been observed by
SCUBA at 450$\mu$m and 850$\mu$m in the early type galaxy NGC 7331
\citep{BianchiMNRAS1998}. The dust emission correlates well with the
observed molecular ring (See Appendix~\ref{n7331} for a complete
discussion). A flat distribution for the B-band optical depth of the 
Galactic disk has been found by \citet{SodroskiApJ1997} from the optical
depth at $240\mu$m derived from DIRBE observation.

Nevertheless, most of the galaxy exhibit a centrally peaked molecular 
gas component, dominant on the atomic gas phase. This is the case of 
the late-type spiral NGC 6946 \citep{TacconiApJ1986}, whose dust 
distribution is the main concern of this thesis (Chapter.~\ref{model}), 
Therefore, I assume the dust to be distributed in a smooth  radial and 
vertical exponential disk, similar to the stellar one (Sect.~\ref{stardisk}). 
The number density of dust grains can be written as 
\begin{equation}
n(r,z)=n_0 \exp (-r/\alpha_d -\mid z\mid/\beta_d),
\label{dustexp}
\end{equation}
with $n_0$ the central number density. The radial and vertical
scalelengths $\alpha_d$ and $\beta_d$ can be selected independently
from the analogous stellar parameters.
Usually it is assumed $\alpha_d\approx\alpha_\star$ and 
$\beta_d\approx 0.5\beta_{\star}$. The choice of the last parameter
is mainly dictated by the impossibility of simulating the extinction
lanes in edge-on galaxies with a dust distribution higher than the
stars \citep{XilourisSub1998}.
Recently there have been suggestions for more extended dust
distributions, both from analysis of extinction of starlight and
FIR emission. \citet{PeletierA&A1995}, from an analysis of the
variation of scalelength ratios in different colours with inclination,
concluded that $\alpha_d/\alpha_\star\ge1$.
\citet{DaviesMNRAS1997} modelled the Galactic FIR emission at 140 $\mu$m
and 240 $\mu$m observed by the satellite $COBE$ with an extended dust
distribution with scalelengths $\alpha_d/\alpha_\star=1.5$ and 
$\beta_d/\beta_\star=2$. 
In all of seven edge-on spiral galaxies, \citet{XilourisSub1998} are
able to fit optical and NIR starlight with their radiative transfer
model only using a larger radial scalelength for dust with
respect to stars:
the mean value is $\alpha_d/\alpha_\star=1.4\pm0.2$.

The dust number density in Eqn.~(\ref{dustexp}) is normalised in the
model from the optical depth. For a model with face-on central optical
depth in the V-band $\tau_V$, the central number density is given by
\begin{equation}
n_0=\frac{\tau_V}{2 \beta_d \sigma_\mathrm{ext}(V)},
\label{prem3}
\end{equation}
where $\sigma_\mathrm{ext}(V)$ is the extinction cross section. 
For a model at a specific wavelength, the absorption coefficient
\begin{equation}
k_\lambda (r,z)=n(r,z) \sigma_\mathrm{ext}(\lambda)
\end{equation}
is integrated along a path to compute the optical depth and there is no 
need to specify the absolute values of $\sigma_\mathrm{ext}(\lambda)$ 
but only the ratio $\sigma_\mathrm{ext}(\lambda)/\sigma_\mathrm{ext}(V)$,
given by the assumed extinction law (Sect.~\ref{assu_ext}).
The dust disk has the same truncations as the stellar, at 6 scalelengths
both in the vertical and radial direction. 

In chapter~\ref{model}, I will explore various values of 
$\alpha_d/\alpha_\star$ and $\beta_d/\beta_\star$ and $\tau_V$, to 
better reproduce the characteristics of the dust FIR emission.

\section{The Monte Carlo code}
\label{montecarlo}

The model used in this thesis is based on the BFG  Monte Carlo code for the
radiative transfer (complete with scattering) in dusty spiral galaxies. 
The BFG code originally
included the radiative transfer of all the Stokes parameters, to 
also model polarisation. Dust properties were computed from the
\citet{DraineApJ1984} dust model, using Mie's theory for spherical
grains. For use in this thesis the BFG code has been simplified:
the radiative transfer is carried out only for the intensity using
empirical dust properties and 
phase functions
(Sect.~\ref{assu_ext}). Clumping \citep{BianchiSub1999} is not yet
included in this work.

Within the Monte Carlo method, the life of a photon (i.e. a unit of
energy in the program) can be followed through scattering and absorption
processes, until the radiation is able to escape the dusty medium.
I'll give here a brief description of the computation scheme of the 
Monte Carlo code, referring the interested reader to the BFG paper. 
The main steps are:
\begin{description}
\item[Emission:] the position of a photon in the 3-D space is derived 
according to the stellar distributions described in Sect.~\ref{stardisk}. 
The photon is emitted isotropically. Photons are emitted with unit
intensity.
\item[Calculation of optical depth:] the optical depth $\tau_T$ through the dust
distribution is now computed from the emission position along the photon 
travelling direction. A fraction $e^{-\tau_T}$ of all the energy
travelling in that direction is able to propagate through the
dust. With the Monte Carlo method it is then possible to extract the optical 
depth $\tau$ at which the 
photon impinges on a dust grain. This optical depth can be computed inverting
\begin{equation}
\int_0^{\tau} e^{-\xi} d\xi=R,
\end{equation}
where $R$ is a random number in the range [0,1]. If the derived $\tau$
is smaller than $\tau_T$, the photon suffer extinction, otherwise
escapes the dusty medium. This process is quite inefficient when the
optical depth of the dust distribution is small, most of the photons
leaving the dust distribution unaffected. To overcame this problem, the
{\em forced scattering} method is used \citep{CashwellBook1959,WittApJS1977}:
essentially, a fraction $e^{-\tau_T}$ of the photon energy is unextinguished 
and the remaining $1-e^{-\tau_T}$ is forced to scatter.
When the optical depth is small ($\tau_T<10^{-4}$) or the photon path is
free of dust, the photon escapes the cycle.
Once $\tau$ is known, the geometrical position corresponding to the
point where the photons collide with a dust grain is computed.
\item[Scattering and Absorption:] a fraction of the photon energy, given
by the albedo $\omega$, is scattered, while the remaining $(1-\omega)$
is absorbed. The scattering polar angle $\theta$, i.e. the angle between
the original photon path and the scattered direction, is computed using
the \citet{HenyeyApJ1941} scattering phase function (Eqn.~\ref{hgphase}),
inverting
\begin{equation}
\int_0^\theta \phi(\theta') \sin\theta'= R,
\label{phasinv}
\end{equation}
with $R$ a random number. The inversion of Eqn.~(\ref{phasinv}) is given
by the analytical formula
\begin{equation}
\theta=\arccos\left[\frac{1}{2g}\left(1+g^2-\frac{(1-g^2)^2}{(1+g(1-2R))^2}
\right)\right],
\end{equation}
with $g$ the asymmetry parameter (Sect.~\ref{extinction}). 
Another angle is needed to define the direction of the scattered
radiation. Since for spherical grains there is no preferential direction
perpendicular to the original photon path, an azimuthal angle is
extracted randomly in the range $[0,2\pi]$.

The original BFG code has been modified to store the information
about absorption, to be used in the derivation of dust temperature.
Using the model symmetries around the vertical axis and about the 
galactic plane to improve the signal-to-noise, a map of the absorbed 
energy is produced as a function of the galactocentric distance and 
the height above the plane.
\item[Exit conditions:] the last two steps are then repeated, using
the new direction of the scattered photon, the coordinates of the 
scattering point and the energy reduced by absorption.
The cycle is repeated until the energy of the photon falls below a
threshold value ($10^{-4}$ of the initial intensity) or until the exit 
conditions on $\tau$ are verified.
\end{description}
After the exit conditions are satisfied, the photon is
characterised by the last scattering point, its travelling direction and
its energy. The two symmetries of the model, planar and axial, are exploited 
to reduce the 
computational time: if the model is supposed to be observed from a point
at infinite distance in the (x,z) plane, each photon position is rotated 
around the symmetry axis until its direction is parallel to that plane;
then, for each photon coming from (x,y,z), another with the same direction
coming from (x,-y,z) is added; two other photons are added coming from (x,y,-z) and (x,-y,-z) and with a direction specular 
to the original one with respect to the galactic plane. A total of 4 photons
are produced from each one. The photons are then classified according to
the angle between their direction and the symmetry axis, to produce maps 
of the galaxy as seen from different inclinations. This is done by dividing
the whole solid angle in $N_B$ latitudinal bands of the same solid angle.
In BFG and here, $N_B=15$. There are therefore 8 independent images at
mean inclinations of 20$^\circ$, 37$^\circ$, 48$^\circ$, 58$^\circ$, 
66$^\circ$, 75$^\circ$, 82$^\circ$ and 90$^\circ$.  
When models are produced for a specific angle
only, as in Chapter~\ref{model}, it means that a band of solid angle
$4\pi/15$ is used, with a mean angle equal to the given one. 

Finally, all the photons in an angle band are projected into the plane of 
the sky according to their point of last scattering. For the models
of this thesis, I have used maps of 201x201 pixel, to cover a region
of 12x12 stellar radial scalelengths around the centre of the galaxy.
Maps of absorbed energy are derived to cover 6 dust scalelengths 
in the radial direction and in the positive vertical direction in 
101x101 pixels.

\section{Normalisation of the radiative transfer output}
\label{SED}
The Monte Carlo model described in the previous section is monochromatic, 
since the optical properties of dust must be specified for a particular
wavelength. Also the geometrical distributions of stars may depend on
$\lambda$ (Sect.~\ref{adopted}).
Therefore the absorbed energy maps contain information on the energy
absorbed by dust from a single wavelength only:  a map of the \emph{total} 
energy absorbed by dust can be produced running several monochromatic models
to cover the Spectral Energy Distribution (SED) of stellar radiation,
the main source of dust heating.

\begin{figure}
\centerline{\psfig{file=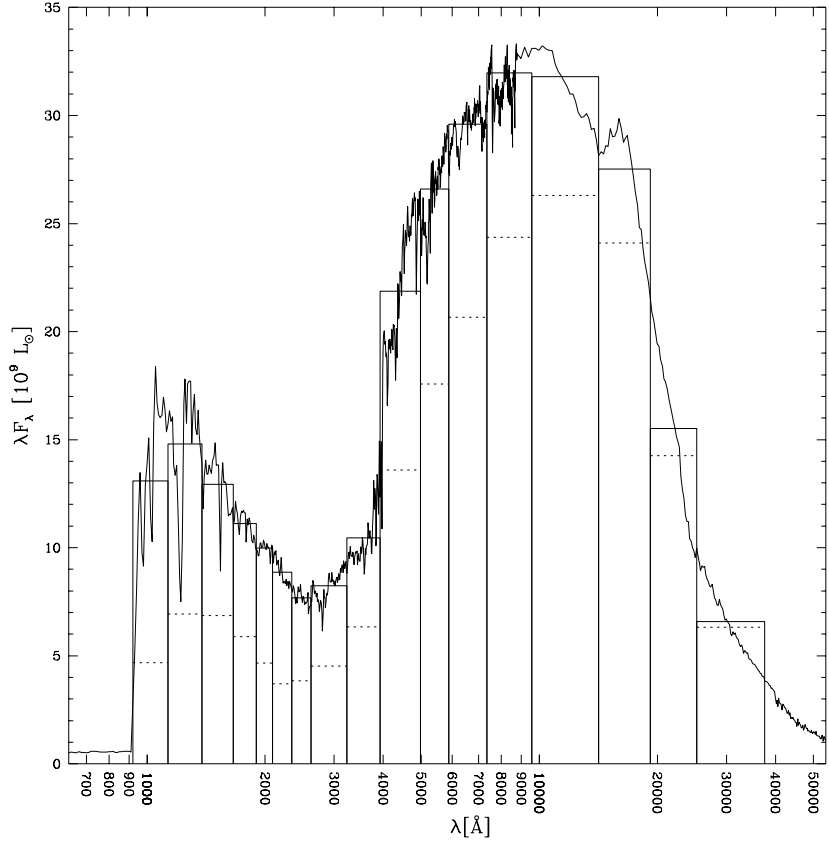,width=12cm}}
\caption{Synthetic spectrum of an unextinguished Sbc galaxy, according to the
spectrophotometric evolution model PEGASE of \citet{FiocA&A1997}. The 
17 wavelength bands used in the model (Tab.~\ref{band_def}) are also
shown. Because of the quantities plotted ($\lambda F_\lambda$ vs 
$\log (\lambda)$ ) the amount of energy emitted by stars in each band is
proportional to the area of the blocks.
For each band, the dashed line shows the amount of energy that
actually escapes the galaxy.}
\label{fig_sed}
\end{figure}

To describe the output of the program, I use in this chapter a model
with a stellar exponential disk of radial scalelength 
$\alpha_\star=2.5 $ kpc (the choice for a model of NGC 6946, see 
Sect.~\ref{n6946_scales}) and $\alpha_\star/\beta_\star=14.4$
(Sect.~\ref{adopted}). The adopted dust disk has the same radial
scalelength as the stars, but a vertical scalelength half the stellar one.
The central face-on optical depth in this model is $\tau_V=5$. Images have been
produced for the 15 contiguous angle bands defined in BFG
(Sect.~\ref{montecarlo}).  I use a synthetic
SED for a spiral galaxy, derived with the spectrophotometric
evolution model PEGASE\footnote{The spectra has been produced using the
parameters for an Sbc galaxy given in \citet{FiocA&A1997}, \emph{without}
taking dust into consideration, the extinction being modelled by the code
described in this thesis. For the spectrum, I have assumed a galaxy of
mass 10$^{11}$ M$_\odot$.} \citep{FiocA&A1997}. The spectrum is shown
in Fig.~(\ref{fig_sed}).

In Table~(\ref{band_def}) the 17 wavelength bands used in the model 
are defined: the
bands from UV1 to K have essentially the same spectral coverage as the
homonymous described by \citet{GordonApJ1997}, for which they provide 
extinction and scattering properties (Sect.~\ref{assu_ext}). Two
bands (namely EUV and LMN) have been added to extend the spectral
coverage in the ultraviolet up to the ionization limit and in the near
infrared. A Monte Carlo simulation is run for each of the band.
The energy emitted in each band is computed integrating the stellar SED
over the band limits (Fig.~\ref{fig_sed}).
In the example of this chapter, models are normalised to the {\em 
intrinsic} energy, i.e. the energy a galaxy would emit in the optical 
without intervening dust. For each band, I show in Fig.~(\ref{fig_sed})
the amount of energy that is able to escape dust, the rest being
absorbed.
When comparing the simulation to observed SEDs, as in the next 
chapter, the models are normalised to the \emph{observed} energy, i.e.
after radiation has been processed by dust. The {\em intrinsic} SED
is then inferred from the radiative transfer model.

\begin{table}
\centerline{
\begin{tabular}{lcc}
\hline
band&\multicolumn{2}{c}{$\lambda_\mathrm{i}<\lambda<\lambda_\mathrm{f}$}\\
    &\multicolumn{2}{c}{[\AA]}\\
\hline
EUV & 912 & 1125 \\
UV1 & 1125 & 1375 \\
UV2 & 1375 & 1655 \\
UV3 & 1655 & 1900 \\
UV4 & 1900 & 2090 \\
UV5 & 2090 & 2340 \\
UV6 & 2340 & 2620 \\
UV7 & 2620 & 3230 \\
U   & 3230 & 3930 \\ \hline
\end{tabular}
~~
\begin{tabular}{lcc}
\hline
band&\multicolumn{2}{c}{$\lambda_\mathrm{i}<\lambda<\lambda_\mathrm{f}$}\\
    &\multicolumn{2}{c}{[\AA]}\\
\hline
B   & 3930 & 4979 \\
V   & 4979 & 5878 \\
R   & 5878 & 7350 \\
I   & 7350 & 9500 \\
J   & 9500 & 14000 \\
H   & 14000& 19000 \\
K   & 19000& 25000 \\
LMN & 25000&37500 \\
    &      &      \\\hline
\end{tabular}
}
\caption{Wavelength ranges of the bands used in the model.}
\label{band_def}
\end{table}

Usually 10$^7$ photons are run for each simulation. Approximatively, 24
hours of computing on a SUN ULTRA 5 workstation are needed for a
complete set of 17 simulations with $\tau_\mathrm{V}=1$.
Optical images are produced in units of surface brightness
(L$_\odot$ kpc$^{-2}$ $\mu$m$^{-1}$ sterad$^{-1}$). 
An example of the images produced by the radiative transfer code in 
the optical (B-band) is given in Fig.~(\ref{fig_opfir}).

\section{The FIR code}
\label{fircode}
For each waveband, a map of the absorbed energy density as a function 
of the galactocentric distance and of the height above the galactic 
plane is produced. Of all the radiation absorbed in a specific waveband,
a fraction is absorbed by grains not emitting at the thermodynamic 
equilibrium. Since I am interested in modelling only the thermal
emission, a correction is applied to the absorbed energy maps, as
described in Sect.~(\ref{sec_desert}).

\begin{figure}[t]
\centerline{\psfig{file=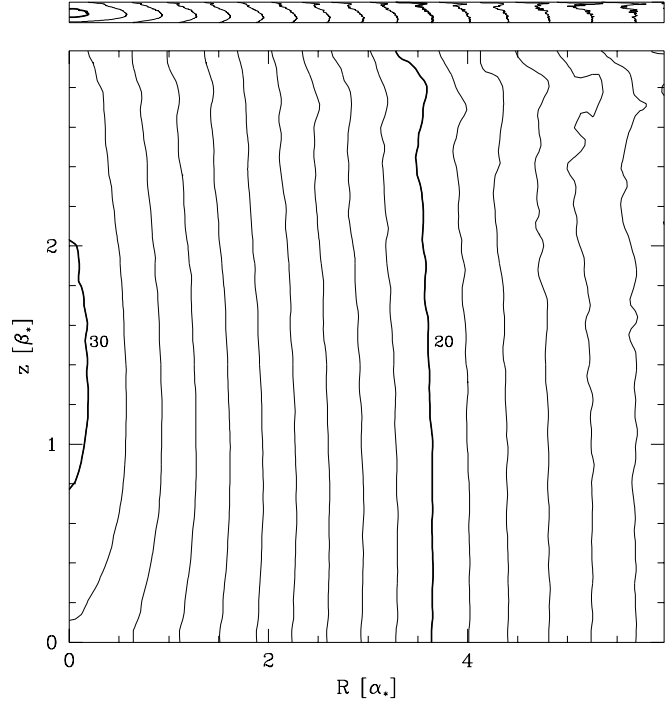,width=9cm,bbllx=100pt,bburx=520pt,bblly=150pt,bbury=610pt}}
\caption{Dust temperature map on a meridian plane for the model of this
chapter. Temperature contours are plotted every 1K. Contours at 20 K and 30
K are highlighted by a label and a thicker line. In the main plot, the scale 
along the z-axis has been expanded for clarity. A plot with the actual
ratio between the axes is shown at the top.}
\label{fig_tempe}
\end{figure}

After all the absorbed energy maps are scaled to the energy input of
each band, they are summed together, to produce a single map of
the absorbed energy that goes into thermal emission.
The final map, $W_\mathrm{abs}(r,z)$, has units of energy per unit time
per unit volume (L$_\odot$ kpc$^{-3}$). Knowing the number density of dust
grains, $n(r,z)$, the power absorbed by a single grain, 
$W_\mathrm{abs}(r,z)/n(r,z)$, can be derived. Equating the absorbed
and emitted radiation, a relation for the temperature $T_{\mathrm{d}}(r,z)$
can be then found. Assuming that dust grains have all the same radius, 
it can be derived from Eqn.~(\ref{eq_wem}) that
\begin{equation}
\frac{W_\mathrm{abs}(r,z)}{n(r,z)}= 4\pi a^2 \int_0^\infty
Q_{\mathrm{em}}(\lambda)\pi B_\lambda(T_{\mathrm{d}}(r,z))d\lambda.
\label{prem1}
\end{equation}
In the case of an exponential dust distribution, the number density
$n(r,z)$ can be obtained by Eqn.~(\ref{dustexp}). The central 
number density of Eqn.~(\ref{prem3}) can be rewritten in terms of the 
grain radius $a$ and the extinction efficiency as
\begin{equation}
n_0=\frac{\tau_\mathrm{V}}{2 \beta_d \pi a^2 Q_\mathrm{ext}(V)}.
\label{prem3b}
\end{equation}
Finally, substituting Eqns.~(\ref{dustexp}) and (\ref{prem3b}) in 
Eqn.~(\ref{prem1})
\begin{equation}
\frac{\beta_d W_\mathrm{abs}(r,z)}{2 \tau_\mathrm{V} \exp (-r/\alpha_d -\mid
z\mid/\beta_d)}= \int_0^\infty \frac{Q_{\mathrm{em}}(\lambda)}{Q_\mathrm{ext}(V)} \pi B_\lambda(T_{\mathrm{d}}(r,z))d\lambda. 
\label{prem4}
\end{equation}
A map of the temperature can thus be derived inverting Eqn.~(\ref{prem4}).
I have used the emissivity law $Q_{\mathrm{em}}(\lambda)$ defined in
Eqn.~(\ref{qema_used}). The temperature contour map is shown in 
Fig.~(\ref{fig_tempe}).

The emission coefficient (Eqn.~\ref{emeff}) can be easily derived from the 
temperature from the formula
\begin{eqnarray}
j_\lambda(r,z)&=&n(r,z) \sigma_\mathrm{em} B_\lambda(T_{\mathrm{d}}(r,z))
\nonumber \\ 
&=& \frac{\tau_\mathrm{V}}{2\beta_\mathrm{d}} 
\exp (-r/\alpha_d -\mid z\mid/\beta_d)
\frac{Q_{\mathrm{em}}(\lambda)}{Q_\mathrm{ext}(V)}
B_\lambda(T_{\mathrm{d}}(r,z))
\label{inteff}
\end{eqnarray}

It is interesting to note that both the determination of the temperature 
and the emission coefficient are independent of the grain radius $a$.
If a dust model was to be used for
$Q_{\mathrm{em}}(\lambda)/Q_\mathrm{ext}(V)$, rather then an empirically 
determined value, the emissivity would have been dependent on the dust grain 
radius and on its composition.
Therefore, the approach used here is analogous to use a mean value over the 
dust distribution of sizes and materials for both the radius $a$ and the
ratio $Q_{\mathrm{em}}(\lambda)/Q_\mathrm{ext}(V)$. 

\begin{figure}[t]
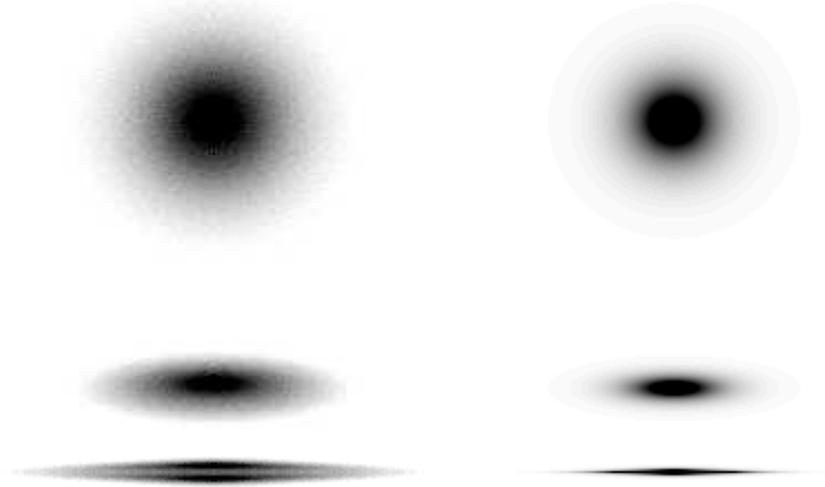

\centerline{
\psfig{file=program/B_20.ps,width=6.0cm,bbllx=0pt,bburx=551pt,bblly=145pt,bbury=648pt}
\psfig{file=program/n100_20.ps,width=6.0cm,bbllx=0pt,bburx=551pt,bblly=145pt,bbury=648pt}
}
\centerline{
\psfig{file=program/B_75.ps,width=6.0cm,bbllx=0pt,bburx=551pt,bblly=320pt,bbury=470pt}
\psfig{file=program/n100_75.ps,width=6.0cm,bbllx=0pt,bburx=551pt,bblly=320pt,bbury=470pt}
}
\centerline{
\psfig{file=program/B_90.ps,width=6.0cm,bbllx=0pt,bburx=551pt,bblly=370pt,bbury=420pt}
\psfig{file=program/n100_90.ps,width=6.0cm,bbllx=0pt,bburx=551pt,bblly=370pt,bbury=420pt}
}
\caption{Example of optical and FIR images produced by the simulation:
B-band (left) and 100 $\mu$m (right) at inclinations of 20$^\circ$, 
75$^\circ$ and 90$^\circ$ (from top to bottom).  The optical images are
noisy because they are the direct output of the Monte Carlo code, while
the FIR images are produced through the analytical integration of
Eqn.~(\ref{inteff}). The edge-on FIR image is thinner than the optical because
the emission comes from the dust disk, rather than from the stellar as
in the optical image.  }
\label{fig_opfir}
\end{figure}

FIR images are created integrating the emission coefficient
along a given line of sight through the dust distribution,
under the assumption that dust is optically thin to FIR radiation. 
Using the emissivity as in Eqn.~(\ref{qema_used}) this is justified
for any model with reasonable $\tau_\mathrm{V}$.
As for the optical images, the far infrared images are produced in units of 
surface brightness (L$_\odot$ kpc$^{-2}$ $\mu$m$^{-1}$ sterad$^{-1}$).
FIR images have the same extent and resolution of the optical images, i.e.
a region of 12x12 stellar radial scalelengths around the centre of the galaxy
is mapped in 201x201 pixels.
As an example, FIR images are shown in Fig.~\ref{fig_opfir}, at
100 $\mu$m.

\begin{figure}
\centerline{\psfig{file=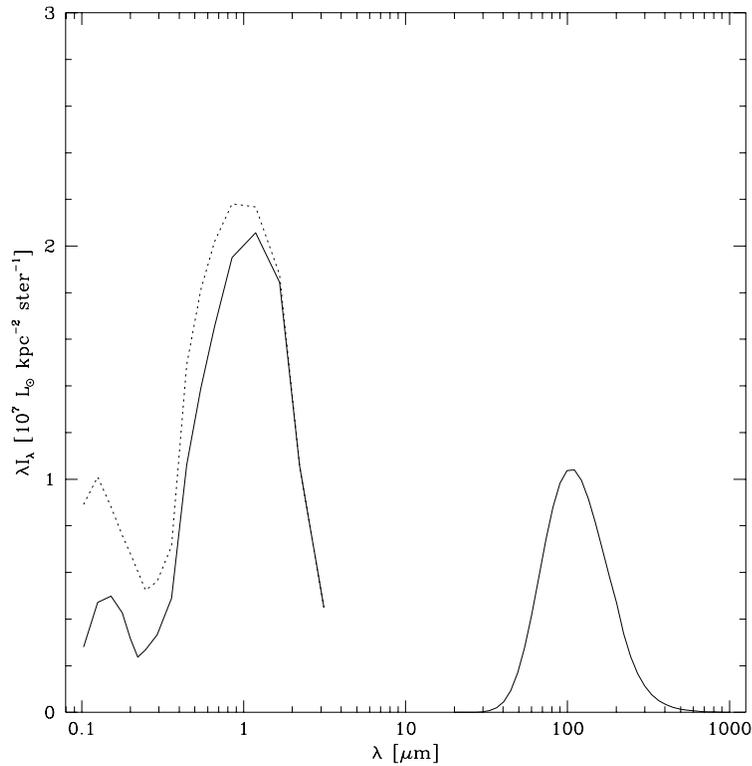,width=10cm,bbllx=55pt,bburx=570pt,bblly=160pt,bbury=700pt}}
\caption{Surface brightness inside the B-band half-light radius for the
model of this chapter, in the optical (using the SED of Fig.~\ref{fig_sed})
and in the FIR (solid line), for images with an inclination of
$20^\circ$. The dotted line shows the same quantity
measured for a transparent model at the same inclination.  }
\label{fig_spectrum}
\end{figure}

Both optical and FIR images can be analysed as real images. In 
Fig.~(\ref{fig_spectrum}) I show the surface brightness
measured over a circular aperture corresponding to the half-light 
radius (in the B-band), for both optical and FIR images at an
inclination of 20$^\circ$ (solid line). The half-light radius
is observed, i.e. it has been derived from the surface brightness
distribution in the B band image.
The surface brightness inside the half light radius for a transparent 
galaxy seen at that inclination is also shown (dashed line).
From Fig.~(\ref{fig_spectrum}) it is not straightforward to derive a relation
between the absorbed and the emitted energy, i.e. it is not possible to
compare the area between the two curves in the UV/optical/NIR (the
absorbed energy) with the area under the FIR curve (the energy emitted by 
dust). This is because we are not looking at the total
energy emitted by the model, but at the energy coming from a specific
inclination: because of the reduced extinction and the influence of
scattering, a model at low inclination looks more transparent
than in the edge-on case. Furthermore, part of the absorbed energy
has been disregarded because is not going into thermal
equilibrium processes.
The sum of the total energy emitted by dust and of the total energy 
observed in the optical, for all the inclination bands, do equal the
total intrinsic unextinguished stellar energy.

\section{Summary}

In this Chapter I have described the code used to model the FIR
emission. A modified version of the BFG radiative transfer code is used
to produce a map of the energy absorbed by dust as a function of the
position within the dust distribution, together with the usual
images of the attenuated emission from the stellar distribution.
The whole spectral range of non-ionising stellar radiation is used.
Using the model for the dust emissivity derived in Sect.~\ref{emission},
a temperature distribution is derived from the absorbed energy map.
Finally, the FIR emission is computed, integrating over the dust
number density and temperature distributions. Images of FIR emission 
for any wavelength and galaxy inclination are produced.

In the next Chapter, I will use the code to model the FIR emission of
the spiral galaxy NGC~6946. For a chosen dust distribution and optical
depth, the output for the stellar emission from the radiative transfer 
code will be normalised to the observed stellar SED. The modelled FIR
emission will be then compared with the observations. Dust parameters
will be changed to match the observed FIR spectrum and the spatial
distribution of FIR images (as defined by a radial exponential
scalelength).

\chapter{Modelling NGC 6946}
\label{model}
NGC~6946 is a large (D$_{25}$=11.5', \citealt{RC3}, RC3) nearby Sc galaxy, 
seen from an inclination of 34$^\circ$ \citep{GarciaGomezA&AS1991}. 
\citet{TullyBook1988} give a distance of 5.5 Mpc, similar to those
obtained by other authors with a variety of methods 
\citep{DeVaucouleursApJ1979,EastmanApJ1996,SchonigerA&A1994}, although
there are estimates as big as 10.1 Mpc \citep{SandageApJ1974}. In this
work I assume a distance of 5.5 Mpc, that gives a scale on the sky
of 27 pc arcsec$^{-1}$.

The optical appearance of the galaxy is characterised by the prominent
spiral arms (Fig.~\ref{fig_images}). Six separate spiral arms can be seen in
optical images \citep{TacconiApJ1990}: the three spiral arms originating
from the northeast quadrant are brighter and more developed than the other 
in the southwest. In his high-resolution extinction maps based on an
energy balance method, \citet{TrewhellaMNRAS1998,TrewhellaThesis1998} 
finds that the interarm regions between the two prominent northeast
spiral arms is a site of strong extinction, rather than being intrinsically
less luminous. Indeed, light polarised by dust is observed in the interarm
regions as well as in the spiral arms \citep{FendtA&A1998}. There is
evidence for recent star formation along the spiral arms and a mild
starburst in the centre \citep{TacconiApJ1990}. The distribution of
atomic and molecular gas in the inner 10' has been studied by
\citet[Sect.~\ref{twodisks}]{TacconiApJ1986}.

The galaxy is marginally resolved in FIR observations from IRAS and KAO
\citep{EngargiolaApJS1991,AltonA&A1998}, with the FIR emission following
the giant HII regions along the spirals and the bright central emission.
The low resolution 200$\mu$m ISO image \citep{AltonA&A1998} shows a
morphology similar to the 100$\mu$m IRAS observation (Fig.~\ref{fig_images}).

In this chapter I will apply the model described in
Chapter~\ref{program} to the optical and FIR observations of NGC~6946,
to determine the spatial distribution and content of dust in the galaxy.

\begin{figure}[t]
{\em
To reduce the size of this file, these figures have been omitted.
The full version of this thesis, including the present figures,
can be found at
{\tt http://www.arcetri.astro.it/$^\sim$sbianchi/tesi/thesis.ps.gz}
}
\vspace{13.5cm}
\caption{Images of NGC~6946 in B-band (top-left), K-band (top-right)
\citep{TrewhellaThesis1998}, at 100-$\mu$m (bottom-left) and 
200-$\mu$m (bottom-right) \citep{AltonA&A1998}. All images have the
same orientation (North to the top, East to the right) and cover the
same area of the sky (9'x9').}
\label{fig_images}
\end{figure}

\section{The stellar Spectral Energy Distribution}
\label{starsed}
The observed Spectral Energy Distribution of the stellar radiation in 
NGC 6946 has been constructed using data available in the literature and 
extrapolating a few data points at the edges of the wavelength range,
where no observations were available. The SED (in units of surface 
brightness) for a circular aperture of 5' (corresponding to the half
light radius in the B-band) is presented in Table~\ref{tab_sed}.
All the data have been corrected for a Galactic
extinction $A_B$=1.73 (RC3) using a standard extinction law
\citep{WhittetBook1992}.

Optical and Near Infrared data comes from \citet{EngargiolaApJS1991},
that provides fluxes for NGC 6946 in the bands U, B, g, V, r, I, J, H, 
K for the half light radius aperture.

In the NIR part of spectrum, the emission could be due in part 
to small grains of dust.  Dust contribution can be computed from 
the \citet{DesertA&A1990} dust model for dust heated by the local 
interstellar radiation field.
According to the model, emission in filter K is 0.004 
times the emission in the IRAS 12 $\mu$m filter.  This is still valid
even if the radiation field is 100 times the local one.
Using NGC 6946 emission at 12 $\mu$m, derived from IRAS HiRes images
(Sect.~\ref{dustsed}), it is therefore possible to compute
the contribution of dust to the emission in K and compare it to
the observed emission in that band. As expected, the K-band 
emission can be considered as purely stellar, the dust contributing only
about 0.3\% to the total.

The stellar emission at longer wavelengths has been extrapolated using 
the synthetic galactic SEDs of \citet{FiocA&A1997}. In this wavelength 
range, the SED of their unextinguished spiral galaxy models is almost
linear when $\lambda F_\lambda$ is plotted versus $\log \lambda$. 
A value at 5 $\mu$m has thus been computed from the K-band flux.

\begin{table}
\centerline{
\begin{tabular}{ccc}\hline
band & $\lambda$ &I$_\lambda$ (5'$\bigcirc$\makebox[0in][r]{/\hspace{0.6ex}})\\
&$\mu$m&10$^6$ L$_\odot$ kpc$^{-2}$ $\mu$m$^{-1}$ sterad$^{-1}$\\ \hline
                & 0.091 & 25             \\
\emph{short}-UV & 0.165 & 14 $\pm$ 4     \\
\emph{medium}-UV& 0.250 &  9 $\pm$ 6     \\
\emph{long}-UV  & 0.315 & 15 $\pm$ 7     \\
U               & 0.360 & 17 $\pm$ 2     \\
B               & 0.435 & 21 $\pm$ 2     \\
g               & 0.495 & 20 $\pm$ 2     \\
V               & 0.554 & 20 $\pm$ 1     \\
r               & 0.655 & 20 $\pm$ 2     \\
I               & 0.850 & 20 $\pm$ 2     \\
J               & 1.250 & 14 $\pm$ 1     \\
H               & 1.650 & 12 $\pm$ 1     \\
K               & 2.200 &  5.5 $\pm$ 0.5 \\
                & 5.000 & 0.25           \\
		\hline
\end{tabular}
}
\caption{NGC 6946 surface brightnesses in several bands for a 
circular aperture of diameter 5' (corresponding to the half light 
radius in the B-band) \citep{EngargiolaApJS1991}.
Data have been derived from \citet{EngargiolaApJS1991} and 
\citet{RifattoA&AS1995b} and corrected for Galactic extinction, 
see text for details.}
\label{tab_sed}
\end{table}
Fluxes for the non-ionising UV radiation shortward of the U-band
have been taken from \citet*{RifattoA&AS1995b}: the authors give total
magnitudes for 2400 galaxies in three photometric bands centred at 
1650\AA\ (\emph{short}-UV), 2500\AA\ (\emph{medium}-UV), 3150\AA\ 
(\emph{long}-UV). Data collected by several satellites (notably IUE), 
balloon and rocket-borne experiments with different apertures and 
sensitivities have been homogenised to a common scale.
After dividing the sample in three morphological bins (E/S0, Sa/Sb,
Sc/Sd), standard luminosity profiles in the B-band \citep{ButaAJ1995}
have been fitted to the data in each of the three photometric band
and the total magnitudes were derived.

Using the luminosity profiles for Sc/Sd galaxies, I have derived 
the UV magnitudes for NGC 6946 inside the aperture used for
Optical and NIR data. Fluxes have been derived from magnitudes 
using the calibration described in \citet*{RifattoA&AS1995a}.
Errors in the fluxes are quite big, mainly arising from the
aperture correction. 

The flux at the Lyman limit (912\AA) has been extrapolated after
observing that in a $\lambda F_\lambda$ versus $\log \lambda$ plot
the observed SED is flat for the \emph{short}- and \emph{medium}-UV.
I have assumed that the same trend is valid down to the ionization
limit. I haven't included the ionising UV in the model: this will be
justified later (Sect.~\ref{noion}).

\section{The dust Spectral Energy Distribution}
\label{dustsed}
The FIR output of the code is compared to the SED of dust emission
as measured from the IRAS and ISO maps of \citet{AltonA&A1998}.
The SED (in units of surface brightness) for a circular aperture 
equivalent to the half light radius is presented in 
Table~\ref{tab_sed_2}, in the same units as for Table~\ref{tab_sed}.

Shortward of 100$\mu$m I have used IRAS High Resolution (HiRes)
images. Original data from the IRAS satellite \citep{NeugebauerApJL1984} 
have a coarse resolution (FWHM $\approx$ 1.5' x 5' at 60 $\mu$m). HiRes
images are produced using a model of the response of IRAS detectors and 
a process called Maximum Correlation Method, to restore a resolution
close to the diffraction limit \citep[][and references therein]{RiceAJ1993}.
As an example, resolution of HiRes images is  45'' x 60'' and 80'' x
100'' (FWHM), at 60 $\mu$m and 100 $\mu$m, respectively
\citep{AltonMNRAS1998,AltonA&A1998}. Derived values have been colour 
corrected using the corrections for NGC~6946 derived by 
\citet{RiceApJS1988} (18\%, -14\%, 7\%, 4\% at 12, 25, 60 and
100 $\mu$m). Integrated values are consistent (within a 20\% error, 
\citealt{EngargiolaApJS1991,AltonA&A1998}) with the analogous data
provided by \citet{EngargiolaApJS1991}, derived on previous
enhanced resolution IRAS images.

Data at 200$\mu$m comes from observations with the ISOPHOT instrument
\citep{LemkeA&A1996} aboard the ISO satellite \citep{KesslerA&A1996},
with a resolution of 117'' (FWHM).
Integrated values have an error of 15\%, but the present calibration of
the instrument may to be overestimated by about 30\%
\citep{AltonA&A1998}. The value in Table~\ref{tab_sed_2} is consistent,
within the errors, with the measurements by \citet{EngargiolaApJS1991} 
on 200$\mu$m images taken by the air-born telescope KAO.

Finally, I have included in Table~\ref{tab_sed_2} data at 160$\mu$m
derived from the KAO telescope \citep{EngargiolaApJS1991}, with a
resolution of 45'' (FWHM).

\begin{table}[t]
\centerline{
\begin{tabular}{cc}\hline
$\lambda$ &I$_\lambda$ (5'$\bigcirc$\makebox[0in][r]{/\hspace{0.6ex}})\\
$\mu$m&10$^6$ L$_\odot$ kpc$^{-2}$ $\mu$m$^{-1}$ sterad$^{-1}$\\ \hline
 12 & 0.35 $\pm$ 0.07     \\
 25 & 0.10 $\pm$ 0.02     \\
 60 & 0.14 $\pm$ 0.03     \\
100 & 0.10 $\pm$ 0.02     \\
160 & 0.05 $\pm$ 0.01     \\
200 & 0.031 $\pm$ 0.005     \\
450 & 2 10$^{-4}$ --  6 10$^{-4}$  \\
850 & 1.2 10$^{-5}$ -- 1.8 10$^{-5}$   \\
\hline
\end{tabular}
}
\caption{NGC 6946 surface brightnesses in the FIR and sub-mm, for a
circular aperture of diameter 5' (corresponding to the half light 
radius in the B-band) \citep{EngargiolaApJS1991}. Data have been 
derived from ISO and IRAS HiRes images \citep{AltonA&A1998} and 
KAO observations \citep{EngargiolaApJS1991}. Data at 450 $\mu$m and 850
$\mu$m come from SCUBA observations (Sect.~\ref{scuobs}).}
\label{tab_sed_2}
\end{table}

\section{High resolution SCUBA observations of NGC 6946}
\label{scuobs}

NGC 6946 was observed by me using the Sub-millimetre Common User Bolometer 
Array (SCUBA) on the 15 m James Clark Maxwell Telescope in Hawaii.
Observations were carried out on April 10, 11 and June 17, 18, 19, 20 1998,
at 450$\mu$m and 850$\mu$m.

SCUBA consists of two bolometer arrays of 91 elements for the short
wavelengths and 37 elements for the long wavelengths. The {\em 
short-wavelength array} is optimised for observing at 450-$\mu$m and the
{\em long-wavelength array} for observing at 850-$\mu$m
\citep{HollandMNRASprep1998}. The camera is mounted on the Nasmyth focus of
the telescope. The arrays have a field of view of about 2.3 arcmin in 
diameter and can be used simultaneously, by means of a dichroic beamsplitter.
The spacing between the bolometers does not produce instantaneously 
fully sampled images. Therefore, it is necessary to move the secondary mirror 
of the telescope according to specific patterns.

For sources smaller than 2.3 arcmin, the {\em jiggle-map} mode is used:
the secondary mirror moves according to a 64-point pattern to fully
sample the selected region of sky at both long and short wavelengths,
and chops to off-source positions to remove
the sky background. The telescope nods to remove slowly varying atmospheric 
gradients. The highly inclined galaxy NGC 7331 \citep[][see also
Appendix~\ref{n7331}]{BianchiMNRAS1998} and the edge-on NGC 891 
\citep{AltonApJL1998} have been observed using this mode\footnote{A more 
extensive description of observations with the {\em jiggle-map} mode is 
given in Appendix~\ref{n7331}.}.
The dimension of the arrays is suitable to contain the extent of the two
galaxies along their minor axis. Chopping to relatively distant positions 
perpendicular to the major axes ensures the sampling of a source-free
portion of the sky.

For sources larger than 2.3 arcmin, like nearby face-on galaxies, the {\em 
jiggle-map} mode is not suitable. The need for source-free observations
would lead to very large chop throws, resulting in a degradation of the
sky background subtraction and of the beam size. The {\em scan-map} mode
is therefore used in this case. The telescope scans the source at a
rate of 24 arcsec per second, along specific angles to ensure a fully
sampled map. Meanwhile the secondary chops with a frequency of 7.8 Hz
within the observed field.
While this ensures a correct subtraction of the sky background,
the resulting maps unfortunately have the profile of the source
convolved with the chop. The profile of the source is restored deconvolving
the chop from the observed map by mean of Fourier Transform analysis.

Scan-maps of NGC~6946 presented here are fully sampled over an area of 8'x8'. 
Each set of observations consisted of six scans, with different chop
configurations: chop throws of 20'', 30'' and 65'' along RA and Dec are
needed to retrieve the final image.
Data have been reduced using the \textsc{STARLINK} package
\textsc{SURF}~\citep{JennessMan1997}. Images were first flat-fielded to
correct for different sensitivities of the bolometers. Noisy bolometers
were masked and spikes from transient detections removed by applying a
$5\sigma$ clip. A correction for atmospheric extinction was applied,
using measures of the atmosphere opacities taken several times during 
the nights of observation. Zenith optical depth varied during the six
nights, with $\tau_{450}=0.4-2.5$ and $\tau_{850}=0.1-0.5$. The 450$\mu$m
opacity on the last night was too high ($\tau>3$) for the source to be
detected and therefore the relative maps were not used for this
wavelength.
Because of the chopping in the source field and not along the scan
direction, each bolometers sees a different background: a baseline,
estimated from a linear interpolation at the edges of the scan, has been
subtracted for each bolometer in each map. 

Images have been corrected for systematic sky variation. Sky fluctuations 
could be derived by observing the time sequence of observations for bolometers 
with negligible source signal. However, for large objects observed in
scan-map mode, it is difficult to disentangle the signal due to the sky
from the source. This problem is overcome by subtracting from the data
a model of the source, obtained from the data themselves. The sky
variation for all the bolometers is then derived and subtracted.

Data taken with the same chop configuration were rebinned together into
a map in an equatorial coordinate frame, to increase the signal to
noise. Six maps with 3'' pixels were finally obtained for each
wavelength, combining 33 and 25 observations, at 850 and 450 $\mu$m,
respectively.  In each of the six maps the signal from the source is 
convolved with a different chop function. The Fourier Transform (FT) of a
map is therefore the product of the FTs of the source signal and of the chop. 
Since the latter is a sine wave, a simple division should retrieve the 
FT of the source and the deconvolved image, after applying an inverse FT
back into image space. In principle, an observation with a single chop
configuration would be needed for this purpose. However, problems arises 
near the zeros of the sine-wave of the chop function FT. 
At these frequencies, the noise is boosted up and the signal to noise of the 
final image is significantly reduced. Therefore, observation taken with
different chop configurations are used, with chop throws selected in a
way that the zeros of one chop FT do not coincide with those of another,
except at zero frequency. The noise introduced in the source FT by
the division with one chop FT are thus smoothed by coadding with another
source FT with noise at different frequencies. This method is known as
the Emerson II technique \citep{HollandMNRASprep1998,JennessProc1998}. The
six chop configurations described above are recommended for this
technique. 

Unfortunately the deconvolution introduces some artifacts in the images,
like curved sky background. This may be due to residual, uncorrected,
sky fluctuation at frequencies close to zero, where all the chops FT 
goes to zero. Work to solve this problem is ongoing
(Jenness, private communication). To enhance the contrast between the
sky and the source, I have modelled a curved surface from the images,
masking all the regions were the signal was evidently coming from the
galaxy. The surface has been then subtracted from the image.

\begin{figure}
\centerline{
\psfig{figure=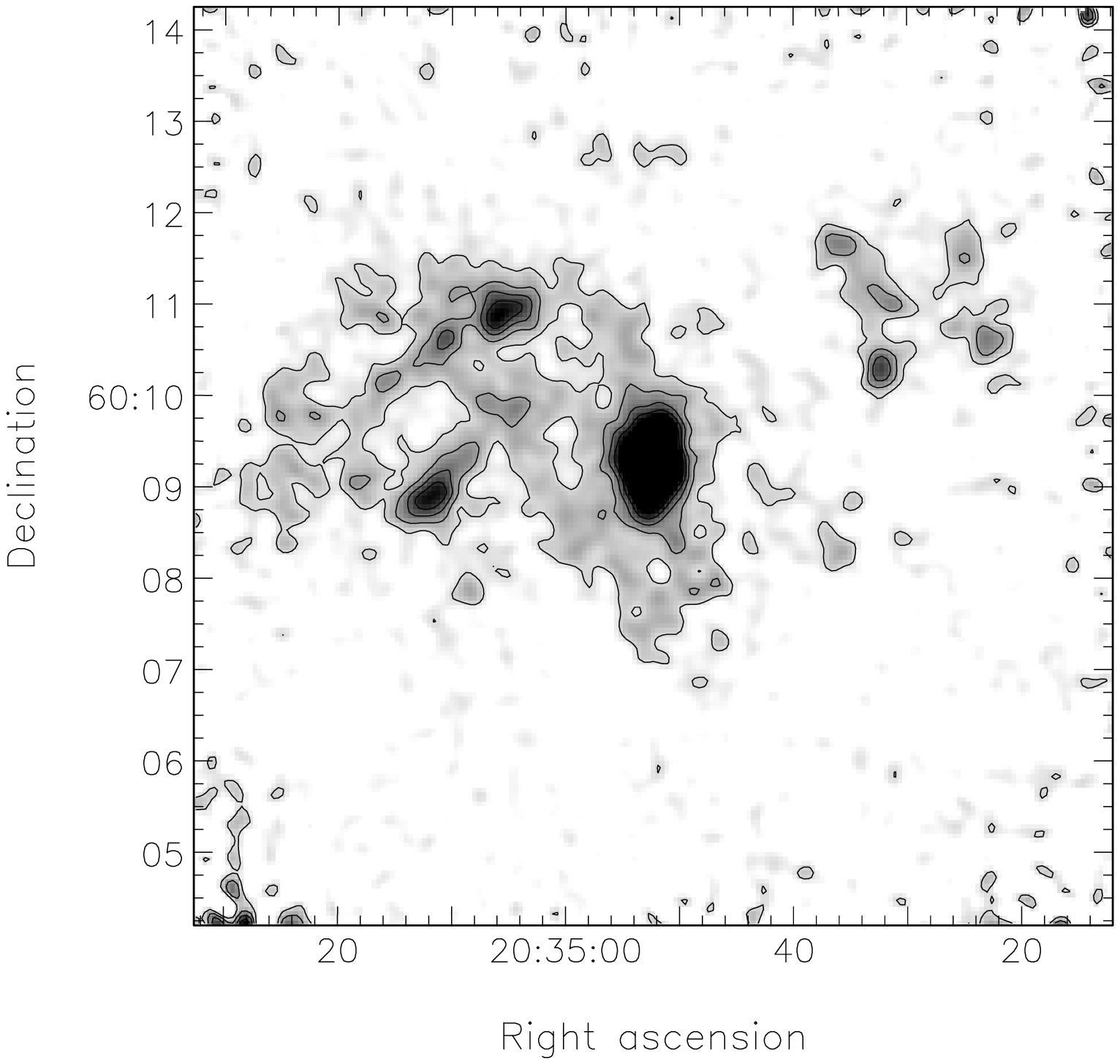,height=7.5cm}
\psfig{figure=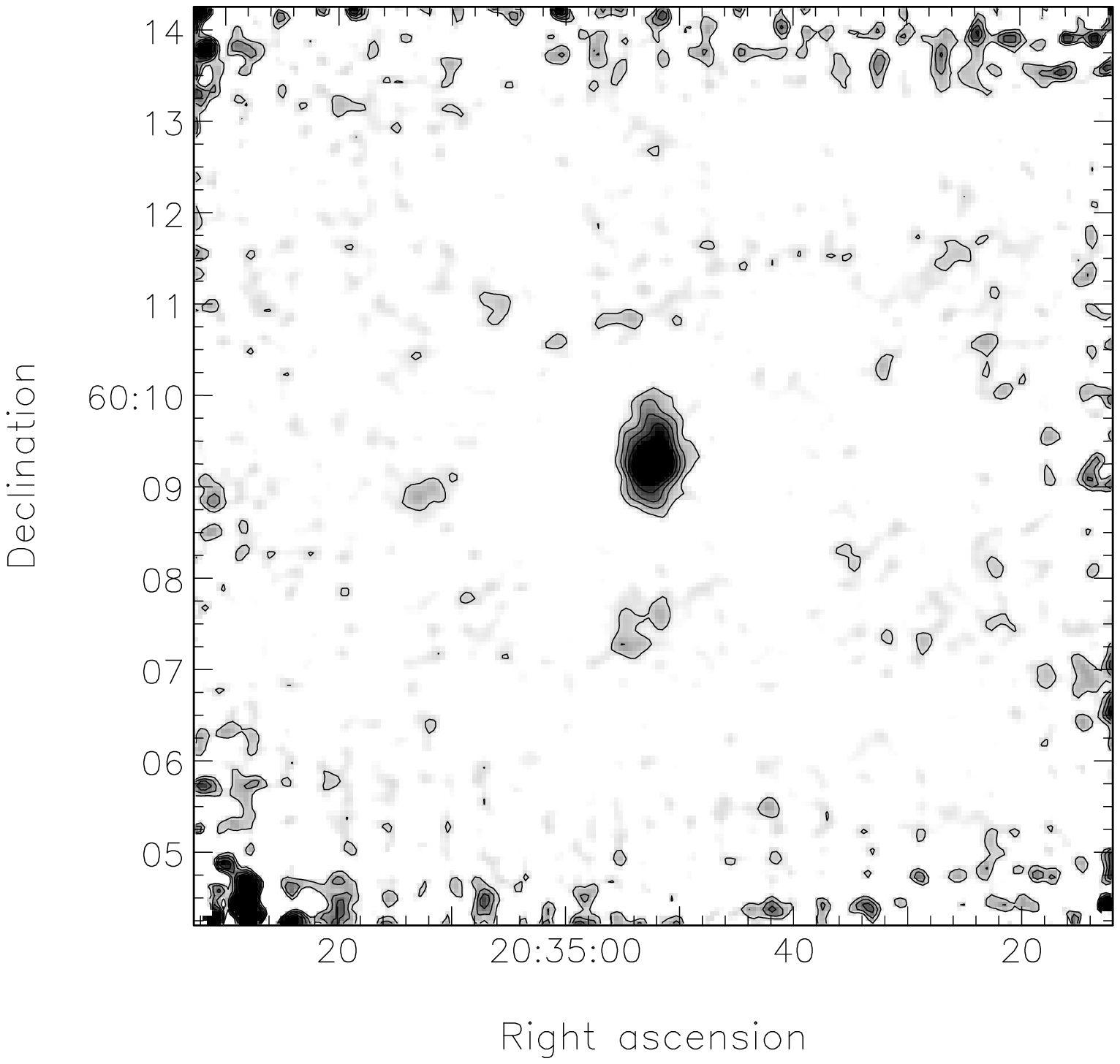,height=7.5cm}
}
\caption{Sub-mm images of NGC 6946, at 850 $\mu$m (left) and 450 $\mu$m
(right). Grey scales show features 1 $\sigma$ above the sky, while
contours starts at $3\sigma$ and have steps of $3\sigma$. Both images
have a beam size FWHM=17.7''. An area of 10'x10' is displayed, although
only the central 8'x8' are guaranteed to be fully sampled. North is on
top, East on the left.  The centre of the galaxy is in RA=20$^h$ 34$^m$ 
52.3$^s$ and Dec=60$^\circ$ 09$'$ 14.21'' \citep[J2000;][]{RC3}.
}
\label{scuba_6946}
\end{figure}

Calibration was achieved from scan-maps of Uranus, that were reduced in
the same way as the galaxy. Comparing data for each night I derived a
relative error in calibration of 8 per cent and 17 per cent, for 850-$\mu$m 
and 450-$\mu$m
respectively. From the planet profile, the beam size was
estimated: a FWHM of 15.2'' and 8.7'' were measured for the beam at
850 and 450 $\mu$m, respectively. To increase the signal to noise, the 
850 $\mu$m image has been smoothed with a gaussian of 9'', thus
degrading the beam to a FWHM of 17.7''. The  450 $\mu$m image has been
smoothed to the same resolution as for the 850 $\mu$m one, to facilitate
the comparison between features present in both. The sky $\sigma$ in the 
smoothed
images is 3.3 mJy beam$^{-1}$ at 850 $\mu$m and 22 mJy beam$^{-1}$ at 
450 $\mu$m. The final images, after removing the curved background and
smoothing are presented in Fig.~\ref{scuba_6946}. For each wavelength,
the grey scale shows all the features $>1\sigma$, while contours starts
at $3\sigma$ and have steps of $3\sigma$.

\begin{figure}
{\em To reduce the size of this file, this figure has been omitted.
The full version of this thesis, including the present figure,
can be found at
{\tt http://www.arcetri.astro.it/$^\sim$sbianchi/tesi/thesis.ps.gz}
}
\vfill

\caption{U-band image of NGC 6946 \citep{TrewhellaThesis1998} with 
overimposed the same 850 $\mu$m contours as in Fig.~\ref{scuba_6946}.
}
\label{over}
\end{figure}

The 850$\mu$m image shows a bright nucleus and several features that 
clearly trace the spiral arms (see Fig.~\ref{over}, where a U-band image 
of the galaxy \citep{TrewhellaThesis1998} is shown with the sub-mm contour
overimposed). As already seen in optical images \citep{TacconiApJ1990},
the spiral arms originating in the northeast quadrant are more pronounced 
than the others, where only regions with bright HII regions have
detectable emission. The 850$\mu$m image presents a striking similarity
to the CO (J=2-1) emission map in \citet{SautyA&A1998}, observed with
the IRAM 30m radiotelescope with a comparable resolution.
This is hardly surprising, since the molecular gas is the dominant
component of the ISM over the optical disk of NGC 6946
\citep[][Sect.~\ref{twodisks}]{TacconiApJ1986}.
The nucleus is elongated in the direction north-south, as observed for
the central bar of molecular gas \citep{IshizukiApJ1990,ReganApJL1995}.

Emission associated with a more diffuse atomic gas component cannot be
detected, for several reasons. First of all the face-on inclination of
the galaxy: since dust is optically thin to its own emission a faint
component can be observed only if the dust column density is large.
This is the case for the highly inclined galaxies observed with the
jiggle-map mode, i.e. NGC 7331 and NGC 891, where higher signal to noise
were obtained coadding a smaller number of observations. The large
face-on galaxy M51 has been observed using the scan-map mode and
confirms the necessity of long integrations (Tilanus, private
communication).  Furthermore, 
chopping inside the source field removes not only the emission from the sky
but also from possible components with a shallow gradient: this may be the
case for dust associated with the flat HI distribution in NGC 6946 
(Sect.~\ref{twodisks}). Finally, a faint diffuse emission could have
been masked by the mentioned artifacts and subtracted together with the
curved background.

The 450$\mu$m image is much noisier than the 850$\mu$m one,
because of the larger sky emission in this wavelength. Only the nucleus
can be clearly detected, although most of the features at a 3-$\sigma$
level corresponds to regions emitting in the long wavelength image.

It is difficult to derive an integrated flux inside the half-light
radius, as in Sect.~\ref{starsed} and \ref{dustsed}, because the
subtraction of the curved background and the chopping within the
observation field are likely to remove faint emission. Assuming that 
the regions without evident signal have an undetected
emission of about 1 $\sigma$, upper limits can be derived for the
integrated values. The nuclear and most of the bright spiral arm
emission at 850$\mu$m is included within the optical half light radius.
Regions brighter then 3$\sigma$ have a flux of 2 Jy at this wavelength.
An upper limit of 3 Jy can be derived for the total emission inside the
half light radius. At 450$\mu$m, only the nucleus is detected above
3$\sigma$, with a flux of 9.3 Jy. Assuming a 1$\sigma$ emission for the
rest of the half light radius aperture, the estimated upper limit is 27
Jy. These ranges, converted into the units used in this thesis, are shown
in Table~\ref{tab_sed_2}.

For the nucleus it is possible to derive a temperature from the two sub-mm
fluxes. The 3-$\sigma$ central region in the 450$\mu$m image has a 
a flux of 1.2 Jy at 850$\mu$m, resulting in a temperature of T=20$\pm$7 K, 
using the emissivity as in Eqn.~(\ref{qema_used}). The large error comes 
from the calibration uncertainties. 
Because of the lower resolution of IRAS and ISO images, it 
is not possible to derive a temperature over the same small area from
the 100 and 200 $\mu$m data. The mean temperature of dust inside the 
half-light radius, derived from IRAS and ISO, is T=26$\pm$3, on the
upper limit of that measured with SCUBA data. A similar
temperature can be derived substituting one of the two FIR observations
with the integrated flux of an 850$\mu$m image smoothed to the same
resolution. 

The gas column density can be derived from the 850$\mu$m flux, if the
gas-to-dust ratio is known.
Assuming a mean dust grain radius $a=$ 0.1$\mu$m and mass density
$\rho$=3 g cm$^{-3}$ \citep{HildebrandQJRAS1983},
the emissivity of Eqn.~(\ref{qema_used})\footnote{Using
Eqn.~(\ref{qema_b2}) (i.e. $\beta=2$ at any wavelength) instead of
Eqn.~(\ref{qema_used}) will result in values 15\% smaller than the one
quoted here.} with  an extinction efficiency
in the V-band of 1.5 \citep{CaseyApJ1991}, a gas-to-dust ratio
of 160 \citep{SodroskiApJ1994} and T=26K, a hydrogen column density
$N(H)=1.5\times 10^{21}$ cm$^{-2}$ can be derived for the 1-$\sigma$ 
level, $N(H)=1.7\times 10^{23}$ cm$^{-2}$  for the central peak
and  $N(H)=2.5\times 10^{22}$ cm$^{-2}$ for the two bright
HII regions in the northeast spiral arms. \citet{IshizukiApJ1990}
observed the CO(J=1-0) emission in the central 65'' of the galaxy,
using the Nobeyama Millimeter Array. They derived a total H$_2$ mass
of $4\pm 2$ 10$^8$ M$_\odot$. Integrating the 850$\mu$m image over
the same area, a column density $N(H)=5.0\times 10^{22}$
cm$^{-2}$ is retrived, resulting in a mass of 9 10$^8$ M$_\odot$. The two
values are quite close, especially if we consider that the gas-to-dust
ratio is supposed to decrease towards the centre of a galaxy with 
respect to the mean value we used \citep{WhittetBook1992,SodroskiApJ1994}.

The derived column densities supports the idea of an optically thick
dust distribution. For instance, the diffuse component of the north-east
spiral arms at a 3-$\sigma$ level would correspond to a V-band optical
depth $\tau_V\approx 2.2$ (Eqn.~\ref{nh_ebv}). The quite high optical
depth corresponding to the sky noise ($\tau_V\approx 0.7$) shows how
difficult is to obtain sub-mm images of dust emission in the 
outskirts of face-on galaxies, even for the high sensitivity of 
instruments like SCUBA.

\section{Working procedure}
\label{wopro}
From the observed Stellar SED of Sect.~\ref{starsed}, a continuous SED 
has been derived and then integrated over the 17 bands used in the model 
(Sect.~\ref{SED}): the energy emitted in each band (inside the B-band 
half light radius) is thus obtained. 

For each of the bands a radiative transfer simulation is run, to produce
an image seen from the same inclination as NGC 6946, i.e. i=34$^\circ$.
The B-band image is then analysed to
measure the half light radius for the simulation. Successively, the 
intrinsic \emph{unextinguished} stellar energy emission for each band 
is derived assuming that the surface brightness measured within the half
light radius aperture in the simulation is equal to the observed one.
Because of this normalisation, all the models will have the same 
observed stellar SED for the images at 34$^\circ$. The intrinsic 
\emph{unextinguished} energy emitted by stars is derived from 
the observed SED using the information about the fraction of
energy absorbed and the distribution of emitted light in the solid
angle, both provided by the Monte Carlo code.
From the intrinsic \emph{unextinguished} stellar energy it is possible
to derive the observed stellar SED in a dust free case, assuming an
isotropic emission and measuring the half light radius for a transparent
model.

After the normalisation and the correction for the MIR emission
(Sect.~\ref{sec_desert}), dust temperatures are computed from the
maps of absorbed energy.  Knowing the temperature distribution in the
galactic model, images of dust emission at FIR wavelength are produced,
seen from the same inclination as the optical (Sect.~\ref{fircode}).
FIR images at several wavelengths are integrated inside the model 
half-light radius and the observed dust emission SED compared to the
observed one described in Sect.~\ref{dustsed}.
When comparing the model results to the observations, it would be more
correct to smooth the images to the resolution of the instruments involved,
to be sure to sample the same areas. This is effectively done when
a single image is analysed. For the spectrum, instead, I have chosen to
present integrations on images at their original
resolution\footnote{As described in Sect.~\ref{montecarlo}, a region of 
12$\alpha_\star$ is covered by 201 pixel, thus giving a
pixel size of 5.7'' (150 pc), for the adopted $\alpha_\star$=95''
(see Sect.~\ref{n6946_scales}). Therefore the ISO beam (117'' FWHM)
can be modelled by a gaussian of FWHM$\approx$20 pixels.}. 
Because of the large area of integration, the difference between
integrating on images at full resolution or smoothed is generally small,
$\approx$10\%, smaller than the errors of observations.

The spatial distribution of dust emission in the models is also
compared to the observed one. 
\citet{AltonA&A1998} measured optical and FIR scalelengths for a 
sample of seven spiral galaxy, after smoothing all the images to the 
poorest resolution of the 200$\mu$m ISO images. 
The ratio between the scalelengths in the B-band, at 100$\mu$m and
200$\mu$m, measured by \citet{AltonA&A1998} on  NGC 6946, is presented in 
Tab.~\ref{tab_paul}. It is interesting to note that the scale length
ratios for the seven spiral galaxies of \citeauthor{AltonA&A1998} sample
are quite similar between each other, thus indicating similar dust heating 
scenarios in different galaxies.
The images used in \citet{AltonA&A1998} are presented in Fig.~\ref{fig_images}.

\begin{table}[b]
\centerline{
\begin{tabular}{ccc}
\hline
B & $\frac{B}{200\mu\mbox{m}}$ & $\frac{100\mu\mbox{m}}{200\mu\mbox{m}}$ \\
145'' & 0.92 & 0.52\\
\hline
\end{tabular}
}
\caption{Exponential scalelengths for NGC 6946, measured between 1.5'
and 3.5', after smoothing the images to the ISO resolution
\citep{AltonA&A1998}.}
\label{tab_paul}
\end{table}

To compare the model results with the observations, I have therefore 
convolved  both optical and FIR simulations with a gaussian of the
dimension of the ISO beam.
After the smoothing, scalelengths were measured in the same range as in 
\citet{AltonA&A1998}, from 1.5' to 3.5' from the galactic centre.

\section{Adopted scalelengths}
\label{n6946_scales}

For the sake of simplicity, the model presented in this chapter have 
all the same geometrical distribution of stars. I will study the 
properties of dust extinction and emission modifying the geometrical 
parameters of the dust distribution only. Only for a few test cases 
will I adopt different stellar distributions.

For the stellar component, I use an exponential disk distribution 
with the same radial scalelength for all the wavebands of the
simulation. As for the vertical scalelength, I scale it using the 
ratio $\alpha_\star/\langle\beta_\star\rangle=14.4$ derived in
Sect.~\ref{adopted}. Again, a single vertical scalelength is used 
for all the wavebands.

The results of this chapter concerning optical, FIR surface 
brightnesses and temperature distributions do not depend on the 
absolute values of the scalelengths, as long as the quantities are 
plotted as functions of scaled galactocentric and vertical distances. 
However, absolute values are needed if we want to derive correct 
values for emitted and absorbed energies in each band.

Radial scalelengths have been derived from the images described
in \citet{TrewhellaThesis1998,TrewhellaMNRAS1998}
(Fig.~\ref{fig_images}). Profiles
averaged over elliptical isophotes have been produced, using a 
position angle $PA=64^\circ$ and an inclination $i=34^\circ$
\citep{GarciaGomezA&AS1991}.
Under the assumption that the stellar radial scalelength is the
same for any waveband, any value measured in the NIR should be 
closer to the intrinsic, unextinguished one, because of the small 
extinction in this spectral range. 
In a K-band image I have obtained $\alpha_K=95''$, that correspond to
2.5 kpc, for an assumed distance of 5.5 Mpc.
As a comparison, B-band scalelength is $\alpha_B=125''$.
Similar values (for the B-band) can be found in the literature
\citep[][and references therein]{EngargiolaApJS1991,AthanassoulaA&AS1993}.

I therefore use $\alpha_\star=2.5$kpc, and $\beta_\star=$170 kpc.

\section{The standard model}
\label{standard}

In radiative transfer models of spiral galaxies, it is usually
assumed that $\beta_{\mathrm{d}}\approx 0.5 \beta_\star$ and
$\alpha_{\mathrm{d}}=\alpha_\star$, i.e. the dust disk is half as thick
as the stellar one 
\citep[][and references therein]{ByunApJ1994,BianchiApJ1996,DaviesMNRAS1997}.
This choice of parameters is motivated by the presence of extinction 
lanes along the major axis of edge-on galaxies, that cannot be explained
without a thin dust disk. Moreover, dust is supposed to form preferentially 
in a young stellar environment, the star distribution of which is thinner
than the old stellar population (Sect.~\ref{adopted}).
Therefore I have first produced models with those geometrical
parameters for the dust distribution, for four different values of the
face-on optical depth in the V-band, $\tau_{\mathrm{V}}=$0.5, 1, 5 and
10.

\begin{figure}[ht]
\centerline{\psfig{figure=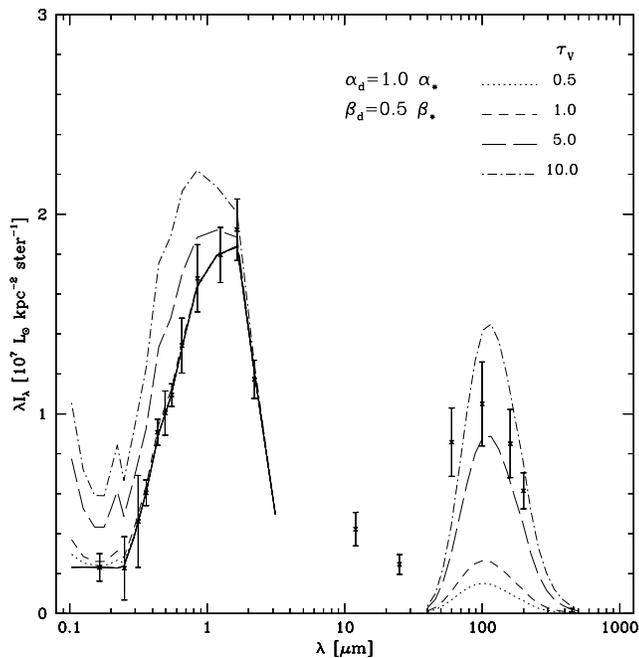,height=9.5cm}}
\caption{Surface brightness inside the B-band half-light radius for
models with $\alpha_{\mathrm{d}}=\alpha_\star$, $\beta_{\mathrm{d}}=0.5
\beta_\star$ and optical depths $\tau_{\mathrm{V}}=$0.5, 1, 5 and 10.
The data points are those described in Sect.~\ref{starsed} and
\ref{dustsed}, for the stellar and dust emission, respectively.
The solid line represent the SED derived from the data of 
stellar emission as described in Sect.~\ref{wopro}. For each model,
the lines in the UV-optical-NIR range represent the surface brightness
measured for a transparent model at the NGC 6946 inclination of
34$^\circ$.}
\label{sed_standard}
\end{figure}

In Fig.~\ref{sed_standard} I present the SED for the four models, both
for the intrinsic unextinguished stellar radiation and for the FIR
emission. The thick solid line represent the stellar output of the 
galaxy as derived from the observed data, that is used as an input
(Sect.~\ref{wopro}). Both in this models and in the ones presented
later, the spike at $\lambda\approx$2000\AA\ in the unextinguished SED 
is produced by the extinction bump at 2175\AA\ characteristic
of the Galactic extinction law (Sect.~\ref{assu_ext}) to which
correspond a flat observed SED.

\begin{table}[t]
\centerline{
\begin{tabular}{cccccc}
\hline
$\tau_{\mathrm{V}}$& intrinsic & emitted & absorbed &fraction  & FIR emitted \\ 
                 & energy    &  energy &   energy & absorbed & energy      \\ 
                 &\multicolumn{3}{c}{10$^{10}$ L$_\odot$} &\%
			           & 10$^{10}$ L$_\odot$\\ 
\hline
0.5   &   3.92   & 3.72 & 0.20 &  5 & 0.13\\
1.0   &   4.07   & 3.72 & 0.36 &  9 & 0.24\\
5.0   &   5.71   & 4.18 & 1.53 & 27 & 1.04\\
10.   &   7.57   & 4.68 & 2.89 & 38 & 1.98\\
\hline
\end{tabular}
}
\caption{Intrinsic stellar emission (intrinsic energy), stellar energy
output (emitted energy), energy absorbed by dust, 
fraction of energy absorbed by dust and energy emitted in the FIR by dust
for the models with $\beta_{\mathrm{d}}\approx 0.5 \beta_\star$ and
$\alpha_{\mathrm{d}}=\alpha_\star$.  }
\label{tab_standard}
\end{table}
 
Values for the total energy emitted by the galaxy
are given in Table~\ref{tab_standard}. Obviously, models with larger 
optical depths have a larger intrinsic energy output to produce the 
same amount of observed light. It is interesting to note that the
stellar energy output changes with the optical depth as well. This 
depends on the fact that the model is normalised on the observed surface
brightness, i.e. on the amount of stellar radiation that escapes the
galaxy along a specific direction from a given aperture.
In a transparent model, stellar emission would be isotropic, the amount
of energy emitted per unit solid angle independent of the inclination
from which the model is observed. In extinguished models, light escapes
preferentially along lines of sight at low inclination, because of the 
reduced path through dust and the effectiveness of scattering with respect 
to the more inclined cases \citep{BianchiApJ1996}. The anisotropy of
the stellar output increases with the optical depth.
Therefore, in a model that emits a given amount of energy when observed 
at low inclination, a larger amount of energy is emitted over the
whole solid angle for an optically thin (i.e. more isotropic) case.
The aperture dimension to compute the surface brightness is defined
photometrically and this makes it dependent on the optical depth of
the model: in an opaque galaxy less radiation comes from the central 
regions with respect to more transparent cases, resulting in a larger
half-light radius.  For a given half-light radius surface brightness,
optically thin models will require a smaller amount of energy emitted 
along the given inclination. 
In models with $\tau_{\mathrm{V}}=$0.5 and 1, this two competing 
effects determine the constancy of the emitted energy.
For the larger optical depths, the increase in 
the half-light radius is the more important, and the emitted energy
increases.

\begin{figure}[t]
\centerline{\psfig{figure=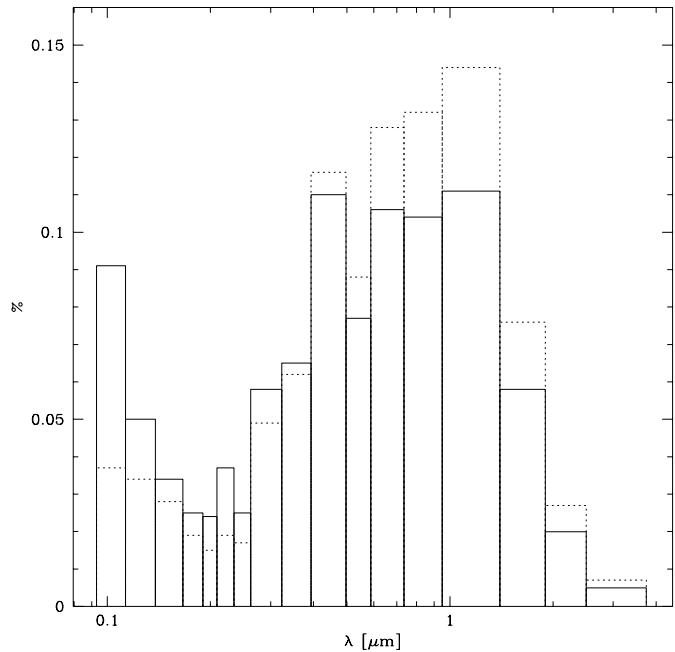,height=9.5cm}}
\caption{Percetage contribution of energy absorbed in each band to the
total absorbed energy (solid line), for a model with 
$\alpha_{\mathrm{d}}=\alpha_\star$, $\beta_{\mathrm{d}}$=0.5 and
$\tau_{\mathrm{V}}=$0.5. The percentage contribution of each band to the
energy that goes into FIR radiation only is shown by the dotted
line.}
\label{frac}
\end{figure}

Table~\ref{tab_standard} gives the total absorbed energy in each
model. The percentage contribution of energy absorbed from each band to 
the total
absorbed energy is shown in Fig.~\ref{frac} (solid line). I show the
values for $\tau_{\mathrm{V}}$=0.5, since results for the other three 
models are quite similar. For these standard models, most of the radiation 
is absorbed from the Optical-NIR wavebands (60\% for light at 
$\lambda>4000$\AA). This is quite in contrast with the results obtained
by \cite{XuA&A1995} carrying out an energy balance on a sample of 134
nearby spirals with available UV, B and IRAS fluxes. They found that
60$\pm 9$\% of the absorbed radiation comes from light in the non-ionising 
UV (912\AA $<\lambda<$3650\AA). It is difficult to compare this work
with \citeauthor{XuA&A1995}, mainly because of the simplistic sandwich
geometry they use. Apart from the geometry, the different result may be 
due to the assumption of isotropic scattering and smaller albedo for 
UV radiation with respect to the one used here, this increases extinction in
UV, or to a selection effect in favour of bright UV galaxies, although 
\citet{BuatA&A1996} dismiss its presence.
\citet{TrewhellaThesis1998} model of NGC6946 emission agrees with this work.

Applying the correction of Sect.~\ref{sec_desert}, approximately
32\% of the total absorbed energy is estimated to go into MIR emission
the remaining 68\% being available for thermal equilibrium processes
and FIR emission. The percentage contribution of each optical band to the FIR
emission only is also shown in Fig.~\ref{frac} (dotted line).
Since small-grains and PAHs responsible for
not-equilibrium processes have an higher absorption efficiency at
shorter wavelength, the contribution of absorption from Optical-NIR
wavebands is higher after the MIR correction: now the
radiation originally emitted at $\lambda>4000$\AA\ contributes 
70\% of the FIR emission. In the models of the next sections,
MIR corrections are quite similar to the one for the standard model
and therefore they will not be discussed separately. I will devote 
Sect.~\ref{mircapp} to a comparison between the estimated
and observed MIR emission.
The total energy emitted in the FIR is given in Table~\ref{tab_standard}.

\begin{figure}
\centerline{\psfig{figure=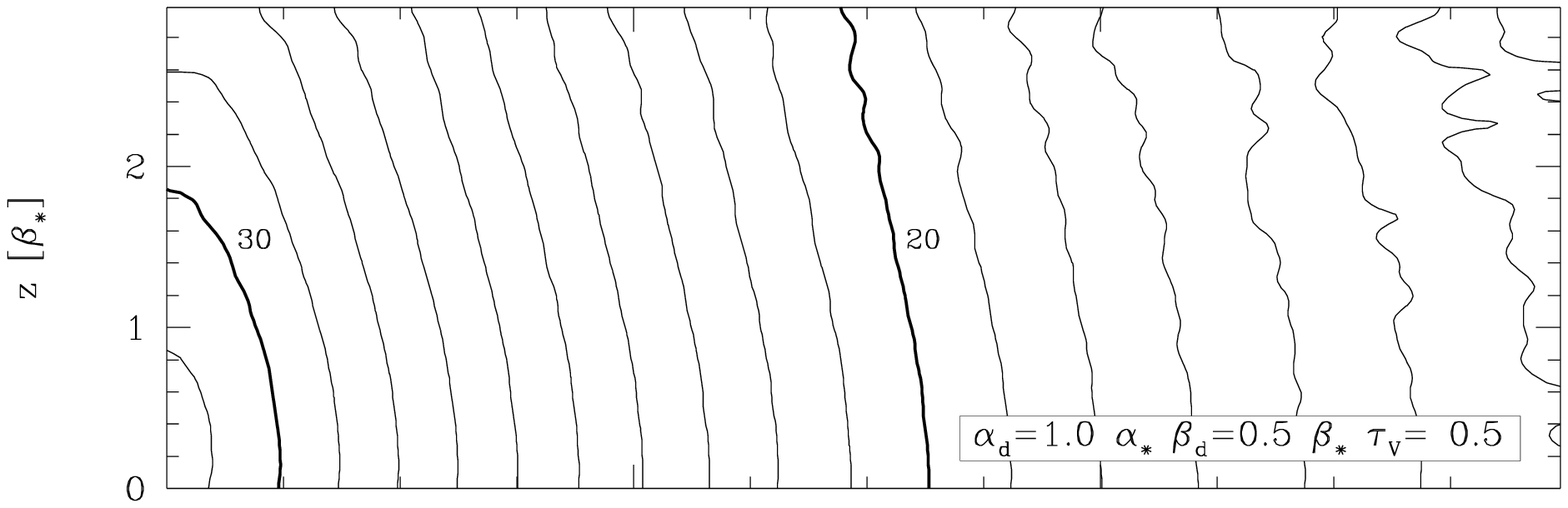,height=4.7cm,bbllx=18pt,bblly=195pt,bburx=592pt,bbury=380pt}}
\centerline{\psfig{figure=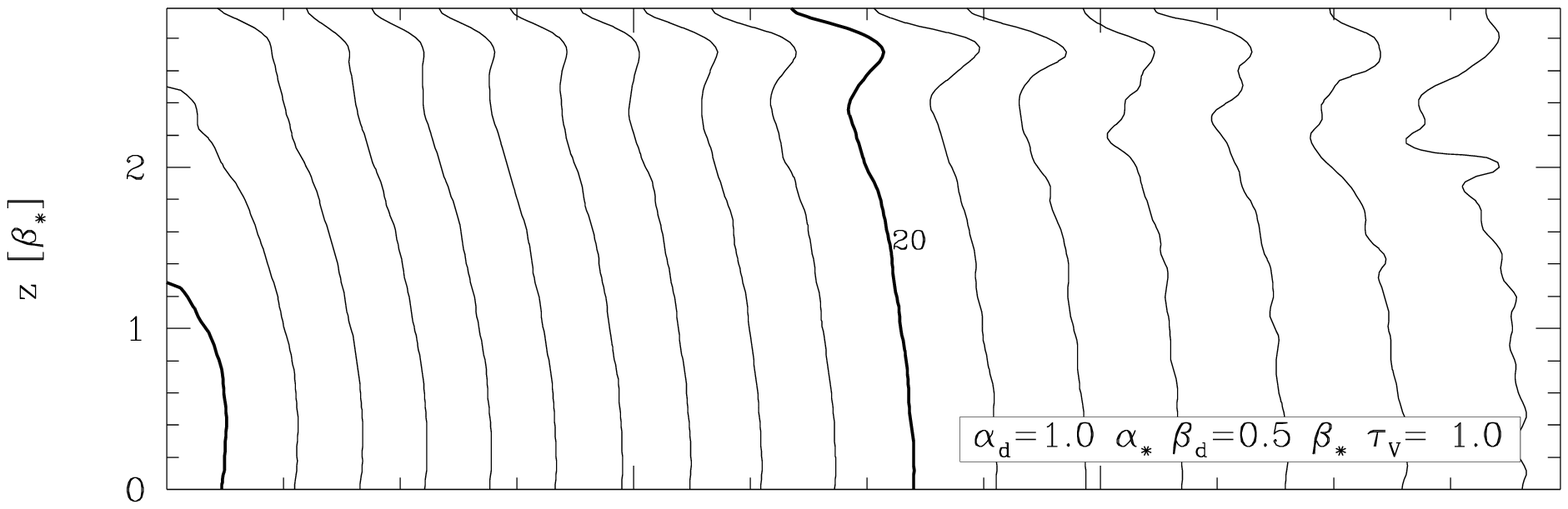,height=4.7cm,bbllx=18pt,bblly=195pt,bburx=592pt,bbury=380pt}}
\centerline{\psfig{figure=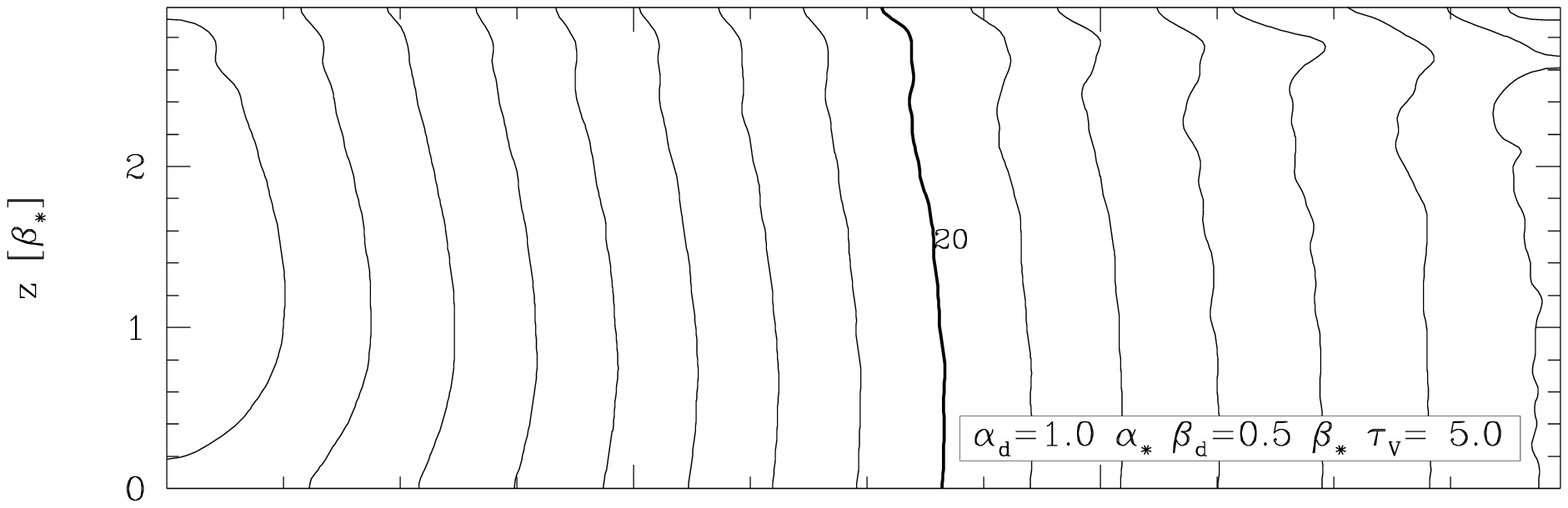,height=4.7cm,bbllx=18pt,bblly=195pt,bburx=592pt,bbury=380pt}}
\centerline{\psfig{figure=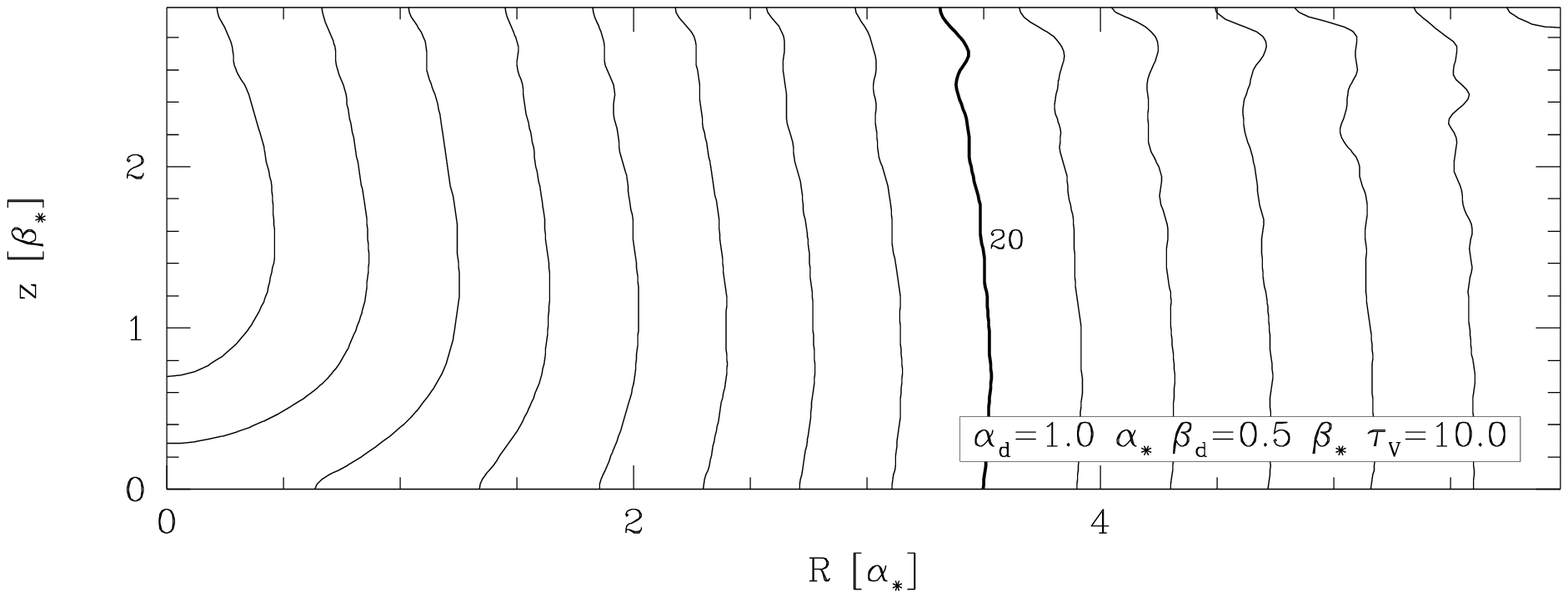,height=5.5cm,bbllx=18pt,bblly=165pt,bburx=592pt,bbury=380pt}}
\caption{Dust temperature map on a meridian plane for models with
$\alpha_{\mathrm{d}}=\alpha_\star$, $\beta_{\mathrm{d}}=0.5 \beta_\star$ and
optical depths $\tau_{\mathrm{V}}=$0.5, 1, 5 and 10, from top to bottom.
Temperature contours are plotted every 1K and highlighted at 20 K and
30K by a label and a thicker line. The scale along the z-axis has been 
expanded for clarity.}
\label{fig_temp}
\end{figure}

The temperature distributions for each model are shown in 
Fig.~\ref{fig_temp}, as a function of the galactocentric radius and 
height above the plane. Apart from the central region,
the distributions are very similar. 
At a galactocentric distance of 3$\alpha_\star$ (essentially the Sun 
position, for $R_\odot$=8.5kpc and $\alpha_\star$=3kpc, see
Sect.~\ref{obscales}), the dust temperature is $\approx$21K, as 
observed towards the poles in our Galaxy (Sect.~\ref{emi_method}).

For a dust distribution thinner than the stellar one, the stellar
radiation field is expected to increase with the height above the plane
in an optically thick model \citep{DraineApJ1984,RowanRobinsonMNRAS1986}, 
because the stars closer to the plane are shielded. This is evident
in the central regions ($R<1.5\alpha_\star$) of the models: when the
optical depth increases, dust at higher temperature is found at higher
positions above the plane. In the models with higher extinction, the
effect can still be followed at bigger galactocentric distances, the region
at higher temperature approaching the galactic plane at large
distances. 
Vertical gradients are very shallow, because of the greater extent of
the galaxy in the radial direction with respect to the vertical and because 
the stellar distribution is smooth. For larger optical depths, the
effect previously described contributes to make them even smaller.

The FIR spectrum produced by these temperature distributions is shown in
Fig.~\ref{sed_standard}. It is evident that only models with optical
depth between $\tau_{\mathrm{V}}$=5 and 10 produce enough energy to match 
the observational data. This is a general property of all the models
we are going to discuss: a substantial extinction is necessary to
produce the observed SED in the FIR. The total amount of energy
emitted in the FIR is 1-2 10$^{10}$ L$_ \odot$, that correspond to a
fraction 0.25-0.42 of the energy emitted by the galaxy in the
UV-Optical-NIR (Table~\ref{tab_standard}). Therefore, $\sim 1/3 $ of the
bolometric luminosity of the galaxy is absorbed by dust. An analogous
result is obtained by \citet{XuA&A1995} for their sample of 134 nearby
galaxies.

The peak of the FIR emission is quite close to 100$\mu$m, corresponding to 
an effective temperatures slightly smaller than 30K. The peak temperature
is thus a reflection of the temperature of the central regions of the
galaxy. Optically thick models have the maximum shifted towards longer
wavelengths with  respect to optically thin cases, because of the smaller
temperatures (Fig.~\ref{fig_temp}).

The ratio between scalelengths, measured on optical and FIR images as 
described in Sect.~\ref{wopro}, are presented in Table~\ref{tab_sc_standard}. 
No one of the models is able to reproduce the observed ratios of
Table~\ref{tab_paul}. Increasing the optical depth, the FIR scalelengths 
increase, because of the smaller temperature in the centre. But at the
same time the optical profiles become flatter, because of extinction. 
The increase in the ratio B/$200\mu\mathrm{m}$ is dominated by the second
effect. 
Because the amount of dust colder than 20K is not large in any 
of the models, the emission at $200\mu\mathrm{m}$ is caused by grains at 
the same temperature as for the emission at $100\mu\mathrm{m}$.
Therefore the ratio ${100\mu\mathrm{m}}/{200\mu\mathrm{m}}$ is not sensitive 
to the optical depth.

\begin{table}[t]
\centerline{
\begin{tabular}{ccc}
\hline
$\tau_{\mathrm{V}}$ & $\frac{B}{200\mu\mathrm{m}}$ &  
		  $\frac{100\mu\mathrm{m}}{200\mu\mathrm{m}}$ \\
\hline
0.5 &  1.44  &  0.81 \\
1.0 &  1.47  &  0.81 \\
5.0 &  1.88  &  0.84 \\
10. &  2.30  &  0.85 \\
\hline
\end{tabular}
}
\caption{Exponential scalelengths for NGC 6946 models of Sect.~\ref{standard}, 
measured between 1.5' and 3.5', after smoothing the images to the ISO 
resolution.}
\label{tab_sc_standard}
\end{table}

To summarise the results of this section, standard models can reproduce 
the observed temperature and the FIR spectrum (as long as the dust disk
is optically thick). On the contrary, it is not possible to reproduce
the spatial distribution of the emission: the FIR scalelengths are too
small compare to the optical, while in the observations they are of the
same order \citep{AltonA&A1998}. The large optical depths required to 
match the observed FIR spectrum concur in aggravating the problem, 
because of the increase of optical scalelengths with extinction.

Larger FIR scalelengths can be obtained with dust disks more extended 
than the stellar: I explore this possibility in the next Sections.
But first two tests are presented, to study how the results of this work
are influenced by the use of different stellar distributions.

\section{Test: two different stellar distributions}
\label{test}

As stated in Sect.~\ref{n6946_scales}, in the models of this thesis 
I analyse emission and extinction changing the parameters of the dust
distribution but keeping the same disk stellar distribution in all 
the cases. Are the results dependent on this assumption? To answer this
question, I present in this Section the results obtained using
two different stellar distribution.

Different stellar populations have different  scaleheights, the younger
stars being distributed in a thinner disk than the older. As a result,
vertical scalelengths at shorter wavelengths are smaller 
(Sect.~\ref{adopted}). The first model has therefore a stellar
distribution whose vertical scalelength changes with wavelength.
From the Galactic vertical scalelengths in B and V as in 
Tab.~\ref{table_beta}, I have derived a mean value 
$\alpha_\star/\langle\beta_\star\rangle=18$ that I use for the
simulations from EUV to the V-band. From the values for J, H and K-band,
I derive $\alpha_\star/\langle\beta_\star\rangle=11.3$, to be used for
the remaining bands in the optical-NIR. As for the radial scalelength,
I use again $\alpha_\star=2.5$ kpc, thus having 
$\beta_\star^{\scriptscriptstyle \mathrm{EUV-V}}=140$ pc and  
$\beta_\star^{\scriptscriptstyle \mathrm{R-LMN}}=220$ pc. 
The dust distribution of the model
has the same scalelengths as the young stellar population.

The second stellar distribution includes a bulge, together with the same
stellar disk as the standard model. In a late type galaxy such
as NGC 6946, the contribution of the bulge to the observed optical
properties is not big. Nevertheless the presence of a concentrated, almost
point-like, source in the middle of the dust disk could alter the
temperature distribution significantly, and the dust emission as well.
For the bulge I have used an exponential distribution :
\begin{equation}
\rho(r)=\rho_0 e^{-r/h},
\label{bulge_exp}
\end{equation}
where $\rho_0$ is the central luminosity density, $r$ is the distance from
the centre and $h$ the bulge scalelength. Actually, the exponential 
has been used to describe the surface brightness of bulges, i.e. the
projection of Eqn.~(\ref{bulge_exp}) on the plane of the sky
\citep{DeJongA&AS1996}. \citet{DeJongA&A1996a} finds that the 
ratio between the observed scalelengths of bulge and disk in a sample of
86 face-on galaxies is approximately 1/10. To produce this observed $h$,
the intrinsic $h$ in Eqn.~(\ref{bulge_exp}) needs to be $\approx 1.2$
times smaller, for a bulge truncated at 6$h$. For the stellar disk
$\alpha_\star=2.5$ kpc, $h=200$ pc. I have assumed a bulge to disk ratio 
of 0.1, suitable for late type galaxy bulges \citep[][G. Moriondo,
private communication]{DeJongA&A1996a}.

\begin{figure}
\centerline{\psfig{figure=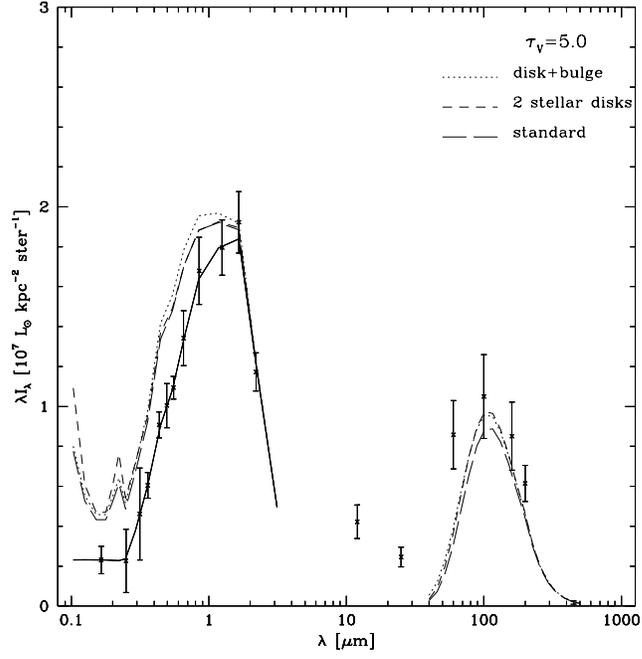,height=9.5cm}}
\caption{Same as Fig.~\ref{sed_standard}, but for the two models
described in Sect.~\ref{test}. Models  are for $\tau_{\mathrm{V}}=$ 5.
One of the standard model of Fig.~\ref{sed_standard} is also included,
for comparison.}
\label{sed_test}
\end{figure}

\begin{figure}
\centerline{\psfig{figure=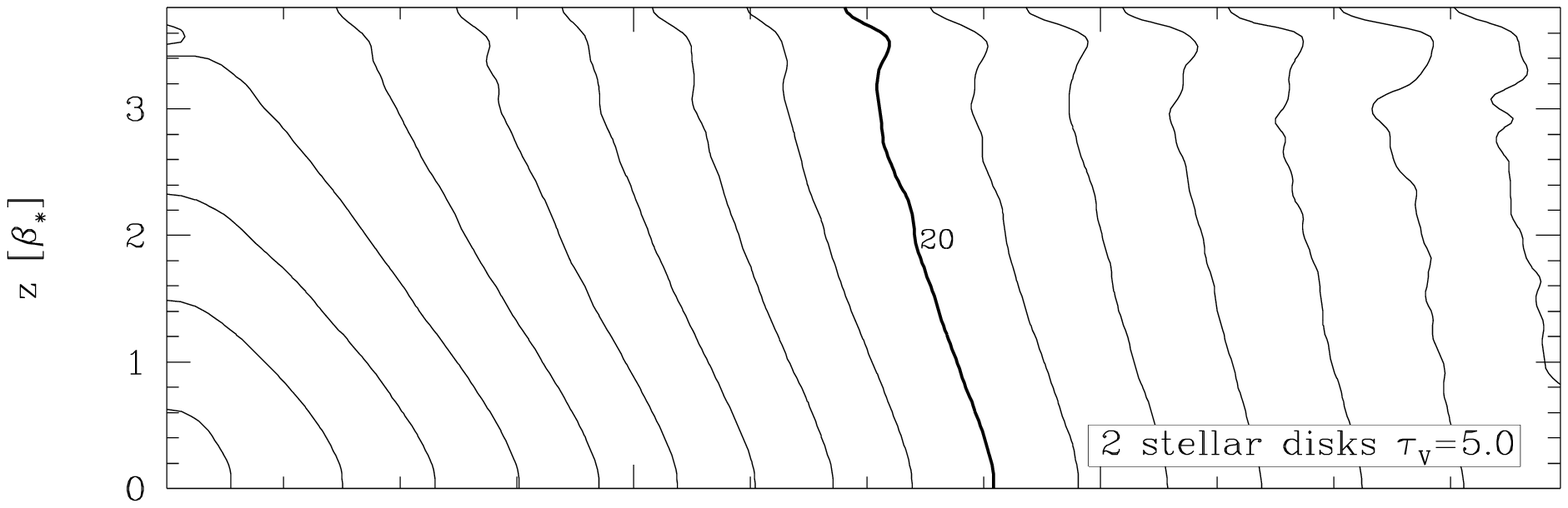,height=4.7cm,bbllx=18pt,bblly=195pt,bburx=592pt,bbury=380pt}}
\centerline{\psfig{figure=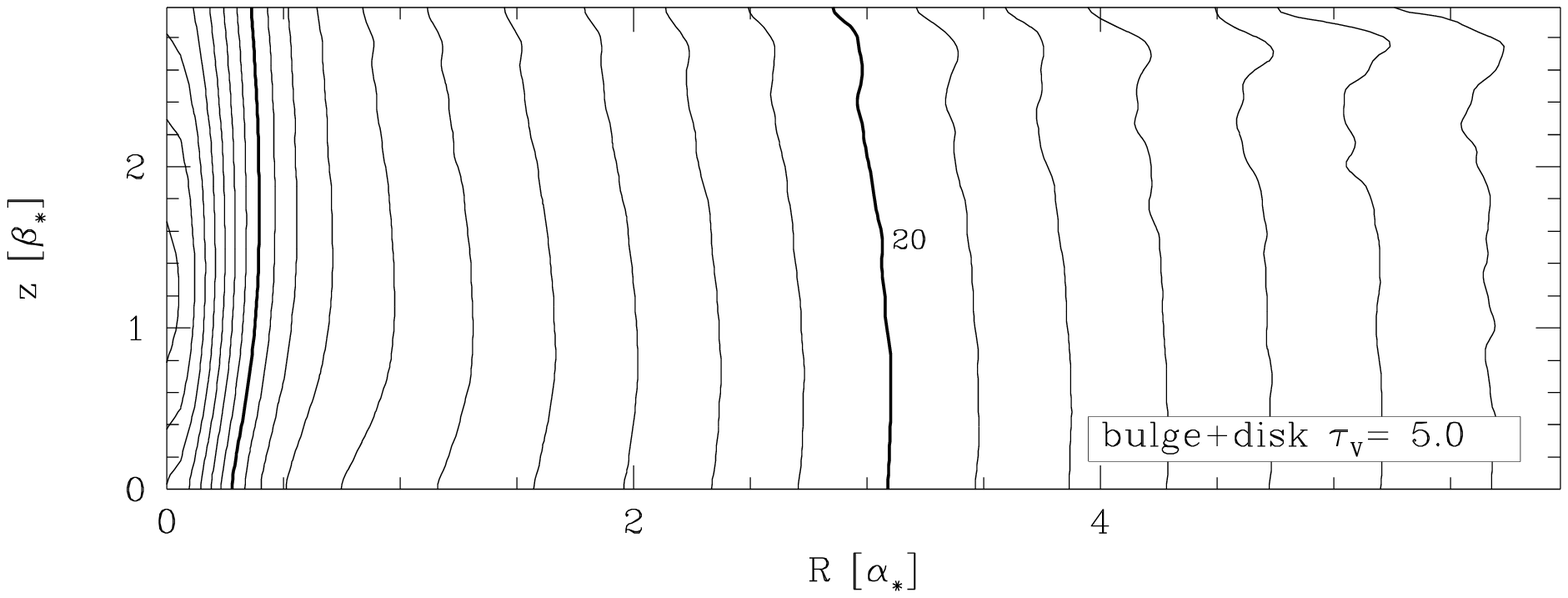,height=5.5cm,bbllx=18pt,bblly=165pt,bburx=592pt,bbury=380pt}}
\caption{Same as Fig.~\ref{fig_temp}, but for the two models
described in Sect.~\ref{test}. Models  are for $\tau_{\mathrm{V}}=$ 5.
In the 2 stellar disk plot, $z$ has been scaled on the larger $z_\star$,
i.e. 220 pc.}
\label{fig_temp_test}
\end{figure}

In Fig.~\ref{sed_test} I present the SEDs of the two test cases, for 
$\tau_\mathrm{V}=5$. The SED of the standard model with the same optical
depth is shown for comparison. The two test models have spectral energy
distributions very similar to the standard one. Intrinsic, emitted and
absorbed energy are quite similar as well, apart from small variation
due to the different spatial distribution of dust and stars. The emitted
FIR energy is 10$^{10} L_\odot$.

The temperature distributions for the $\tau_\mathrm{V}=5$ case are shown
in Fig.~\ref{fig_temp_test}. For the case with bulge, the pattern of the
distribution out of the centre is very similar to the one observed for
the standard case: this is not a surprise, since both have the same
stellar and dust disk, with a bulge that extends only up to 0.5
$\alpha_\star$. In the centre, the presence of the bulge introduce a
stronger gradient, because of the more concentrated light emission.
Temperatures in the core region reach higher values than for the
standard model. As for the standard model, the shielding caused by dust
makes central regions at higher $z$ hotter than those in the plane.

In the case of a disk with different scale heights, the shielding
effect is not present, because for higher optical depths (shortward of
the V-band) the stellar distribution is  co-spatial with the
dust. The centroidal pattern, with hotter regions closer
to the centre of the galaxy, is typical of dust distributions with scale
heights equal to or larger than the stellar (see Sect.~\ref{extended}).
Temperature values are of the same order as for the
previous models.

I have also tried models with smaller optical depths, the results
for the energy output and temperature distributions showing the same 
trend as for the standard model.

The temperature distributions of the test models produce a FIR emission
quite similar to the standard model, and similar ratios of optical and 
FIR scalelengths. 
Again, the B/$200\mu\mbox{m}$ is smaller for smaller optical depth, the 
change with optical depth being dominated by the increase of the B 
scalelength with the optical depth. For the two stellar scaleheights 
model, B profiles are flatter, because in the B-band dust and stars have 
the same thickness, while in the model inclusive of bulge, B profiles are 
steeper, the bulge affecting the profile after smoothing to the ISO resolution. 
Nevertheless, these changes in the B profile are small.
Similar results to the standard model are found also for the 100/200
scalelength ratio. The value for the disk model is the same as for the
standard, while for the bulge, the presence of an hotter centre 
reduce the ratio by 6\% only, not enough to justify the observed one.

In summary, the two (more realistic?) stellar distribution do not produce
results significantly different from the stellar disk described in
Sect.~\ref{n6946_scales}, this is therefore used in the rest of the thesis.

\section{Extended disk models}
\label{extended}

As outlined in Sect.~\ref{extdust} and \ref{dustdistri} there is evidence 
for dust distributions more extended than the stellar, both from optical and 
FIR observations. To analyse the effect of a large dust disk on the FIR
emission, I have first ran models with a larger dust radial scalelength,
$\alpha_{\mathrm{d}}$, keeping the other parameters as for the standard model 
of Sect.~\ref{standard}. 

\begin{figure}
\centerline{\psfig{figure=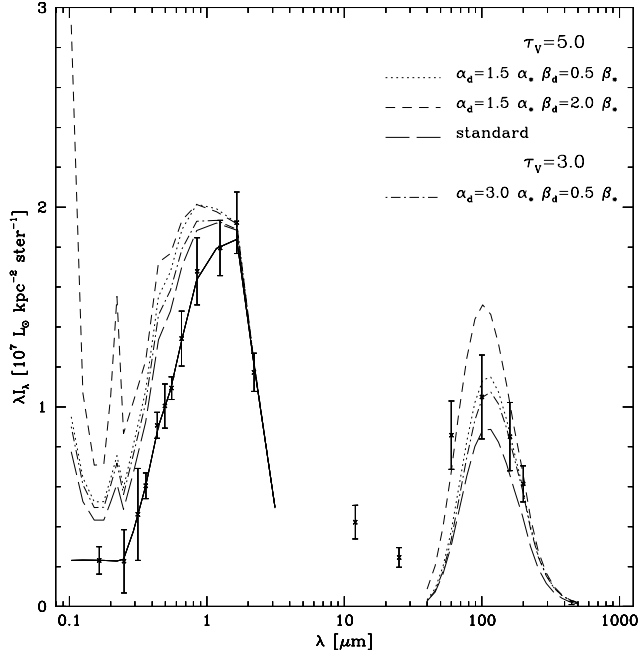,height=9.5cm}}
\caption{Same as Fig.~\ref{sed_standard}, but for the extended
models described in Sect.~\ref{extended}.  The SED for a standard model with
the same optical depth is also included, for comparison.}
\label{sed_ext}
\end{figure}

The SED of an extended model with optical depth $\tau_\mathrm{V}=5$ and
$\alpha_{\mathrm{d}}=1.5\alpha_\star$ \citep{DaviesMNRAS1997,XilourisSub1998}
is shown in Fig.~\ref{sed_ext}, together with the standard model of the
same optical depth.
As for models of Sect.~\ref{standard}, only optically thick cases are
able to match the observed SED. For the same optical depth the extended
model has an higher extinction (e.g. 44\% of the energy is absorbed in
the V-band against the 34\% of the standard model). This is evident in 
the SED of FIR and intrinsic optical radiation shown in Fig.~\ref{sed_ext}.

\begin{figure}
\centerline{
\psfig{figure=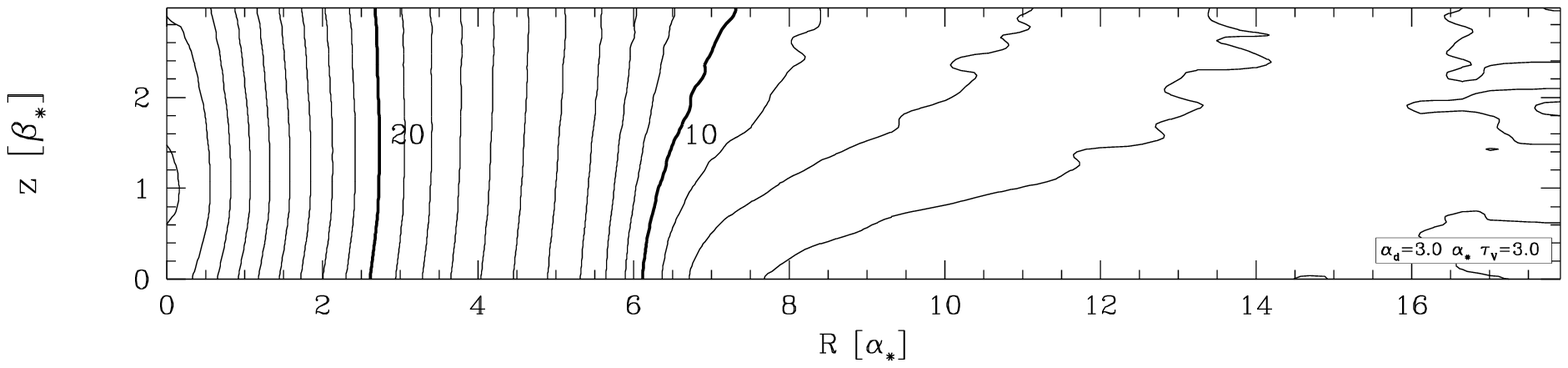,width=5.09cm,bbllx=18pt,bblly=170pt,bburx=592pt,bbury=310pt,angle=-90}
\psfig{figure=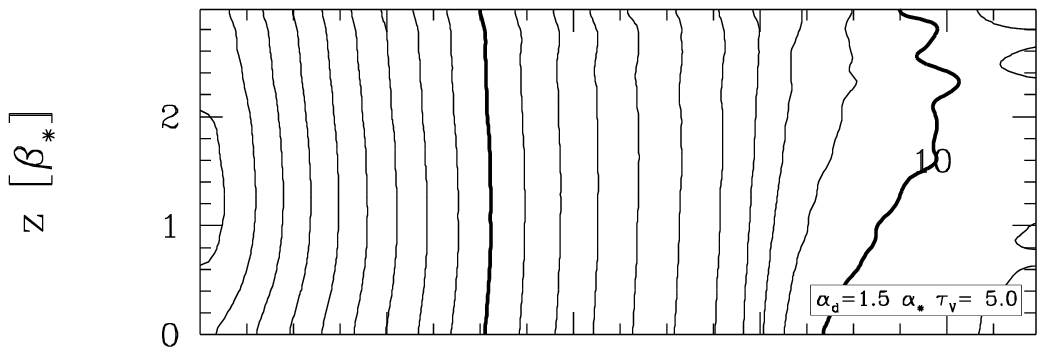,width=4.0cm,bbllx=18pt,bblly=200pt,bburx=592pt,bbury=310pt,angle=-90}
\psfig{figure=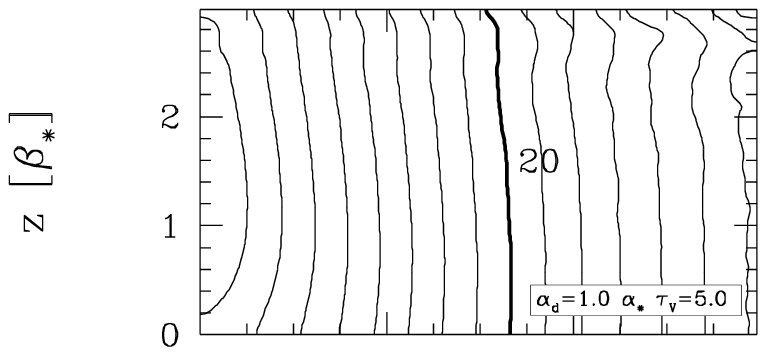,width=4.0cm,bbllx=18pt,bblly=200pt,bburx=592pt,bbury=310pt,angle=-90}
}
\caption{Same as Fig.~\ref{fig_temp}, but for the extended models with
$\alpha_{\mathrm{d}}=1.5\alpha_\star$ and $\tau_{\mathrm{V}}=$5 (central 
panel) and
$\alpha_{\mathrm{d}}=3.0\alpha_\star$ and $\tau_{\mathrm{V}}=$3 (left
panel). The temperature distribution for a standard model with 
$\tau_{\mathrm{V}}=$5 is shown in the right panel, for
comparison. All the models have the same scale along the radial axis.}
\label{fig_temp_ext}
\end{figure}

The temperature distribution for the extended model is shown in 
the central panel of Fig.~\ref{fig_temp_ext}. For ease of comparison, the 
temperature distribution of the standard model is shown again (in the
right panel), with the same scale as for the new model. Within a radius of 
6 $\alpha_\star$ (the extent of the stellar disk) the temperature pattern of 
the extended model is quite similar, apart from a small difference due to the 
normalisation. This is reflected in the peak of the FIR SED, that is 
essentially the same in both the models. Outside of the 6$\alpha_\star$,
the truncation of the stellar distribution, dust is colder and it does not 
modify the shape of the SED.
The steeper gradient in the temperature distribution that can be 
observed in this and the other extended models at 6$\alpha_\star$
is obviously due to the truncation of the stellar disk. A truncation is
indeed suggested by counts of faint stellar sources in the Galaxy
(Sect.~\ref{adopted}). I have run a few tests with stellar distributions
truncated at the same distance as the dust disks, to avoid having dust
in regions without local stellar emission. The steeper gradient disappears
and a larger distance is needed to reach the same temperature.
However, the changes are small. The general trend in the temperature 
distributions is conserved and the maps of FIR emission are not modified
sensibly.
It is interesting to note that for $R>6\alpha_\star$, dust is colder
on the plane than above, because starlight is seen through higher optical
depths along the plane. The temperature distribution of
extended models at smaller optical depths has the same general
characteristics as for the case presented here.

With regard to scalelengths, the same problem as for the standard model
is present in the $\alpha_{\mathrm{d}}=1.5\alpha_\star$ extended model. 
Despite the increase of FIR scalelengths in a model with extended dust 
distribution, the effect is not as large to compensate for the increase of
the B scalelength in optically thick models. Therefore, an optically thick
models are necessary to have a FIR energy output that matches observations, 
while optically thin models have ratios of scalelengths closer to the
observed. For models with $\tau_\mathrm{V}=0.5$, the ratio B/200 is
1.15 and it increases with the optical depth. The ratio 100/200 is 0.75,
almost independent of the optical depth.

A further increase in the dust scalelength can solve the problem.
The SED for a extended model with $\alpha_{\mathrm{d}}=3\alpha_\star$ and
$\tau_\mathrm{V}$=3 is also presented in Fig.~\ref{sed_ext}. As already seen, 
extending the dust scalelength results in a larger fraction of energy
absorbed, for the same optical depth. Therefore a smaller optical
depth, but still in the optically thick regime, can provide the right
amount of absorbed energy. The new model gives a fit to the SED as good
as the previous one. The assumed dust scalelength is the same as the one
that can be derived from the distribution of atomic gas, although a
smaller optical depth is derived from HI observation, if the local
gas-to-dust mass ratio is assumed (Sect.~\ref{twodisks}). The temperature 
distribution is presented in the left panel of Fig.~\ref{fig_temp_ext}.
The behaviour of  the  temperature values is analogous to the one
for the $\alpha_{\mathrm{d}}=1.5\alpha_\star$ model, although a larger quantity of
colder dust is present. The resulting FIR scalelengths are larger.
This, together with the reduction in the optical depth, leads to a B/200
scalelengths ratio of 0.98, close to the observed value. The 100/200
scalelengths ratio is again bigger than the observed, with a value of
0.66.

Fitting the Galactic FIR emission observed by the instrument DIRBE aboard
the satellite COBE, \citet{DaviesMNRAS1997} find that also the vertical
scalelength of dust should be larger than the stellar. Following their
results, I have studied the effects of increasing the ratio 
$\beta_{\mathrm{d}}/\beta_\star$ from 0.5 (standard model) to 2, for
the extended  model with $\alpha_{\mathrm{d}}=1.5\alpha_\star$.
A model with a dust vertical scale height larger that that of the stars
obviously suffers a bigger extinction (now 60\% of the energy is
absorbed in the V-band), because part of the dust distribution act as a
screen in front of dust, i.e. the most effective distribution in
extinguishing radiation. Despite the increase in extinction, optically
thick models are still required to match the observed FIR emission
(Fig.~\ref{sed_ext}).

\begin{figure}[t]
\centerline{\psfig{figure=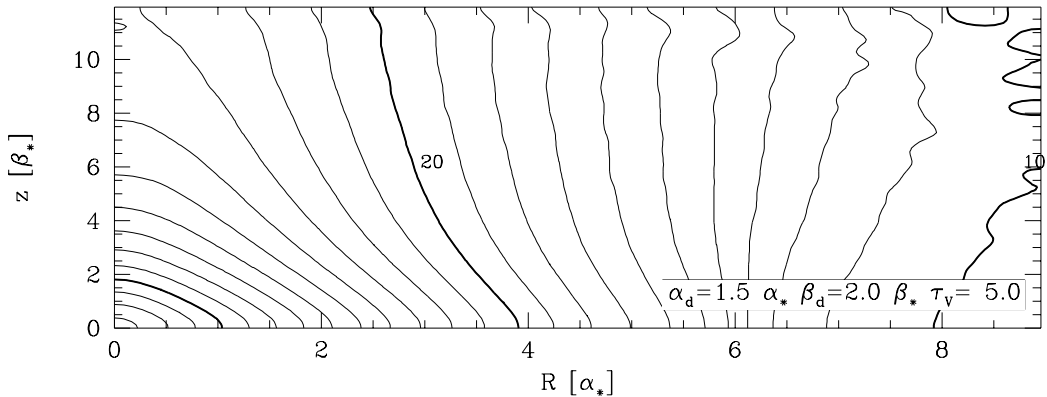,height=5.5cm,bbllx=18pt,bblly=165pt,bburx=592pt,bbury=380pt}}
\caption{Same as Fig.~\ref{fig_temp}, but for the model with
$\alpha_{\mathrm{d}}=1.5 \alpha_\star$,
$\beta_{\mathrm{d}}=2.0 \beta_\star$ and $\tau_{\mathrm{V}}=$ 5.}
\label{fig_temp_ext2}
\end{figure}

The temperature distribution for this model is shown in 
Fig.~\ref{fig_temp_ext2}. Temperatures are generally higher than those
for the previous models and this is reflected in the small displacement
in the peak of the FIR SED in Fig.~\ref{sed_ext}. The temperature pattern 
is centroidal, typical of source being less concentrated spatially than
the absorbers. Outside of the stellar region, above 6$\alpha_\star$, the
vertical gradient reduces, becoming more similar to the models with a
smaller dust scaleheight. This is because in this region scattered light
contributes to the radiation field, and it is distributed with the
dust. 

For optically thin models, the extended vertical scalelength does not
introduce changes in the ratio between optical and FIR scalelengths. In
the optically thick case, though, the B-band scalelength is increased as
an effect of extinction and the ratio B/200 becomes larger.

In conclusions, models with dust distribution more extended radially than
the stars have ratios between optical and FIR scalelengths smaller than
in standard models. As for standard models, optically thick cases are
required to match the observed FIR SED. For the scalelengths ratio to be
close to the observed value in optically thick model, it is necessary to
have a dust distribution with $\alpha_{\mathrm{d}}\approx 3\alpha_\star$.
Such a scalelength is similar to that derived from HI observations
(Sect.~\ref{twodisks}).
The model with $\alpha_{\mathrm{d}}=1.5 \alpha_\star$ deduced from the
works of \citet{DaviesMNRAS1997} and \citet{XilourisSub1998}, instead,
can provide small values of the scalelength ratio only in the optically
thin case, while the optically thick case is still required to match the
observed FIR SED. Assuming $\beta_{\mathrm{d}}=2.0 \beta_\star$
does not improve the modelling.

\section{A model with two dust disks}
\label{twodisks}

\citeauthor{TacconiApJ1986} observed NGC~6946 both in the HI 
(\citeyear{TacconiApJ1986}) and in CO emission (\citeyear{TacconiApJS1989}).
From the observations they derived the column densities of atomic and
molecular hydrogen as a function of the galactocentric radius
\citep{TacconiApJ1986}. 
H$_2$ column density was observed to have a steep profile, with an
exponential scalelength quite close to that of the optical emission
($\approx$ 90'', i.e. $\approx$ 1 $\alpha_\star$, according to the values
adopted in Sect.~\ref{n6946_scales}).
The atomic gas, instead, presents a dip in the centre, the column density 
reaching a maximum at $\approx$ 180'' ($\approx$ 2 $\alpha_\star$), then 
declining exponentially with a scalelength of 300'' ($\approx$ 3
$\alpha_\star$). Both molecular and atomic gas have the same column
density ($\sim 10^{21}$ H atoms cm$^{-2}$) at $\sim$250''. 
HI and H$_2$ have almost the same masses but the atomic gas is distributed 
on a much broader
distribution. The gas morphology observed in NGC~6946, with a central
peak and monotonically decreasing H$_2$ column density and a much
shallower HI profile, is typical of late type spiral galaxies
\citep{YoungARA&A1991}.

Since the H$_2$ distribution follows closely the emission detected by
IRAS, \citet{DaviesMNRAS1999} suggest that cold dust associated with 
the atomic gas component could be responsible for the broader $200\mu$m
profile. As shown in Sect.~\ref{extended}, broad distributions of dust 
have colder temperature than the standard one in the outer skirts:
this affects the emission at longer wavelength producing broader profiles. 
Motivated by the observations, I have therefore introduced a second dust
disk in the model, to represent the dust associated with HI. To mimic
the column density of atomic gas, the dust density of this disk falls off
exponentially with a scalelength of 3 $\alpha_\star$ at R $>2\alpha_\star$, 
being flat for smaller radii. A standard disk as described in
Sect.~\ref{standard} has been used for the dust associated with the molecular 
component. Following the results of Sect.~\ref{extended}, a larger
vertical scalelength does not improve significantly the modelling: I
have thus used $\beta_\mathrm{d}=0.5\beta_\star$ for both the disks.

As for the optical depth of the model, it has been derived from the
gas column density of \citet{TacconiApJ1986}, using the relation between
E(B-V) and total hydrogen (atomic + molecular) column density
of Eqn.~(\ref{nh_ebv}), together with the mean Galactic extinction law
(Sect.~\ref{extinction}). Extrapolating from the column density at
$\sim$250'', we derived a central face-on optical depth 
$\tau_\mathrm{V}^m\sim 10$ for
the dust associated with the molecular disk and $\tau_\mathrm{V}^a\sim
0.5$ for the atomic gas disk. Using the central value given by
\citet{TacconiApJ1986} a larger central optical depth would have been 
derived ($\tau_\mathrm{V}^m\sim$ 18): the central region may be the site 
of a moderate starburst \citep{EngargiolaApJS1991} and have different 
characteristics with respect to the smooth medium. I have therefore preferred 
the
extrapolated value. A high optical depth through the central region of
the galaxy has also been derived by \citet{EngargiolaApJS1991} and
\citet{DevereuxAJ1993}. \citet{EvansThesis1992} derived a central
face-on optical depth of $\tau_{V}=6-7$, using a TRIPLEX model 
(Sect.~\ref{modelet}) and
the energy balance method. The optically thick central region is also
confirmed by the high resolution extinction maps of 
\citet{TrewhellaThesis1998,TrewhellaMNRAS1998}.

\begin{figure}[t]
\centerline{\psfig{figure=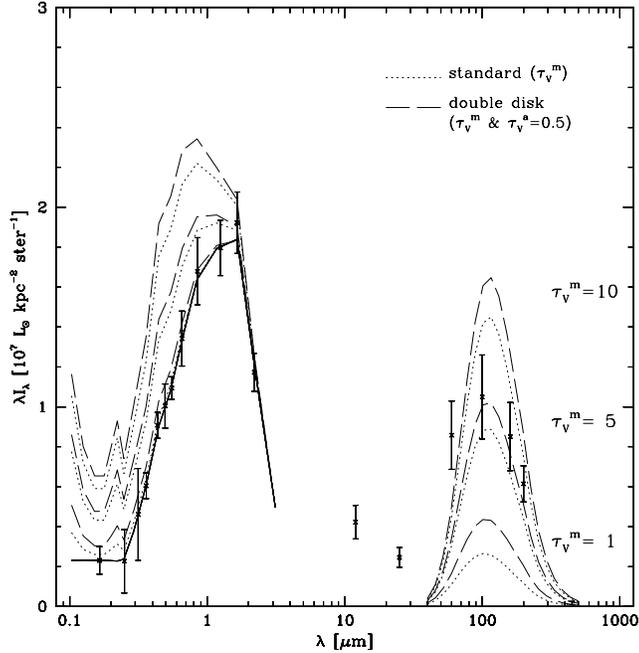,height=9.5cm}}
\caption{Same as Fig.~\ref{sed_standard}, but for the double disk
models of Sect.~\ref{twodisks}, with optical depth for the dust 
associated with the molecular component $\tau_{\mathrm{V}}^m=1, 5, 10$
and optical depth for the dust associated with the atomic component
$\tau_{\mathrm{V}}^a=0.5$.  The SED for standard model with
the same optical depth as for the dust associated with the molecular 
component in the double disk models,
are also included, for comparison.}
\label{sedP}
\end{figure}

The SED for this new model is presented in Fig.~\ref{sedP}, together
with the SED for the standard model with the same optical depth as the
dust disk associated with the molecular component ($\tau_\mathrm{V}=10$). 
The FIR output is higher that than of the standard model, because of the
extra dust disk. The temperature distribution inside the stellar disk
is similar in the two models (as the peak of FIR emission shows) while
the behaviour at larger radii is close to that of the extended model
of Sect.~\ref{extended} in the same region. 

As seen in Sect.~\ref{extended}, an extended model with
$\alpha_{\mathrm{d}}=3.0\alpha_\star$ can produce large FIR
scalelengths. However, in the model of this Section, the disk
associated with the molecular component dominates the 
FIR emission. The behaviour of the model is close to that of
a high optical depth standard model rather that to
an extended model, with a B/200 ratio 1.86.

I have also ran double disk models reducing the optical depth of the
disk associated with the molecular component. The simulations for
$\tau_\mathrm{V}^m=$ 1 and 5 are also presented in Fig.~\ref{sedP}.
Like in the case of Sect.~\ref{extended}, a model with 
$\tau_\mathrm{V}^m=5$ matches quite well the FIR emission,
but the scalelengths ratio is higher than for an extended disk.
The same for the model with $\tau_\mathrm{V}^m=1$, that is not able to
fit the observed SED.

Therefore, it is not possible to explain the observed ratio of FIR and
optical scalelengths with a double disk model inferred from the observed
distributions of molecular and atomic gas.

\section{Discussion}

In this section I discuss a few aspects of the models presented
up to this point, to provide an ensemble view of the variation of
FIR scalelength, emission and dust temperature with the varying
parameters of the dust distribution.
Also discussed is the validity of the approximations of the model,
namely,  the assumption of smooth distributions for stars and dust, the
neglect of the ionising UV in the stellar SED and the MIR correction, 
used to derive the FIR emission from the total energy absorbed by dust.

\subsection{FIR scalelengths and SEDs}
\label{discales}

Once a model for the radiative transfer in a spiral galaxy is 
available, it is relatively simple to derive an optical depth 
comparing the stellar SED and the FIR emission from dust. This 
{\em energy balance} method has been applied on NGC 6946 by
\citet{EvansThesis1992} and \citet{TrewhellaThesis1998,TrewhellaMNRAS1998}, 
using essentially the same set of data as in this work.
Both the authors used the TRIPLEX model of \citet{DisneyMNRAS1989},
i.e. an analytical approximation for the radiative transfer, neglecting
scattering, in a standard model. The neglect of scattering results
in an underestimate of the optical depth, since the effective
opacity decreases when dust is allowed to diffuse radiation
\citep{BianchiApJ1996}: \citet{TrewhellaThesis1998,TrewhellaMNRAS1998} 
corrected for this effect assuming that a model inclusive of scattering
can be simulated by a pure absorption TRIPLEX with optical depth reduced
by a factor (1-$\omega$). Their results are compatible with NGC~6946
being optically thick through its centre.

When I constrain the fit to the energy balance only, I obtain analogous
results for the standard model: optically thick models with face-on optical 
depth $\tau_\mathrm{V}\sim$5 are necessary to match the observed SED.
From a sample of 134 nearby spirals with the same ratio of bolometric 
luminosity absorbed by dust ($\sim$ 1/3) as in NGC~6946,
\citet{XuA&A1995} derived a mean optical depth $\tau_B=0.60$, compatible
with the sample being optically thin. This is mainly because they used a 
plane parallel sandwich model and their optical depth is more
representative of a mean opacity over the whole galactic disk rather
than the central value.

However, the main aim of the model of this thesis is to describe
both the SED of the FIR emission {\em and} its spatial distribution.
\citet{AltonA&A1998} measured a value close to unity 
for the ratio between the scalelengths of images in the B band and at 
200$\mu$m, in a sample of seven galaxies including NGC~6946.
The FIR scalelength increase with the optical depth, for any model
and dust emission wavelength.  Unfortunately, for the optically thick 
model necessary to match the SED, the B-band scalelength increases with 
the optical depth, because of the extinction. Therefore, the ratio B/200 
is smaller in the optically thin cases.

Besides, the scalelength ratio measured on standard models is always
larger than the observed, even for small values of the optical depth.
Values close to unity can be explained only if the dust distribution is more
extended radially than the stellar one \citep{DaviesMNRAS1999}.
Fits of surface brightness in edge-on galaxies \citep{XilourisSub1998}
and models of FIR emission \citep{DaviesMNRAS1997}, suggest that the
dust scalelength is $\approx$ 1.5 times the stellar.
When such an extended model is adopted (Sect.~\ref{extended}) FIR 
scalelengths are indeed larger than those for the standard model, for any 
value of the optical depth and of the wavelength of emission.
Now values closer to the observed can be reached, but still the increase
of the B-band scalelength with the opacity is high enough to
compensate for the increase in the FIR scalelengths.
We are therefore in front of two mutually exclusive situations:
optically thin extended models that have both B and 200$\mu$m consistent
with observations, but with a FIR output smaller than the observed; and
an optically thick model with the required SED, but with 200$\mu$m emission
more concentrated than the optical one.

To obtain a value for the B/200 scalelengths ratio close to the observed,
still being able to produce an adequate energy output in the FIR, it is
necessary to increase further the dust scalelength. A optically thick
model with $\tau_\mathrm{V}=3$ and $\alpha_{\mathrm{d}}=3.0\alpha_\star$
is indeed able to provide a good fit of both the B/200 scalelengths
ratio and the dust emission SED. The dust distribution in this case has
a scalelength similar to that observed for the atomic gas in NGC~6946,
although a smaller optical depth is derived from HI observations, under
the assumption of a gas-to-dust ratio as in the solar neighborhood
(Sect.~\ref{twodisks}).  In Sect.~\ref{twodisks} I produced a model
with two dust distributions, associated to both the gas components.
As already said, the atomic gas requires an optically thin, extended
disk, while the molecular gas suggest an optically thick standard
distribution. Such a model fails to reproduce the observed properties,
since it is the optically thick disk associated with the molecular component 
that dominates the model behaviour.

For NGC~6946 the scalelength derived from the 100$\mu$m IRAS image is 
$\approx 1/2$ of that for the 200$\mu$m ISO observation 
\citep{AltonA&A1998,DaviesMNRAS1999}. For the models presented here, 
the ratio 100/200 is larger, always bigger than 0.7. As already
outlined, this ratio depends mainly on the geometric properties of the
dust distribution; for a specific model, the ratio 100/200 is almost
constant with the optical depth, both the scalelengths increasing at the
same rate with increasing opacity. This is because both emissions are due 
to dust at the same temperature. A larger scalelength for the 200$\mu$m 
emission would be possible if there is a large amount of extended cold dust
emitting in this wavelength but not at 100$\mu$m. Indeed, when extended
models are considered, the ratio 100/200 is smaller than for the standard
model, because dust at temperatures lower than 15K is present.
Nevertheless, the amount of colder dust is never enough to match the
observed values.

On the other hand, the small value for 100/200 could also be due to
more concentrated hot dust, this time contributing most to the
100$\mu$m than to the 200$\mu$m emission. This would be the case 
if dust is heated preferentially from hot stars in HII regions (See
Sect.~\ref{heating}). Due to this uncertainty, it is better to
use the B/200 scalelength ratio, rather than the 100/200, to discriminate
between models.

\subsection{Temperatures}
\label{tempecon}
\begin{figure}[t]
\centerline{\psfig{figure=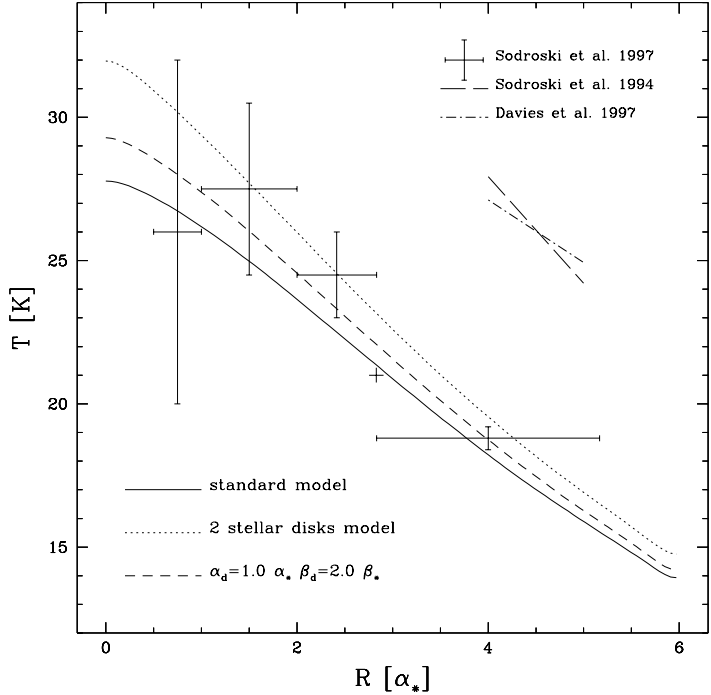,height=10cm}}
\caption{Temperature along the galactic plane as a function of the
galactocentric distance for three different models: a
standard model (Sect.~\ref{standard}), a model with
dust vertical scalelength changing from the UV to the NIR
(Sect.~\ref{test}) and a model with $\beta_\mathrm{d}=2\beta_\star$.
All models have $\tau_\mathrm{V}=5$ and $\alpha_\mathrm{d}=\alpha_\star$.
Data points are derived from \citet{SodroskiApJ1997} as described in
Sect.~\ref{tempecon}. The temperature gradient at the Sun distance
derived by \citet{SodroskiApJ1994} and \citet{DaviesMNRAS1997} are also
shown. The cross marks the temperature of 21K at the Sun distance from
the galactic centre, derived from \citet{SchlegelApJ1998} as described
in Sect.~\ref{tempecon}.}
\label{sodcon}
\end{figure}

In this section I compare the temperature distribution in the model
with temperatures actually measured. As shown previously, the temperature 
distribution does not change substantially from one model to the other. 
In Fig.~\ref{sodcon} I plot, for three representative models, the 
temperature along the galactic plane as a function of the galactocentric 
distance. I chose a standard model (Sect.~\ref{standard}), a model with
dust vertical scalelength changing from the UV to the NIR
(Sect.~\ref{test}) and a model with $\beta_\mathrm{d}=2\beta_\star$.
In all the three models $\tau_\mathrm{V}=5$ and 
$\alpha_\mathrm{d}=\alpha_\star$. Models with parameters different from
the plotted ones have similar spans in the temperature distributions.

The best determination of temperatures for dust heated by the diffuse
ISRF are those obtained for the Galaxy, because of the availability of 
high signal-to-noise spectra and images at $\lambda>100\mu$m (mainly
from instruments aboard the satellite COBE). Shorter wavelengths may
trace hotter dust associated with star-forming regions and are certainly 
contaminated by non-thermal emission from very small grains.
The latter seems to be the explanation of the relative constancy with
longitude of the temperature along the Galactic plane derived from the
use of 60$\mu$m and 100$\mu$m IRAS data only \citep{SodroskiApJ1989}.
When 140$\mu$m and 240$\mu$m images from the instrument DIRBE on board
of COBE are used \citep{SodroskiApJ1994,SodroskiApJ1997}, derived
temperatures decrease as a function of the Galactocentric distance, as 
predicted for the ISRF (that shows the same behaviour).

Determination of temperature in the Galaxy (as well as in other edge-on
galaxies) is affected by a projection effect: when a single temperature
is assumed to fit a spectrum or a flux ratio along a specific line of
sight through the Galaxy, results are biased towards higher values of
the temperature \citep{SodroskiApJ1994}. The bias is not strong, since 
the radial gradients derived on the Galaxy are very shallow:
\citet{SodroskiApJ1994} compared the observed variation with longitude
and latitude of the temperature derived from 140$\mu$m and 240$\mu$m
DIRBE data and conclude that it is consistent with a model where the 
temperature varies exponentially with a radial scalelength of 21 kpc;
a scalelength of 35.7 is derived by \citet{DaviesMNRAS1997}, after
fitting the temperature variation with longitude and latitude and 
the DIRBE fluxes at 140$\mu$m and 240$\mu$m with an extended dust model 
(Sect.~\ref{extended}). The slopes for these two exponentials at the
Sun distance from the centre of the Galaxy are shown in
Fig.~\ref{sodcon}.

The models previously described present very small variation of
temperature with the height above the galactic plane. Therefore it 
is straightforward to derive a temperature for dust at the Sun distance,
observing FIR emission at high latitudes.
Using the temperature maps derived by
\citet{SchlegelApJ1998} from 100 $\mu$m  and 240 $\mu$m DIRBE images,
the mean temperature for a region of 20$^\circ$ diameter
around the Galactic north pole is 21K, when the data are corrected for
the emissivity law used in this work (Eqn.~(\ref{qema_used}); see also
Sect.~\ref{emi_method}). After an analogous correction, this temperature
is consistent with the temperature of the warm component derived by 
\citet{ReachApJ1995} using high latitude spectra at $\lambda>104\mu$m
observed by another instrument aboard COBE, the spectrophotometer FIRAS. 
This value is also presented in Fig.~\ref{sodcon}.

\citet{SodroskiApJ1997} use three-dimensional HI and H$_2$ maps
(assuming a Galactic rotation curve to convert velocities into distances) 
and thermal radio-continuum observations to decompose the Galactic FIR 
emission observed by DIRBE into three components, associated with the atomic, 
molecular and the ionised gas phases. Several properties, among which the
temperature, are retrieved for each of the dust component in four
annuli at different distances from the centre. The component associated
with the atomic gas is supposed to be heated by the diffuse ISRF.
I plot in Fig.~\ref{sodcon} the temperature of the dust associated with 
the atomic gas for each annulus, after scaling the
distances to a Galactic scalelength $\alpha_\star=3$ kpc and correcting
the temperature for the emissivity law used in this work.
The vertical error bar represent the error in the temperature
determination, while the horizontal the width of each annulus.
From Fig.~\ref{sodcon} it seems that models with a larger ratio between
the vertical scalelength of dust and stars describe better the temperature 
profile. However, the decomposition of \citeauthor{SodroskiApJ1997} has 
large errors.  Furthermore, the mean temperatures for each annulus are 
biased towards the higher values and thus are more representative of a 
point inner in the circle, rather than the mid-point for which they have 
been plotted.
Therefore, on the basis of their temperature profiles, none of the models in 
Fig.~\ref{sodcon} can be excluded, all of them being consistent with the
observed data.

More difficult is to derive temperature gradients in external galaxies,
because of the lack of resolution and of the low signal-to noise.
When IRAS fluxes at 60$\mu$m and 100$\mu$m are used, the temperature
distributions are flat \citep{DevereuxAJ1992,DevereuxAJ1993}, as observed
in the Galaxy. This is the case also for NGC~6946: 60$\mu$m and
100$\mu$m radial profiles have the same gradient up to 3 scalelength
from the centre \citep{AltonA&A1998,DaviesMNRAS1999}.

\begin{figure}[t]
\centerline{\psfig{figure=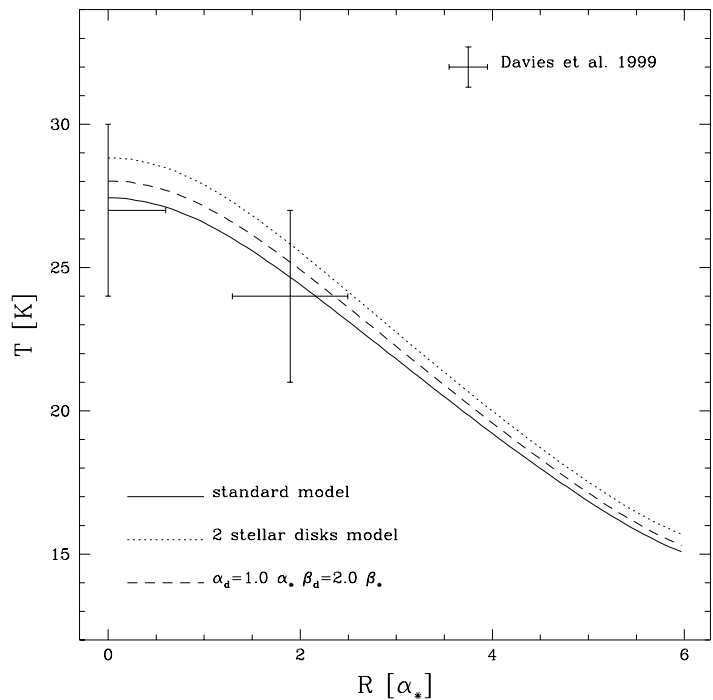,height=10cm}}
\caption{Temperature as a function of the projected galactocentric 
distance for the same models as in Fig.~\ref{sodcon}, derived from the
simulated images at 100$\mu$m and 200$\mu$m.
Data points are from \citet{DaviesMNRAS1999}.}
\label{n6946_tpro}
\end{figure}

Deriving a gradient for the temperature of the dust component in thermal
equilibrium with the ISRF is even more difficult, because of the poorer
resolution at 200$\mu$m. To compare model results with observations, I
have smoothed to the ISO resolution (Sect.~\ref{wopro}) both 100$\mu$m 
and 200$\mu$m images for the same models as in Fig.~\ref{sodcon}.
The temperature profiles derived from the ratio of fluxes at this two
wavelengths are shown in Fig.~\ref{n6946_tpro}.
\citet{DaviesMNRAS1999} derived the temperatures from 100$\mu$m IRAS and
200$\mu$m ISO fluxes at two different position on NGC~6946, in the
centre and on the disk, at a distance of 3 arcmin from the centre
($\sim 2 \alpha_\star$). I have plotted in Fig.~\ref{n6946_tpro} their
data derived for $\beta=1$; this temperature should be close to those 
that would have been derived using the emissivity law of this thesis, since 
in this spectral range it has a slope $\beta=1$, turning to a steeper
$\beta=2$ for $\lambda>200\mu$m (Eqn.~\ref{qema_used}). The errors on
the temperature have been derived from the errors on ISO and IRAS fluxes
quoted in Sect.~\ref{dustsed}, while the error bars on the positions 
reflect the wide aperture (of the dimension of an ISO resolution
element), for which the temperature have been derived.

Again, all the models are consistent with the observations,
within their large errors.

\subsection{The dust heating mechanism: ISRF vs hot stars}
\label{heating}

The present models make use of smooth distributions for stars and dust.
Therefore, the model is appropriate to describe dust heated by a smooth,
diffuse ISRF. Instead, dust close to hot stars in  star-forming regions 
is heated by a more intense radiation field and reaches higher temperatures. 

There is a debate about which of these two heating mechanisms is dominant 
in a spiral galaxy. If dust heated by hot-star in star forming regions 
contributes to the majority of the FIR emission, then it is possible to
use FIR fluxes to measure the recent star-formation in a galaxy.
\citeauthor{DevereuxApJL1990} in a series of papers
\citep{DevereuxApJL1990,DevereuxApJ1991,DevereuxAJ1992,DevereuxAJ1993}
suggest that the FIR luminosity is dominated by warm dust that absorb 
radiation from OB stars. They argue that the ISRF can only heat dust to
temperatures of 15-20K, rather than the 30-40K usually observed in spiral 
galaxies \citep{DevereuxApJ1991}. I have shown here that higher
temperatures are compatible with realistic models of the ISRF.
Moreover, their warm dust temperatures are derived from 60 and 100$\mu$m
IRAS fluxes, that may be contaminated by small grains \citep{SodroskiApJ1989}.
\citet{XuApJ1996a} derived the fraction of the total FIR luminosity associated
with star-formation, using observations of IRAS-resolved bright HII regions. 
A value of 30$\pm$14\% is found.

When COBE observations of the Galaxy at 140$\mu$m and 240$\mu$m are
used, a different picture emerges: most of the FIR emission (70\%)
arises from dust associated with the atomic gas \citep{SodroskiApJ1994}. 
The longitude and radial gradients \citep{SodroskiApJ1997} of the 
temperature are consistent with this dust being heated by the ISRF. 
20\% of the FIR is emitted by dust associated with the molecular
component. Dust associated with the molecular gas is heated primarily by
embedded OB stars and secondarily by the ISRF \citep{SodroskiApJ1997}.
Nevertheless its temperature is similar to that of the dust associated with 
the atomic gas. Finally, only 10\% is due to hot dust associated with the
HII phase.

It is not easy to evaluate how the results of this chapter depend on
the assumption about smooth distributions. If hot stars emitting mainly
in the UV are subject to a larger and localised extinction, as it would
be the case for newly born stars still located in HII regions embedded 
in molecular clouds, their contribution to the FIR output would be higher 
than in my model, both because of the larger absorbed energy and higher
temperature of the dust. Therefore the derived optical depths for the
smooth medium may be overestimated in this work.

\begin{figure}[t]
\centerline{\psfig{figure=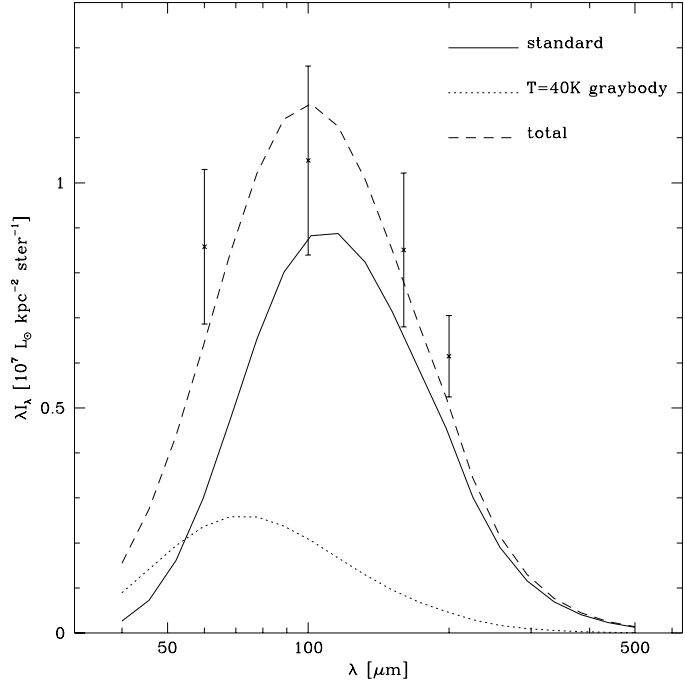,height=10cm}}
\caption{FIR spectral energy distribution for a standard model with
$\tau_\mathrm{V}$=5 (solid line) and for a grey-body with T=40K (dotted
line). The sum of the two SEDs is plotted with a dashed line. The
hot dust emission has been normalised to contribute to the 30\% of the 
total FIR emission.}
\label{obadded}
\end{figure}

As a test, I can try to add the emission from hot-stars to the FIR
spectrum derived by the model. In Fig.~\ref{obadded} I plot the FIR
spectrum for a standard model with $\tau_V=5$. The radiation from
dust heated by OB stars can be simulated by a grey-body spectrum with
a temperature of 40K, as derived from the temperature of dust associated
with the HII phase in the inner Galaxy \citep{SodroskiApJ1997} using
the emissivity of Eqn.~\ref{qema_used}. The hot dust spectrum has been 
normalised to contribute to the 30\% of the total FIR (as from 
\citet{XuApJ1996a} work). 

Adding the hot component emission obviously results in an stronger FIR
emission, especially around the peak of hot dust emission. Nevertheless, 
the difference between the new SED and the original is within the
errors in the observation. Thus, for the accuracy with which FIR fluxes 
are presently observed, the determination of the optical depth is not 
severely affected by the smooth distribution assumption.
Moreover, the contribution of the hot dust is overestimated,
the spectrum of the standard model already including the 
contribution to the FIR output of light absorbed from 
the UV, although absorbed by the smooth medium only. 

A proper description will be possible only with the inclusion of clumping 
in the model, necessary
both for dust and stars. The core of a dense clump of dust, in fact, is
shielded from the stellar radiation and the dust temperature is lower
than for dust in the smooth medium. On the other hand, it is inside 
denser regions of gas (and dust) than star formation occurs. Therefore 
a clumping model should take into account the presence of sources
embedded in dust. In this case dust in the clump would be at higher
temperatures than in the smooth medium and the contribution of hot stars
to the FIR emission could be evaluated.

\subsection{The ionising ultraviolet}
\label{noion}

As outlined in Sect.~\ref{starsed}, I do not consider the contribution
of absorption of ionising UV photons ($\lambda\le912$\AA) to the 
dust emission. In this section I evaluate the impact of this assumption.

The shape of the spectrum heavily depends critically on the assumed
parameter for the evolution model \citep{FiocA&A1997}, I estimate it
from H$\alpha$ observations, the H$\alpha$ flux being related to the
strength of the ionising radiation.
\citet{KennicuttAJ1983} measured a H$\alpha$ flux for NGC 6946
\begin{equation}
f(\mbox{H$\alpha$})=3.2\; 10^{-11} \mbox{erg cm$^{-2}$ s$^{-1}$}.
\end{equation}
The flux is contaminated by [NII] lines, but they derived a correction
from a sample of galaxies with available spectroscopy: for spiral
galaxies, $\approx$75\% of that emission is due to H$\alpha$ only.
Similar values for the flux are measured by \citet{DevereuxAJ1993}. 
Galactic extinction in the direction of NGC 6946 is $A_\mathrm{V}=1.31$
corresponding to $A_{\mathrm{H}\alpha}=1.06$.
Applying those corrections the flux is
\begin{equation}
f(\mbox{H$\alpha$})=6.4\; 10^{-11} \mbox{erg cm$^{-2}$ s$^{-1}$}.
\end{equation}
The intrinsic H$\alpha$ flux can be derived if the internal extinction
in the galaxy is known: for a standard model with $\tau_V=5$, 30\% of
the radiation is absorbed in the R-band (in whose spectral range the
H$\alpha$ line is located), the intrinsic flux being therefore
\begin{equation}
f(\mbox{H$\alpha$})=9.1\; 10^{-11} \mbox{erg cm$^{-2}$ s$^{-1}$}.
\end{equation}
From the H$\alpha$ flux, assuming a standard ionisation condition in HII
regions, the luminosity in the ionising (Lyman) continuum can be found
\citep{LequeuxProc1980}. 
Following \citet{XuA&A1995}, the Lyman continuum flux can be derived as
\begin{equation}
f(\mbox{Lyc})=33.9 f(\mbox{H$\alpha$})
=3.1\;10^{-9} \mbox{erg cm$^{-2}$ s$^{-1}$},
\end{equation}
where 75\% of the ionising radiation is assumed to be
absorbed by gas and converted in emission lines at larger wavelengths
\citep[see also][]{MezgerA&A1978,DeGioiaEastwoodApJ1992}.
If the remaining 25\% is entirely absorbed by dust, the ionising
flux converted into infrared radiation is
\begin{equation}
f^{\mathrm{abs}}(\mbox{Lyc})=7.7\;10^{-10} \mbox{erg cm$^{-2}$ s$^{-1}$}.
\label{fabs}
\end{equation}
In the Far UV the extinction law is dominated by small grain absorption,
therefore most of the energy absorbed from the ionising flux goes into 
MIR radiation from stochastically heated grains (Sect.~\ref{sec_desert}).
Assuming that 28\% of the energy absorbed goes into thermal
radiation in the FIR\footnote{This is an upper limit, derived from the MIR
correction of the EUV band, Table~\ref{optpar}. In the ionising UV the
contribution of absorbed photons to the non-equilibrium heating is
higher}, the ionising flux contributes to the FIR emission
with a luminosity
\begin{equation}
L^{\mathrm{FIR}}(\mbox{Lyc})=2.2 \;10^{8} \mbox{L}_\odot,
\end{equation}
where a distance of 5.5 Mpc has been assumed.
In a standard model with $\tau_V$=5 the total FIR luminosity is
\begin{equation}
L^{\mathrm{FIR}}= 1.0\;10^{10} \mbox{L}_\odot,
\end{equation}
and the contribution from ionising UV would then be $\approx$ 2\%. 
As a comparison, the contribution to FIR from the EUV band for the
same model is 3.6\%. I have used in this section the standard model
because it provides the same amount of FIR energy as the observed.
In models with higher extinction, like those of Sec.~\ref{extended},
the contribution of the ionising UV is smaller, the FIR energy
output increasing faster with extinction than the infrared emission 
arising from Lyc photons (Eqn.~\ref{fabs}).
On the contrary, models with smaller extinction have higher ionising
UV contribution. In a standard model with $\tau_\mathrm{V}$=1, for 
example, 6\% of the total absorbed energy is coming from ionising
photons, while only 4\% comes from the EUV band. In any case, the 
energy output of this model does not match the observed one
(Sect.~\ref{standard}).

Therefore, I conclude that disregarding the contribution 
of the ionising UV to the total absorbed energy does not modify
substantially the results obtained in this chapter.
\citet{XuA&A1995} argue that the ionising UV contributes as much as 
20$\pm 1$\% to the total FIR emission in a sample of 23
late type galaxies. Their UV contribution includes direct absorption 
of Lyc photons and indirect (via emission lines). It is 
difficult to compare this result to the one derived here, since in the
present model the absorption of emission line photons is taken care
of in the spectral band where the emission occurs (e.g. in the R-band as
in this section for the H$\alpha$ line), the eventual contribution of 
re-combination summed up to the stellar SED for each band.
Nevertheless, the ratio between Lyc emission and total absorbed energy 
(their FIR, including MIR radiation) is similar to the one derived
for the standard $\tau_V$=5 model.

\subsection{Estimated and observed MIR emission}
\label{mircapp}

To compute the dust temperature at the thermal equilibrium 
(Sect.~\ref{fircode}) I have excluded from the total absorbed energy
the fraction that goes into non-equilibrium heating. This energy is
essentially re-emitted in the MIR spectral range. In this section I
compare the derived MIR energy output with the one observed in NGC~6946.

The fraction of absorbed energy that goes into MIR emission depends 
essentially on the absorption of light from the short wavelength spectrum, 
since the absorption efficiency of small grains responsible for
non-equilibrium processes is higher in the UV (Sect.~\ref{sec_desert}).
For models where the dust scaleheight is smaller than the stellar, the 
amount of energy absorbed from UV bands does not increase very much with the 
optical depth (The {\em saturation effect}, \citealt{BianchiApJ1996}). 
The MIR corrections for these models are therefore quite constant, 
$\sim$32\% of the total absorbed energy being re-emitted in the MIR. 
In optically thin models this fraction is still constant when more extended 
vertical distributions are concerned. 
Optically thick models with vertically extended distributions
have a higher efficiency in extinguishing radiation (they are closer to
a screen model) and therefore the fraction of energy absorbed in bands
with higher optical depth is larger. As an example, for a model with
$\tau_\mathrm{V}=5$ and $\alpha_\mathrm{d}=1.5\alpha_\star$
(Sect.~\ref{extended}), the amount of energy absorbed in the band UV2
changes from 58\% to 69\% when the vertical scalelength is doubled,
while the absorption in the J band changes only from 24\% to 25\%.
The MIR correction therefore increases, but not by a large amount, being
41\% for the model with both radial and vertical extended dust
distributions.

For a local Interstellar Radiation Field \citep{DesertA&A1990}
the contribution of small grain emission to the 60 $\mu$m IRAS band
is $\sim$62\%, while at 200 $\mu$m it is only 14\% and at 200 $\mu$m
4\%. Therefore, the fraction of energy emitted in non 
equilibrium heating can be roughly estimated measuring the MIR emission 
shortward of 60 $\mu$m. After integrating a continuous SED interpolated 
from the data points in Table~\ref{tab_sed_2}, the MIR energy is derived
to be 34\% of the total infrared energy emitted by dust. The same
value is found when the data provided by \citet{EngargiolaApJS1991} for
the whole galaxy, rather than for the half light radius, are used.

The value derived from observation is very close to the model one.
This justifies the use of \citet{DesertA&A1990} dust model as
described in Sect.~\ref{sec_desert}. It is interesting to note that the 
infrared galactic spectrum used in the \citet{DesertA&A1990} model is 
different from the one of NGC~6946.  As an example, the ratio between 
fluxes at 60 $\mu$m and 100 $\mu$m is 0.2, while it is 0.5 with our NGC~6946 
data.  This does not necessarily mean that the dust model of
\citet{DesertA&A1990} cannot be applied to NGC~6946. The different ratio
could be due to the different heating condition in the local interstellar
radiation field, with respect to the mean radiation field in NGC~6946.
Larger ratios between 60 $\mu$m and 100 $\mu$m can be derived 
from \citet{DesertA&A1990} model when the ISRF is larger than the local.

\section{A halo of dust}
\label{halo}

Yet another possible distribution for dust remain to be explored:
a spherical halo. A reasonable amount of cold dust at large distance 
above the galactic plane can provide a FIR emission at larger
scalelengths than those obtained for the extended disk seen in 
Sect.~\ref{extended} and \ref{twodisks}. 

A significant fraction of the total mass of dust produced by a galaxy
during its lifetime can be injected into the halo because of the 
imbalance between the radiation pressure and the galactic gravitational 
force \citep{DaviesMNRAS1998}. Unfortunately, information
about the density and distribution of a putative dusty halo are by far
more uncertain than those for the dusty disks. A dust halo would
act as a screen distribution for the galaxy and therefore will not
produce a substantial differential extinction on different parts of the
galaxy, unless it has a steep gradient. Thus it would be impossible to
detect it fitting the optical appearance with a radiative transfer
model, as in the works of
\citet{XilourisA&A1997,XilourisA&A1998,XilourisSub1998}.
\citet{ZaritskyAJ1994}
analyses the difference in colours of background galaxies between fields 
at different distances from the centre of two nearby galaxies; he find 
that fields at a projected galactocentric distance of 60 kpc have 
a B-I colour excess of 0.067 with respect to fields at a distance of 220 kpc.
This suggests the presence of a halo of dust, although a better
statistical determination is required since there is only a 2$\sigma$
difference. Comparing his result with observations of the mean opacity
through the centre of spiral galaxies ($A_V$=1.0) he derives a halo
scalelength of 31$\pm$8 kpc. Since the halo dust component may be
unrelated to the dusty disk, he argues that this leads to a lower
limit for the scale length, provided the central optical depth is not
severely underestimated. 

Due to the lack of reasonable constrains, the parameters with which I
describe the halo dust distribution in this section do not have a
physical justification. However, the models I will present can be 
regarded as an exercise, to show what the influence of such a
distribution could be on the FIR emission. The parameters are chosen on the 
basis of the results of the model presented earlier.
Obviously the halo cannot be the only dust component in a galaxy,
since observations of edge-on galaxies clearly show the existence of a
flat dust distribution that produces the extinction lane. In this
section I will use a standard dust disk. As for the dusty halo, I use
a constant density spherical distribution, that extends up to the
boundaries of the dust disk. This means a radius of 6$\alpha_\star$=15kpc,
the maximum radius of the dust disk. 

\begin{figure}
\centerline{\psfig{figure=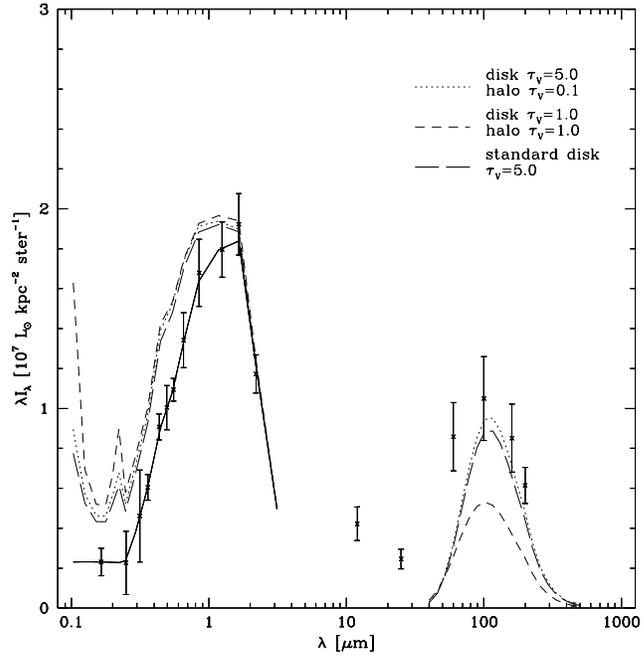,height=9.5cm}}
\caption{Same as Fig.~\ref{sed_standard}, but for the standard disk
+ halo models of Sect.~\ref{halo}.
The SED for standard model with $\tau_\mathrm{V}=5$
is also included, for comparison.}
\label{sedH}
\end{figure}

\begin{figure}
\centerline{\psfig{figure=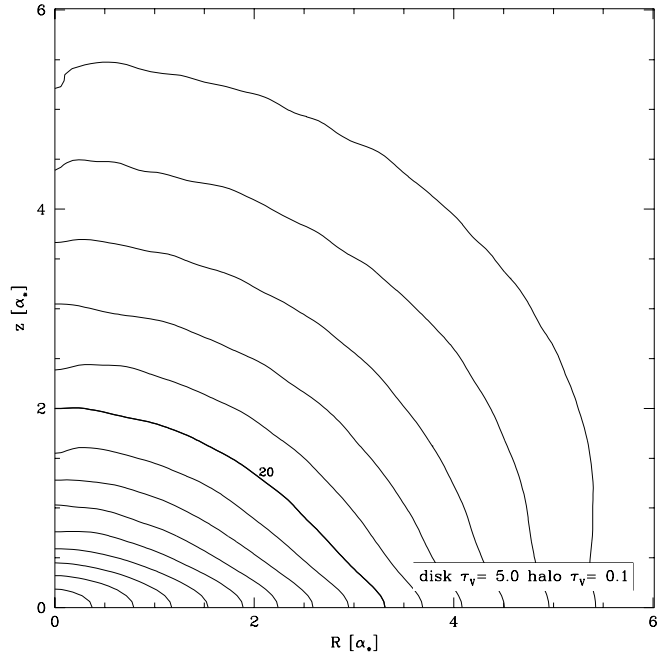,height=9.5cm}}
\caption{Same as Fig.~\ref{fig_temp}, but for a model with a standard
disk of $\tau_{\mathrm{V}}=$ 5.0 and a constant spherical halo of 
$\tau_{\mathrm{V}}=$ 0.1.}
\label{fig_temp_halo}
\end{figure}

As seen in the previous sections, an optical depth $\tau_\mathrm{V}$=5
for the dusty disk is necessary to produce the same amount of FIR emission 
as observed. A constant dust halo, acting as a screen, will be very effective
in extinguishing radiation, therefore I chose for this structure an
optically thin status, $\tau_\mathrm{V}$=0.1, not to alter significantly the
energy output. The SED in the FIR is shown in Fig.~\ref{sedH}; as
predicted, the dust halo does not introduce a big difference in extinction 
and emission. The temperature distribution for this model
is shown in Fig.~\ref{fig_temp_halo}. It shows a centroidal pattern, as
already seen for models with stellar emission completely inside the dust
distribution (Sect.~\ref{test} and \ref{extended}). For the inner part
of the disk, the temperature has the same gradient as for a standard
disk only, although the actual scaling depends on the different
normalisation of the model. Since the halo has a constant density,
there is now a larger quantity of dust at lower temperature. However,
for the chosen geometrical and optical parameters, the mass of the halo,
has only 1/4 of the mass of the dust in the disk, and the latter
dominate the behaviour of the scalelength ratio, that is similar to a
standard model.

A possible solution would be to use a more massive dust halo,
responsible for a larger part of the dust extinction and emission
and a more transparent disk. An increase in the FIR scalelengths
would thus correspond to only a marginal increase in the optical
scalelength, and the ratio would be closer to that observed. 
For a model where both halo and disk have the same optical depth
$\tau_\mathrm{V}=1$, the mass of dust in the halo would be 12 times 
larger than for the disk. The SED for such a model is also displayed 
in Fig.~\ref{sedH}, while the temperature distribution has a quite
similar pattern as for the previous model and is not shown. It is 
interesting to note that, despite the total 
FIR output of the model with halo is nearly the same as for a standard
model with $\tau_\mathrm{V}=5$ (also plotted for reference in
Fig.~\ref{sedH}), the surface brightness is about 1/2. This is because 
the FIR scalelength are larger than for the standard model, and a larger
part of the FIR radiation is emitted outside the half-light radius.
Indeed, within this model the scalelength ratios are remarkably close to 
those observed, being 0.9 for B/200 and 0.6 for 100/200.

Would such a halo be observed in edge-on galaxies? \citet{AltonMNRAS1998}
used resolution enhanced HiRes IRAS maps to study the FIR emission in 24
edge-on galaxies, including starburst and quiescent objects. None of the
object was found to be resolved along the minor axis. HiRes images at 
100$\mu$m  have a resolution of $\approx 90''$ and a typical 3$\sigma$
level of 0.75 MJy sterad$^{-1}$. When the disk+halo model with
$\tau_V=1$ is observed at 90$^\circ$, almost all of the emission,
extending up to 5 IRAS resolution elements from the centre, is at a
flux larger then the 3$\sigma$ level. 

Therefore a dust halo model that could explain the scalelength ratio
and marginally the FIR emission would be easily resolved in IRAS images.
It is not observed.

\section{Summary}

The radiative transfer and dust emission model described in
Chapter~\ref{program} has been applied to the spiral galaxy NGC~6946.
The stellar SED for the galaxy has been derived from literature data,
requiring the UV, Optical and NIR radiation, after being processed through 
dust, to be the same as observed, for any dust distribution.
Various aspects of the dust FIR emission have been simulated,
i.e. temperature distributions, FIR spectra and images at specific
wavelengths. I have explored several optical and geometrical parameters 
for the dust distribution, to reproduce the observational results in
the FIR.

It was relatively easy to find a model with a FIR spectrum able to match 
the observations. While the temperature distribution (and therefore the
peak of spectral emission) does not vary substantially for any dust
disk parameters and optical depth, an optically thick dust distribution
is required to reproduce the emitted energy ($\tau_\mathrm{V}\sim 5$,
the exact value depending on the geometrical details of the dust
distribution).
Using a dust scalelength $\alpha_{\mathrm{d}}=1.5\alpha_\star$, as
suggested by models of surface brightness in edge-on galaxies and other
FIR simulations, it is not possible to produce optically thick models that 
simulate the observed spatial distribution of optical and FIR light. 
Under this assumption, only optically thin models have the observed ratio 
between optical and FIR scalelengths, but they do not have the required 
energy output.

To produce a good fit, both to the FIR energy output and to the B/200
scalelenths ratio, with an optically thick model ($\tau_\mathrm{V}\sim
3$), it is necessary to extend further the dust scalelength to
$\alpha_{\mathrm{d}}=3.0\alpha_\star$. The dust distribution would be
similar to that of the HI, althougth the atomic gas column densities
suggest a lower optical depth. A model with a dust component for each 
gas phase, however, fails to reproduce the observed properties.

It is possible, in principle, to reproduce scalelength ratios and
(although only marginally) the energy output of the galaxy by including a
halo of dust to a disk model. However, such an halo would be easily
detected in FIR images, which it is not.

A further discussion will be presented in Chapter~\ref{conclu}.

\chapter{Conclusions}
\label{conclu}

In the last Chapter I have applied the radiative transfer and FIR
emission model described in Chapter~\ref{program} to the spiral galaxy
NGC~6946. Several models for different dust geometries and optical depths 
were tested against the observations, to explain the fluxes and spatial
distribution of the FIR emission. This Chapter presents a resume of the
work done for the Thesis. A brief summary of the model characteristics
and of the results on NGC~6946 is presented in Sect.~\ref{rep_model}
and ~\ref{rep_result}. A discussion of the implication of the findings
is given in Sect.~\ref{discuss}. Finally, a summary of the Thesis can
be found in the last Section.

\section{Outline of the model}
\label{rep_model}

The model of this Thesis has been derived from the Monte Carlo radiative 
transfer code for spiral galaxies of \citet*{BianchiApJ1996}. The main 
quality of the Monte Carlo technique consists in the exact treatment of
multiple scattering in the radiative transfer. The original code
included polarisation, a distribution of sizes and materials for the
dust grains and optical properties, like albedo and phase functions,
derived from the Mie theory for spherical grains.  In this
work, the code has been simplified, omitting the polarisation
calculations and assuming albedo and phase functions derived empirically
from observations of reflection nebulae (Sect.~\ref{assu_ext}).
The code has then been made able to store the amount of energy absorbed
from dust, as a function of the distance from the galactic centre
and height above the galactic plane (Sect.~\ref{montecarlo}).

Once the geometry of the dust distribution relative to the stars and
its optical depth in the reference V-band are chosen, the (monochromatic)
radiative transfer code is run for 17 different photometric bands, to 
cover the spectral range of stellar emission. For each band, the output of 
the code (i.e. optical image and absorbed energy map) is scaled on the 
observations of the galaxy to be modelled, in such 
a way that the flux measured inside a specific aperture (I use the half-light 
radius) in the simulated image matches that really observed. As a result,
the SED of the stellar radiation in the model is the same as the
observed (for the chosen aperture). Finally, the 17 maps of energy
absorbed by dust from starlight in each band are summed together, to 
produce a map of the total energy absorbed by dust.
For each position inside the galaxy I therefore know the amount of
energy that is absorbed by dust illuminated by an ISRF that is
consistent with the radiative transfer itself, without any other
assumption (Sect.~\ref{SED} and \ref{fircode}). This is one of the
original points in the present code. To my knowledge, only the 
Monte Carlo radiative transfer code of \citet{WolfA&A1998} combines this
characteristic with a proper treatment of the radiative transfer. 
However, their model has been implemented only for star formation 
environments and not for spiral galaxies.

Absorbed radiation can go into heating of small grains, a process not
occurring at the thermal equilibrium resulting in MIR emission.
Since the models of this Thesis are restricted to the FIR emission
from grains emitting at thermal equilibrium, a
correction to the absorbed energy maps is applied (Sect.~\ref{sec_desert}).
Using the dust emissivity derived in Sect.~\ref{qemi_new}, the corrected
map of absorbed energy is converted into a map of temperature. 
Hence, a final map of FIR emission can be easily obtained for any wavelength,
integrating along a specific line of sight (Sect.~\ref{fircode}).
The FIR scalelengths and SED are derived from the maps, and
compared to the observed data. The dust distribution parameters are then
modified and the procedure repeated until a match is achieved
between simulated and real data.

The model devised for this work is essentially a sophisticated version
of the energy balance technique (Sect.~\ref{firmod}). Not only the amount 
of energy emitted in the FIR is compared to the stellar radiation to
derive the galactic opacity, but also the spatial information is used,
to see if the chosen star-dust geometry is consistent with the FIR
emission.

\section{Summary of the results}
\label{rep_result}

Several models have been explored in Chapter~\ref{model}. Most of the
models are able to fit the SED inside the half-light radius for NGC~6946, 
if the dust distribution has a face-on optical depth $\tau_V\sim$5.
The amount of energy absorbed by dust depends on the dust geometry.
For a standard model with optical depth $\tau_V=5$, 27\% of the
total stellar radiation is absorbed, with a V-band extinction $A_V=0.45$
(Sect.~\ref{standard}). A model with the dust distribution more extended
radially than the stellar ($\alpha_d=1.5\alpha_\star$) has an higher
extinction $A_V=0.62$, and 36\% of the intrinsic starlight is absorbed
(Sect.~\ref{extended}).

However, the models of this Thesis were also required to describe the
observed spatial distribution of the FIR emission.
In a sample of seven galaxies, including NGC~6946,
\citet{AltonA&A1998} find that the 200$\mu$m radial scalelength is larger
than the B-band one, by a mean factor of 1.3 for the whole sample. 
For NGC~6946 the scalelength ratio 200/B is $\approx 1.1$
(Sect.~\ref{wopro}).
As already foreseen \citep{AltonA&A1998,DaviesMNRAS1999}, the standard 
model is not able to provide a FIR scalelength larger than the optical.
\citet{AltonA&A1998} and \citet{DaviesMNRAS1999} proposed an extended
dust distribution. Extended dust models derived from the surface-brightness
of edge-on galaxies \citep[$\alpha_{\mathrm{d}}=1.5\alpha_\star$;]
[]{XilourisSub1998} were tryed.  But even with such extended dust 
distributions the scalelength ratio is different from the observed, if 
the dust disk is optically thick (Sect.~\ref{extended}). 
An increase in the dust radial
scalelength indeed increase the FIR scalelengths with respect to the
standard model. The FIR scalelengths increase also with the optical depth, 
although by a minor amount. Unfortunately, the optical scalelengths
increases as well with $\tau_V$. Therefore, for the optically thick case
required by the match with the observed SED, the increase in the 200$\mu$m
scalelength is compensated by the  increase in the B-band one, and the
ratio 200/B is always smaller than the observed. Only for optically thin
cases, the scalelengths ratio approaches unity (Sect.~\ref{extended}).
When the dust scalelength is extended further, to the values observed for
the atomic gas, a fit to the 200/B ratio can be provided, in the
optically thick case necessary to the match the energy output.
The model, with $\alpha_{\mathrm{d}}=3.0\alpha_\star$ and
$\tau_{\mathrm{V}}=$3, have an extinction $A_V=0.66$, with 37\% of the 
intrinsic stellar radiation re-processed by dust.

The temperature distributions are quite similar, for any of the dust disk 
models. Temperature values in the models are compatible with those observed
in the Galaxy and in NGC~6946 as well (Sect.~\ref{tempecon}).
\citet{AltonA&A1998} also measured the ratio between scalelenghts at
100 and 200$\mu$m. For NGC~6946 the scalelengths ratio 100/200 is
$\approx 0.5$ (Sect.~\ref{wopro}). The 100/200 scalelengths ratio
for the disk models of Chapter~\ref{model} varies less than the 200/B
ratio, being always in the range 0.7-0.9, for any model, with the lower
value for extended dust distributions. 
Extended distributions of dust could in principle decrease the ratio
to the observed values, if large amounts of cold dust emitting at
200$\mu$m but not at 100$\mu$m are present in the external
part of the galaxy. This does not happen in the disk models explored
here, and the emission at 100$\mu$m and 200$\mu$m is essentially
due to dust at the same temperature (Sect.~\ref{discales}). 

An optically thick disk with a homogeneous spherical halo of optical
depth $\tau_V\approx 1$ could fit the data, emitting enough
FIR radiation to match the observed SED and having both 200/B and
100/200 scalelengths ratio similar to the observed (Sect.~\ref{halo}).
However, such a halo would be easily detected in FIR observations of
edge-on galaxies, even with the poor resolution of instruments like IRAS
and ISO. It is not.

\section{Discussion}
\label{discuss}

As already said, an optically thick disk with $\tau_V \sim 5$ is
necessary to explain the SED observed in the FIR for NGC~6946. 
\citet{EvansThesis1992} and \citet{TrewhellaThesis1998} apply the 
energy balance method to the stellar and dust emission of NGC~6946, 
using a TRIPLEX model with dust scaleheight half of the stellar.
This is the same as using a standard model (Sect.~\ref{standard})
and limiting the match to the observations to the SED only.
They both derived high optical depths for the disk, using the data 
inside the half light radius. \citet{EvansThesis1992} measured
$\tau_V=6-7$, while \citet{TrewhellaThesis1998} $\tau_V=4\pm1$.
A high optical depth is also suggested by the high-resolution sub-mm
images from SCUBA described in Sect.~\ref{scuobs}: the diffuse 
inter-arm emission in the NE spiral arms at a distance of 2'
($\approx\alpha_\star$) is compatible with $\tau_V=2.2$.

The high optical depth of NGC~6946 contrasts with the recent
determination of optical depth of \citet{XilourisSub1998}, based on
fits of the surface brightness of edge-on galaxies using a suitable
radiative transfer model for spiral galaxies. For a sample of 
seven edge-on spirals they find optically thin dust disks, with 
a mean central face-on optical depth $\tau_V$=0.5.
The higher opacity of NGC~6946 may be a result of the galaxy being very
gas-rich \citep{AltonSub1999}; Or it may be due to the clumping of the
ISM, affecting in a different way FIR and optical determination of the
optical depth. While FIR observations would detect all of the dust
(at least when the temperature of the clump and inter-clump medium are
similar), optical observations may be affected preferentially by
the extinction of the smoother, lower density (and optical depth)
inter-clump medium.

Only two models available in literature include clumping in a proper
radiative transfer for spiral galaxies. \citet{KuchinskiAJ1998} use 
preliminary results from a Monte Carlo model to derive opacities of
edge-on galaxies from their colour gradients along the minor axis.
After dividing the space occupied by dust in a three-dimensional grid,
some cell are assigned randomly a clumping status, assuming a constant
filling factor for the high density cells all over the galaxy, and
adopting a ratio of 100 between densities in clumps and in the nearby
smooth medium (thus following the formalism developed by
\citet{WittApJ1996} for clumping in an homogeneous sphere illuminated by
a central point source). Although a few aspects of
the inclusion of clumping are presented in the work, the authors defer
a detailed discussion to a forthcoming paper.

The other model is that of \citet{BianchiSub1999}, based, as the work of
this Thesis, on a simplified version of the Monte Carlo radiative
transfer code of \citet{BianchiApJ1996} for spirals (Sect.~\ref{montecarlo}). 
Simulations are conducted in the V-band, for a stellar disk similar to
the one adopted here. Dust is described with two components: a smooth
one, associated with the neutral gas, with a double exponential
distribution and parameters as for a standard model; a clumpy one, 
associated with molecular clouds.  A few values of the fraction of gas 
(dust) mass distributed in clumps are explored. 
A three-dimensional grid covering the whole dust volume is first filled
with the homogeneous dust distribution, then clumpy cells are randomly
selected according to the radial and vertical distribution of 
molecular gas in the Galaxy. Cell dimensions and the mass of each clump
are chosen to match those observed for Giant Molecular Clouds. As a
result of the choice of the parameters, the cubic cells have a fixed 
optical thickness $\tau_V=4$ through each side. As already found for
clumpy models in simpler geometries, the main effect of clumping consists
of reducing the amount of energy absorbed by dust, with respect to a
homogeneous model of the same dust mass. The increase in the fraction of
energy that can escape the galaxy is moderate, resulting in surface
brightness profiles that are less than one magnitude brighter than those
for homogeneous models. Minor and major axis profiles of the simulated
disks reveal that clumping effects are higher in the edge-on case.
This contrasts with the claims of \citet{KuchinskiAJ1998} for edge-on
profiles not being modified by the clumpy structure. It is shown how the
difference in the models behaviour results from the different
parametrisation adopted for the dust distribution.
This is unfortunate, however, as it indicates a strong dependence of
the observed brightness profiles on the detailed internal and spatial
distribution properties of clumps which makes the interpretation of the
data very difficult.

Since the Giant Molecular Clouds simulated by each cell host star-forming
regions, it is logical to assume that part of the galactic stellar
radiation comes from within the clouds. \citet{BianchiSub1999} study this
possibility allowing a fraction of the stellar radiation to be emitted from 
inside the clumpy cells. When embedded stellar emission is considered, 
extinction increases with respect to the case with only dust distributed in
clumps. Extinction in a model including clumping also for stellar radiation
can be even higher than that for a homogeneous case with the same dust
mass.

It is instructive to compute the gas-to-dust mass ratio for the models
of this Thesis.  NGC~6946 has a total gas mass of $9.0\cdot10^9$ M$_\odot$
\citep[][rescaled to the distance of 5.5 Mpc used in this Thesis]
{DevereuxApJ1990}. 
The dust mass of an homogeneous disk can be easily computed from the radial 
scalelength and the V-band face-on optical depth using the formula given by 
\citet{BianchiSub1999}. Optically thick models with
$\alpha_{\mathrm{d}}=1-1.5\alpha_\star$ have gas-to-dust
mass ratios of the same order of the Local Galactic value
\citep{SodroskiApJ1994} of 160 (the $\tau_V$=5 standard model has 360,
while the $\tau_V$=5 $\alpha_d=1.5\alpha_\star$ model 160).
The optically thin model with the same scalelengths, necessary to explain 
the optical-FIR scalelengths ratio, contains less dust, with a gas-to-dust 
ratio higher than the local of an order of magnitude, 1600 
($\tau_V$=0.5 $\alpha_d=1.5\alpha_\star$).
On the other hand, the only model able to provide a simultaneous fit
to the SED and the scalelength ratio, that with
$\alpha_{\mathrm{d}}=3.0\alpha_\star$ and $\tau_{\mathrm{V}}=$3, has a
smaller gas-to-dust ratio of about 70. This may suggest that, despite
the better fit, the dust quantity is overestimated.
Therefore, if the Galactic gas-to-dust mass ratio is to be considered a 
common value for spirals, optically thick models with
$\alpha_d=1-1.5\alpha_\star$ not only are necessary to 
provide a good fit to the FIR SED, but they also have the right amount
of dust. 

A model with two disk distributions was tested in Sect.~\ref{twodisks}.
The model included a standard, optically thick distribution of dust,
derived from the radial profile of the H$_2$ column density, and an
extended ($\alpha_d=3\alpha_\star$) optically thin disk associated to
the atomic component. Adopting $\tau_V=5$ for the standard disk and 
$\tau_V=0.5$ for the extended distribution, a good fit was provided 
for the FIR SED. It was hoped that such a model would have provided a
good fit to the scalelength ratio as well, because of the extended
distribution. However, the dust emission is dominated by the dust
associated with the molecular disk, the amount of colder dust at larger
radii being insufficient to modify the scalelengths.
Both the standard and extended distribution have a similar dust mass 
for the parameters listed before, leading to a gas-to-dust ratio of 180.

What would happen if the dust associated with the H$_2$ gas
component were to be distributed in clumps? Predictions are not easy.
\citet{BianchiSub1999} results cannot be easily used in this case, 
because they have been derived for a different configuration,
with the H$_2$ and its associated dust being distributed in a ring like 
structure and not in an centrally peaked distribution, like in NGC~6946
\citep{TacconiApJ1986}. Furthermore, the disk of diffuse dust in
\citet{BianchiSub1999} is a standard one, while the dust associated with
HI in the double disk model has a radial scalelength three times the stellar.
One may hypothesise the following scenario: the diffuse extended dust is
responsible for the behaviour of the scalelength, and for the optically thin 
face-on $\tau_\mathrm{V}$ derived from edge-on profiles, as measured by
\citet{XilourisSub1998}. The clumpy dust associated with H$_2$
may be responsible for the bulk of the FIR emission, if stellar sources
are present within the clouds. This scenario favours
the hypothesis of a substantial contribution of localised sources to
the dust heating, rather than a diffuse ISRF.
Would this scenario be valid? For the models of \citet{BianchiSub1999}
it is difficult to make a dust disk appear optically thin, if the dust
mass is high, since clumping does not modify a great deal the shape of the
profiles. In the conclusion they show that a clumpy model with optical depth
unity may look the same as an homogeneous distribution with $\tau_V$=0.5, 
when seen edge-on.  
For the models here discussed, a dust mass corresponding to a distribution 
with $\tau_V\approx$5, would have to look as if it has $\tau_V$=0.5.
This is not possible for the models of \citet{BianchiSub1999}, although
the hypothesis cannot be in principle disregarded for NGC~6946, because of
its different distributions of atomic and molecular gas.
I have then based the predictions of the FIR emission on the assumptions that 
the behaviour of the diffuse dust disk is unchanged when high density clumps 
are interspersed. A simple test in Sect.~\ref{heating} showed that it is
unlikely that, under these assumptions, the dust optical depth is
overestimated when carrying out the energy balance.
However, the temperature distribution, the shape of
the SED and the spatial distribution of emission are likely to change
in a way difficult to predict.
Clumps with embedded stellar emission would act as hot spots in the dust
distribution. The way this would influence the FIR emission depends on
the clumps distribution. If clumps are more concentrated towards the
centre of the galaxy, the FIR emission at shorter wavelength may be
steeper. This could explain why the observed 100/200 scalelength ratio
is smaller than any value derived from the models
(Sect.~\ref{discales}).

Two recent models include clumping of dust and embedded stellar
emission to describe the radiative transfer and FIR emission of NGC~6946 
\citep{SilvaApJprep1998,SautyA&A1998}. However, it is not easy to 
use these works to address the problems raised in the last paragraph.
The large number of parameters involved in the modelling prevents
an isolation of the effect of the dust distribution on the FIR heating.
Furthermore, the complexity of the models forces the authors to sacrifice 
a correct treatment of the scattering \citep{SilvaApJprep1998} or a 
complete description of the ISRF in the whole wavelength range of stellar 
radiation \citep{SautyA&A1998}.

The energy balance is used by \citet{SilvaApJprep1998} to 
calibrate the results of a complex photometric evolution model for
galaxies. The galactic disk is described by three exponential distributions:
a distribution of spherical molecular clouds with embedded stars, a 
distribution of stars that have escaped molecular clouds, and a
distribution of diffuse dust and atomic gas. The stellar SED is derived
from a spectral synthesis and galactic evolution model. For each
evolution stage, the residual gas mass is used to find the dust mass
of the galaxy. The radiative transfer for the diffuse medium is carried 
out in an approximate way, rigorous only for an infinite homogeneous 
medium and isotropic scattering. A specific radiative transfer model
is adopted for the molecular clouds. The dust emission is predicted
from the computed ISRF. The several parameters of the model are 
chosen to fit the stellar and dust SED. The code does not produce maps
but only total integrated values for the luminosity.

Among the other objects, they apply the model to NGC~6946. Assuming
that radial and vertical scalelengths are the same in each component,
they derive a total extinction in the B-band $A_B=0.13$ (as derivable
from their ``average'' optical thickness), with 60\% of dust residing in
molecular clumps and the rest in diffuse ISM. For an homogeneous medium
with the same scalelengths, their B-band extinction would correspond to
a face-on optical depth $\tau_V=1$. Since part of the emission occur
inside regions of higher opacity (the molecular clouds), the diffuse 
medium in the \citeauthor{SilvaApJprep1998} model may have $\tau_V<1$.
A comparison with the results of the \citet{BianchiApJ1996} code,
as presented in the database of \citet{FerraraApJS1999}, shows how the
extinction is likely to be overestimated by the \citeauthor{SilvaApJprep1998} 
approximate treatment of radiative transfer of a 10-15\% in the B and V
band, for inclinations close to face-on (G. L. Granato, private communication).
The choice of dust vertical scalelengths equal to the stellar increase
the extinction too, with respect to the standard model with the same dust
mass.  However, it is unlikely for their geometry and radiative transfer model
to underestimate substantially the extinction in the galaxy.
Therefore, their model is consistent with an optically thin ($\tau_V\le1$)
diffuse dust component.  They also claim that young stars soon
escapes from their parent clouds, thus contributing considerably to the 
FIR radiation from diffuse dust in spiral galaxies. Thus the
star-formation rate could be derived from FIR radiation.

A complex model for the radiative transfer within NGC~6946 is also
presented by \citet{SautyA&A1998}. They describe the ISM as a two phase
medium, constituted by molecular clouds and a diffuse constant density
distribution associated with the atomic gas. The molecular clouds have a
distribution in space and in size derived from models of the
gravitational potential of the galaxy and of cloud formation. OB
association are created inside the molecular gas. The dust is scaled on
the gas, using a constant dust-to-gas mass ratio. The radiative transfer
is carried out with a Monte Carlo method, but only for wavelengths
smaller then 2000$\mu$m, to avoid the contribution of stars not included
in their simulations. The ISRF is estimated on each cell of a
three-dimensional grid. For cells not reached by UV radiation, like
in the interarm regions, the ISRF at $\lambda>2000$\AA\ is used, derived
from a R-band map of the galaxy, scaled on the Galactic local ISRF.
The UV flux is scaled on the observed. Dust emission is computed and
maps can be produced. A fit is achieved for the integrated emission in the
four IRAS bands and at 200$\mu$m (KAO observation).

It is difficult to compare their results to the present work, because
they use a different dust distribution from the exponentials, both for
the atomic and molecular gas. They derive a total extinction at
2000\AA\ ~~ $A_{2000}$=0.76. As a comparison, a $\tau_V=5$ standard model 
in the UV4 band (1900\AA$<\lambda<$2090\AA; Sect.~\ref{SED}) has
$A_{2000}=0.83$, while for the $\alpha_d=1.5\alpha_\star$ $\tau_V=0.5$
model $A_{2000}$=0.28. The work also support the hypothesis of dust heated
preferentially by young stars, the UV radiation contributing to 72\% of
the total FIR. However, the greater care in the treatment of the UV 
radiation at $\lambda<2000$\AA\, dictated by the desire to model the 
UV excited H$\alpha$ and C$^{+}$ lines, rather than for the Optical and
NIR ISRF, may have biased their results towards the shorter wavelengths. 
Unfortunately, no radial profiles are presented for wavelengths
$\lambda>100\mu$m. It is therefore impossible to test the model output 
with the observation suggesting an extended dust distribution.

A proper analysis of the effects of clumping on the FIR emission needs, 
however, a model focussed mainly on dust extinction and heating
mechanisms. A correct treatment of the radiative transfer in the whole
spectral range of stellar emission is necessary. Such a model would
help not only to ascertain if the observed optical/FIR scalelength ratio
can be produced by model with the required FIR energy output, but will
also answer the debated question of which is the main contributor to the 
dust heating (Sect.~\ref{heatmech}). If indeed UV radiation from young
objects is the main contributor to the dust heating, recent star
formation rates can be derived from the FIR emission, for a large number
of galactic objects.  Unfortunately, results from a models including
clumping are highly dependent on the description chosen for the clumps
distribution \citep{BianchiSub1999}. The advent of high-resolution 
FIR and sub-mm instrument, will surely help to provide a better
description of the dust distribution. Current instrumentation permit
such observation only on large nearby objects, like the Galaxy and M31.
For those objects, the diffuse ISRF is the main contributor to
the heating (Sect.~\ref{heatmech}).

One of the points raised in favour of young stars as the major source of
dust heating is the impossibility of the diffuse ISRF to heat dust
to temperatures higher than 20K \citep{DevereuxApJ1991}.
I have shown in this Thesis how temperatures higher than 20K are
compatible with models with only smooth distributions of dust and
stars, i.e. with ISRF heating only. Temperatures in the galactic centre
can reach values $\sim 30$K. When the temperature is derived from
integrated FIR fluxes, values are smaller, because of the temperature
gradient with the galactocentric distance. However, values are always 
biased towards hotter dust.
As an example I can use the extended model with $\alpha_d=1.5\alpha_\star$ 
and $\tau_V$=5, to compute the temperature from fluxes integrated inside
the half light radius. The temperature distribution goes from 30K, in
the centre, to 18K, at 1.6$\alpha_\star$, the half light radius
(Fig.~\ref{fig_temp_ext}). When the integrated 100$\mu$m and 200$\mu$m 
fluxes are used, the derived temperature is 26K, still higher than 20K.
Temperatures higher than 20K are therefore compatible with ISRF heating
of dust.

The radial gradients of the temperature in the models are compatible 
with those observed in the Galaxy and in NGC~6946 (Sect.~\ref{tempecon}).
Values of the temperature at a distance corresponding to the Sun
Galactocentric distance are quite close to the values measured from 
Galactic high-latitude FIR emission \citep{ReachApJ1995,SchlegelApJ1998}.
Using the 60$\mu$m DIRBE image as a template for the FIR emission from 
diffuse cirrus clouds in the Galaxy, \citet{LagacheA&A1998} isolate
high-latitude regions of sky with excess emission at the longer wavelengths
observed by DIRBE. For regions with no FIR excess, they measured a
temperature of 17.5K (using $\beta=2$; for the emissivity used in this
work, Eqn.~\ref{qema_used}, the temperature would be of $\approx 21$K;
see Sect~\ref{emi_method}). A significant excess is measured in regions
covering 3.7\% of the sky. Two temperature components are necessary to 
describe the emission in those regions, a warmer component with T=17.8,
analogous to the one measured in the absence of the long wavelength
excess, and a colder component with T=15.0. The coldest component
measured on the Galaxy has T=13K. Again the temperature of the warm
component is similar to those of the models of the Thesis. The colder
component is identified with dense molecular clouds. Regions with
negative excess are present as well, indicating hotter dust in 
molecular clouds with embedded massive young stars.
However, the temperature variations are quite small, corresponding to
local variation of the ISRF of only 30\%. This may suggest that the
effect of clumping on the ISRF and on the dust heating is not heavy.

In the smooth distribution of the current models, colder temperatures
can result from the shielding of stellar radiation along the plane in
optically thick models (Sect.~\ref{standard}). However this does not
produce appreciable differences in the temperature distributions.
Colder temperatures can be obtained also for dust at large distance 
from the galactic centre ($R>6\alpha_\star$). At such distance, in fact,
dust is immersed in a reduced (or null, as in the extended models of this 
Thesis) local ISRF, the  heating coming mostly (if not completely) from 
stars in the inner galaxy (Sect.~\ref{extended}).
For the standard models and for extended models in regions where dust
and stars are mixed, the minimum temperature is T$\approx$14K.
In extended models dust at larger radii is colder. This is shown also by
the larger scalelengths of the FIR emission. Another geometrical
configuration with dust present in region of scarce or null stellar
emission is the one including the halo (Sect.~\ref{halo}).

After reviewing the little indirect evidence for the existence of unseen
baryonic matter, \citet{GerhardApJ1996} consider the possibility that 
part of the dark matter in spiral galaxies may be in the form of gas. 
They produce a model for an extended flattened halo of cold gas clouds 
and defined the parameter range that will permit the halo to be unseen
and stable. Two mechanisms are suggested to provide stability to each 
cloud against collapse and star-formation: the self-gravity of the gas
may be reduced by the presence of a minicluster of particles inside the
cloud; or dust grains associated with the gas and heated by the galactic
and intergalactic radiation field may provide sustain through collisions
with the gas. They propose several empirical tests to verify this
hypothesis, among which deep FIR and sub-mm observations of dust halo
emission. Such a halo of baryonic matter would play a major role in the
galaxy star-formation history. Indeed, large amounts of baryonic matter have
been recently observed by  \citet{ValentijnApJL1999}, from ISO
Short-Wavelength Spectrometer (SWS) observations of H$_2$ rotational
line emission along the disk of NGC~891.
A cool molecular component is found to dominate the emission at larger 
radii, outweighting the atomic gas by a factor of $\sim10$. Such gas
could account for the dark mass in the galaxy.

Unfortunately, the parameters of a putative halo are poorly
constrained. In Sect.~\ref{halo} I tried a few halo configuration,
together with a disk distribution of dust, choosing the parameters on the
basis of the observed behaviours in optically thin and thick models
previously explored. I found a disk+halo model able to provide a
SED reasonably close to the observed and the required scalelength ratio.
However, the halo would have had a FIR emission easily detectable
in the available FIR and sub-mm observations, which it is not.
The limited number of models explored obviously does not rule out
the presence of a dust halo of lower density. Indeed, an optically
thin halo would be currently undetected. However, it will not be able
to explain the observed FIR properties, that are dominated by the disk
distribution (Sect.~\ref{halo}).

In this Thesis I have assumed that the scalelength of the intrinsic
stellar distribution does not change as a function of the wavelength,
dust extinction being responsible for the shallower profile observed in
the optical bands. However, different scalelengths at different $\lambda$ 
may be caused by intrinsic colour variation of the stellar distribution 
with the distance from the galactic centre. Bluer light at larger
distances from the centre may trace a population of stars younger than in
the central part of the disk.
\citet{PeletierA&A1995} studied the variation of the
ratio of B and K-band disk scalelengths with inclination
for a sample of 37 Sb-Sc galaxies. Larger scalelength ratios are
expected in edge-on galaxies, with respect to face-on objects, if dust
is responsible of the observed colour gradient. On the other hand, if
the colour gradient is caused by a gradient in the stellar population,
the ratio should remain constant, regardless of the galaxy inclination.
They find that $\alpha_B/\alpha_K$ goes from 1.3 for face-on galaxies to
1.7 for edge-on, while a change in the stellar population, estimated
from the observed metallicity gradients, can produce a
ratio of 1.17 only. These results suggest that dust extinction is the
main contributor to the observed colour gradients.
An opposite conclusion is drawn by \citet{DeJongA&A1996b}, on a sample
of 86 face-on galaxies. Comparing the observed colour gradients in 
colour-colour plots with those derived from a Monte Carlo radiative
transfer model, he concludes that, for {\em reasonable} models, the observed 
ratio between optical and NIR scalelengths is caused by a change in stellar 
population. This definition does not seem to include  models with dust disks 
more extended that the stellar, that are shown to produce such large
gradients as the observed. This may be considered a further hint for the
existence of extended dust. 

Because of the uncertainty in the intrinsic stellar colour gradient,
it is difficult to evaluate its effect in the present modelling. 
The stellar output is calibrated on the observed SED inside the
half-light radius. If the shallower B profile is caused mainly by a
change in stellar population, the SED at larger distance may be stronger
in the short-wavelength side than the central one. Since extinction is
larger for small $\lambda$, a smaller optical depth will be sufficient 
in the external part to produce the same FIR output. However, the
effect is likely to be small, because most of the absorbed radiation 
comes from the higher extinction regions inside the half-light radius,
for which I use the correct stellar SED.

The mean dust properties derived in the Local Galactic environment have 
then been used for any position in the dust distribution. Dust grain
composition and size distributions may vary along the disk. Given the
high uncertainties on the properties of the local dust distribution
itself, it is not easy to assess the influence of such variation on the
model. \citet{DaviesMNRAS1998} constructed a numerical model to study
the expulsion and retention of dust grains in galactic disks, as a
result of the radiation pressure, gravitational potential and friction
with the gas. They conclude that larger grains (0.1$\mu$m) are likely to 
move from the outer galaxy to the centre, for reasonable disk opacities.
On the contrary, smaller grains (0.01-0.001$\mu$m) remain relatively
close to their formation sites. A reservoir of large grains would
therefore accumulate in the central regions.
Smaller grains can be heated to higher temperatures, for the same ISRF 
\citep{DraineApJ1984}. An overabundance of smaller grains at larger radii 
with respect to the centre would result therefore
in a shallower temperature gradient and FIR profiles. However, the
effect is likely to be small and it is difficult that it can account for
the observed large FIR scalelengths.

As already stated, an extended distribution of dust is needed to explain
the FIR scalelengths. If this distribution is confirmed, our
observations of the universe may be severely affected by dust
extinction. \citep{AltonSub1999} assessed the impact of this hypothesis,
with a universe populated by galaxies with dust distributed as in
a model of NGC~6946. The chosen model is the same as in
Sect.~\ref{twodisks}, with a standard optically thick disk derived from
the H$_2$ distribution and an extended optically thin disk associated to
the atomic gas. The fraction of light emitted at redshift $z$ that fails
to reach instruments observing in the B-band, as a result of the
intervening disks extinction, is then computed. It is found that
30-40\% of the light emitted at $z=2$ would fail to reach us.
Most of the extinction would be due to the extended optically thin disk, 
rather than the one associated with the molecular gas. This because the
geometric cross section (9 times bigger for the dust disk associated with
the atomic gas) is a more critical parameter than the central optical
depth. Therefore the result will hold even if the molecular gas is
distributed in clumps, thus reducing its extinction.
However, note the authors caveat on NGC~6946 being very gas rich. When dust
modelled on the gas distributions and gas-to-dust ratio of the two
separate phases of NGC~891 is used, only a fraction of 5\% of the light 
will not reach us, for the same redshift of emission as before.

Clearly, the knowledge of the dust distribution in a spiral galaxy is
still at its preliminary stages. More work need to be done, both on the
observational and theoretical side. From the discussion above, it
emerges that the inclusion of clumping in proper radiative transfer and
emission model is necessary. This will clarify the impact of ISM and
stellar dishomogenities of the FIR emission and solve the debated
problem of the source of the dust heating. However, it has been shown
that clumpy models depend heavily on the assumptions about the clumps
distribution. Future high resolution and sensitivity instrumentation will
therefore be essential to define the dust distribution and limit the
number of parameters in the model.

For this Thesis, I have tried to fit the FIR emission of one galaxy
only, to assess the feasibility of the method and test the model.
However, NGC~6946 may have characteristics different from those of
a `mean' galaxy. Our group possess a lot of optical and FIR data
for several spirals \citep{TrewhellaThesis1998}. Furthermore,
new FIR data is becoming available through the ISO mission Data
Archive. When this Thesis analysis has been conducted on other
spirals, more general conclusions about the dust distribution and 
its extinction could be drawn.

\section{Summary of the Thesis}
In this Thesis I have devised a model to simulate the FIR emission in
spiral galaxies and I have applied it to the observations of NGC~6946.
An introduction to the influence of dust in astrophysical observations,
a review of the studies of extinction and FIR emission in spiral
galaxies, and a list of the observational evidences for extended
dust distribution is presented in Chapter~1. The basic dust properties
used in the modelling are given in Chapter~2. The model is described
in Chapter~3, with details on the adopted stellar and dust geometries
and on the implementation of the radiative transfer. Chapter~4
is devoted to a fit of the observed FIR properties of NGC~6946.
Conclusions have been drawn in this Chapter.
\vskip 0.4cm

In the following list I summarize the major works done for this Thesis,
together with a few other ancillary projects carried out during my PhD:
\begin{itemize}

\item Modification of a Monte Carlo radiative transfer code previously
developed by myself and creation of FIR emission model for spiral
galaxies (Chapter~\ref{program} and \ref{model}). The main features of
the model are:
\begin{itemize}
\item full treatment of multiple scattering in realistic geometries;
\item use of empirically derived dust properties;
\item derivation of the absorbed energy and temperature from an ISRF
consistent with the radiative transfer;
\item creation of maps of FIR emission and dust temperature distribution.
\end{itemize}

\item Application of the model to NGC~6946 (Chapter~\ref{model}). Main
findings are:
\begin{itemize}
\item observed temperatures are consistent with ISRF heating;
\item optically thick disks are necessary to explain the observed SED of
FIR emission;
\item optical-FIR scalelengths ratio can be explained by extended disks;
\item a model with face-on optical depth $\tau_\mathrm{V}=3$ and spatial
distribution similar to that of the atomic gas provide a good fit to
both the observed SED and the optical-FIR scalelengths ratio.
\end{itemize}

\item Original derivation of dust emissivity from maps of Galactic extinction 
and FIR emission \citep[Sect.~\ref{qemi_new};][]{BianchiA&A1999}.

\item SCUBA observations of dust associated with spiral arms and diffuse
dust in NGC~6946  (Sect.~\ref{scuobs})

\item SCUBA observations of the dust ring in NGC~7331
\citep[Appendix~\ref{n7331};][]{BianchiMNRAS1998}.

\item INT observations and analysis of fields around edge-on galaxies,
to detect extinction of a possible dusty halo through the colour of
background objects (Appendix~\ref{unlucky}); discovery of a faint luminous 
halo in NGC~891 (Sect.~\ref{lumhalo}).

\end{itemize}

\appendix
\chapter{SCUBA imaging of NGC 7331 dust ring}
\label{n7331}

To reduce the size of this file, this section has been omitted.
However, its content has been published integrally in

Bianchi S., Alton P.~B., Davies J.~I., Trewhella M., 1998, MNRAS,
298, L49.

\chapter{Search for dust in the halos of spiral galaxies}
\setcounter{page}{86}
\label{unlucky}

To reduce the size of this file, this section has been omitted.
The full version of this thesis, including the present appendix,
can be found at

{\tt http://www.arcetri.astro.it/$^\sim$sbianchi/tesi/thesis.ps.gz}

\setcounter{section}{3}
\setcounter{subsection}{2}

~\label{lumhalo}
\newpage 
\thispagestyle{empty}
~\newpage
\setcounter{page}{98}

\end{document}